\documentclass[%
reprint,
amsmath,amssymb,
aps,
prx,
]{revtex4-2}

\usepackage[utf8]{inputenc}
\usepackage[T1]{fontenc}
\usepackage{lmodern}
\usepackage{mathrsfs}
\usepackage[english]{babel}
\usepackage[autostyle=true]{csquotes}

\usepackage{graphicx}
\graphicspath{{figures/}}
\usepackage{dcolumn}
\usepackage{bm}
\usepackage{hyperref}
\hypersetup{
	pdftitle={From non-Hermitian linear response to dynamical correlations and fluctuation-dissipation relations in quantum many-body systems},
	pdfauthor={Kevin T. Geier, Philipp Hauke},
}

\usepackage[dvipsnames]{xcolor}
\usepackage{lipsum}
\usepackage[separate-uncertainty=false]{siunitx}
\usepackage{mathtools}
\usepackage{dsfont}
\usepackage{braket}
\usepackage[caption=false]{subfig}

\usepackage[
capitalise,
]{cleveref}

\usepackage{acro}
\acsetup{%
	first-style=long-short,
}

\DeclareAcronym{bec}{%
	short = BEC,
	long = Bose--Einstein condensate,
}

\DeclareAcronym{fdt}{%
	short = FDT,
	long = fluctuation--dissipation theorem,
	short-indefinite = an,
}

\DeclareAcronym{fdr}{%
	short = FDR,
	long = fluctuation--dissipation relation,
	short-indefinite = an,
}

\DeclareAcronym{lrt}{%
	short = LRT,
	long = linear response theory,
}

\DeclareAcronym{bh}{%
	short = BH,
	long = Bose--Hubbard,
}

\DeclareAcronym{mi}{%
	short = MI,
	long = Mott insulator,
	short-indefinite = an,
}

\DeclareAcronym{sf}{%
	short = SF,
	long = superfluid,
	short-indefinite = an,
}

\DeclareAcronym{sde}{%
	short = SDE,
	long = stochastic differential equation,
	short-indefinite = an,
}

\DeclareAcronym{1d}{%
	short = 1D,
	long = one-dimensional,
}

\DeclareAcronym{2d}{%
	short = 2D,
	long = two-dimensional,
}

\DeclareAcronym{3d}{%
	short = 3D,
	long = three-dimensional,
}

\newcommand{\mean}[1]{\overline{#1}}

\newcommand{\diff}{\mathop{}\!\mathrm{d}}
\newcommand{\etothepowerof}[1]{\mathrm{e}^{#1}}
\newcommand{\kb}{k_{\mathrm{B}}}
\newcommand{\commutator}[3][]{#1[ #2, #3 #1]}
\newcommand{\Commutator}[2]{\left[#1, #2\right]}
\newcommand{\anticommutator}[3][]{#1\{ #2, #3 #1\}}
\newcommand{\Anticommutator}[2]{\left\{#1, #2\right\}}
\DeclareMathOperator{\Tr}{Tr}
\newcommand{\doublestruck}[1]{\mathds{#1}}

\DeclareMathOperator*{\argmin}{arg\,min}

\DeclareMathOperator{\sinc}{sinc}

\newcommand\ketbra[2]{{|{#1}\rangle}{\langle{#2}|}}
\newcommand{\waitingtime}{\ensuremath{t_{\mathrm{w}}}}
\newcommand{\finaltime}{\ensuremath{t_{\mathrm{f}}}}
\newcommand{\ensembleaverage}[1]{\ensuremath{\langle\!\langle #1 \rangle\!\rangle}}
\newcommand{\pprojector}{\ensuremath{\mathcal{P}}}
\newcommand{\qprojector}{\ensuremath{\mathcal{Q}}}
\newcommand{\cpl}{\ensuremath{{\mathrm{cpl}}}}
\newcommand{\sys}{\ensuremath{{\mathrm{S}}}}
\newcommand{\anc}{\ensuremath{{\mathrm{A}}}}
\newcommand{\eff}{\ensuremath{\mathrm{eff}}}
\newcommand{\chiprimenh}{\ensuremath{\chi^{\prime \, \mathrm{(NH)}}}}
\newcommand{\chitwoprime}{\ensuremath{\chi^{\prime \prime}}}

\begin{document}

\title{From non-Hermitian linear response to dynamical correlations and fluctuation--dissipation relations in quantum many-body systems}

\author{Kevin T. Geier}
\email[]{kevinthomas.geier@unitn.it}
\affiliation{INO-CNR BEC Center and Dipartimento di Fisica, Universit\`a di Trento, 38123 Povo, Italy} 
\affiliation{Institute for Theoretical Physics, Ruprecht-Karls-Universität Heidelberg, Philosophenweg 16, 69120 Heidelberg, Germany}
\affiliation{Kirchhoff Institute for Physics, Ruprecht-Karls-Universität Heidelberg, Im Neuenheimer Feld 227, 69120 Heidelberg, Germany}

\author{Philipp Hauke}
%\email[]{philipp.hauke@unitn.it}
%\homepage[]{https://hauke-group.physics.unitn.it}
\affiliation{INO-CNR BEC Center and Dipartimento di Fisica, Universit\`a di Trento, 38123 Povo, Italy}
\affiliation{Institute for Theoretical Physics, Ruprecht-Karls-Universität Heidelberg, Philosophenweg 16, 69120 Heidelberg, Germany}
\affiliation{Kirchhoff Institute for Physics, Ruprecht-Karls-Universität Heidelberg, Im Neuenheimer Feld 227, 69120 Heidelberg, Germany}

\date{\today}

\begin{abstract}
Quantum many-body systems are characterized by their correlations.
While equal-time correlators and unequal-time commutators between operators are standard observables, the direct access to unequal-time \emph{anti-commutators} poses a formidable experimental challenge.
Here, we propose a general technique for measuring unequal-time anti-commutators using the linear response of a system to a non-Hermitian perturbation.
We illustrate the protocol at the example of a Bose--Hubbard model, where the approach to thermal equilibrium in a closed quantum system can be tracked by measuring both sides of the fluctuation--dissipation relation.
We relate the scheme to the quantum Zeno effect and weak measurements, and illustrate possible implementations at the example of a cold-atom system.
Our proposal provides a way of characterizing dynamical correlations in quantum many-body systems with potential applications in understanding strongly correlated matter as well as for novel quantum technologies.
\end{abstract}

%\keywords{dynamical correlations, fluctuation--dissipation relations, linear response theory, non-Hermitian physics, quantum Zeno effect, engineered dissipation}

\maketitle

\section{\label{sec:introduction}Introduction}

Dynamical correlations involving observables at unequal times encode many fundamental properties of quantum many-body systems.
They are at the basis of ubiquitous phenomena ranging from
optical coherence~\cite{Glauber1963,Scully1997} and transport phenomena~\cite{Jensen1991,Coleman2015},
over far-from-equilibrium universality~\cite{Berges2015,PineiroOrioli2015,Schmied2019,Boguslavski2020,Chatrchyan2021}, glassy dynamics and aging~\cite{Sciolla2015,Halimeh2021,Arceri2020},
as well as dynamical topological transitions~\cite{Zache2019},
to thermalization, integrability, and quantum chaos~\cite{Deutsch1991,Srednicki1994,Gogolin2016,DAlessio2016,Deutsch2018}.
Historically, a groundbreaking role has been played by the \ac{fdr}~\cite{Callen1951,Kubo1957,Kubo1966}, which can be viewed as a generalization of the famous Einstein relation for Brownian motion~\cite{Einstein1905}. In essence, the \ac{fdr} connects unequal-time anti-commutators and commutators: in thermal equilibrium, fluctuations of an observable at any given frequency are intrinsically connected with the energy dissipated when the system is perturbed at that same frequency.
As it is governed by a single global parameter, the temperature, the \ac{fdr} is an excellent probe for thermalization of closed quantum systems~\cite{Foini2011,Foini2012,Torre2013,Khatami2013,Lenarcic2014,Rossini2014,PineiroOrioli2019,Schuckert2020}.
Certifying that a given quantum state is thermal can also be valuable in novel quantum technologies, e.g., for applying dynamical protocols to detect entanglement~\cite{Hauke2016,Brenes2020,CostadeAlmeida2021} --- a key resource for quantum-enhanced metrology~\cite{Degen2017,Pezze2018}.

Notwithstanding its fundamental importance, both sides of the \ac{fdr} have thus far only been measured for classical systems~\cite{Grigera1999,Bellon2002,Netz2018}.
For quantum systems, only one side, the unequal-time commutator, is easily accessible thanks to Kubo's celebrated linear response theory~\cite{Kubo1966,Coleman2015}, which has been extensively used to characterize quantum systems out of equilibrium~\cite{Foini2011,Foini2012,Torre2013,Khatami2013,Lenarcic2014,Rossini2014,PineiroOrioli2019,Schuckert2020}.
A main difficulty regarding the measurement of dynamical correlations stems from the fact that a projective von Neumann measurement at a particular time collapses the quantum state~\cite{Neumann1932}, which prevents an unperturbed measurement at a later time and thus hinders a measurement of the time--time correlation with respect to the initial state.
Various pioneering proposals for measuring unequal-time correlations on various platforms exist~\cite{RomeroIsart2012,Knap2013,Pedernales2014,Uhrich2017,Kastner2018,Uhrich2019,Roggero2019,Yang2020,Castrignano2020,Schuckert2020}, but attempts to overcome the inherent difficulties of such a measurement are often specific to certain setups or apply only to a limited set of observables. As of today, an experimental observation of the unequal-time anti-commutator in a quantum many-body system remains elusive.

\begin{figure}
	\subfloat{\label{fig:fdr:response}\includegraphics[width=\columnwidth]{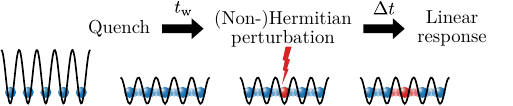}}\\
	\vspace{5pt}%
	\subfloat{\label{fig:fdr:spectrum}}%
	\subfloat{\label{fig:fdr:earlylate}\includegraphics[width=\columnwidth]{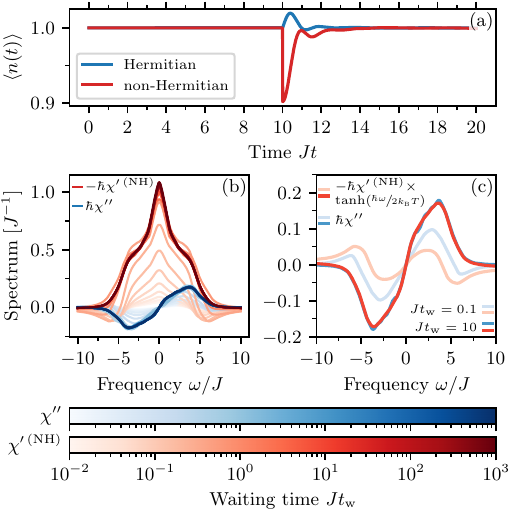}}%
	\caption{\label{fig:fdr}%
		(Non\protect\nobreakdash-)Hermitian linear response protocol for measuring \protect\acfp{fdr}, exemplified for a Bose--Hubbard chain.
		(a)~Schematic illustration of the protocol and response of the density $\braket{n(t)}$ to an (anti\protect\nobreakdash-)Hermitian perturbation $H_1(t) = -(i) \hbar s \delta(t - \waitingtime) n$ of strength ${s = \num{0.05}}$, applied at the waiting time ${J \waitingtime = \num{10}}$.
		(b)~Thermalization dynamics of the dissipative part of the \enquote{Hermitian} dynamic susceptibility~$\chi_{n n}^{\prime \prime}(\waitingtime, \omega)$ (\enquote{commutator}) and the reactive part of the \enquote{non-Hermitian} dynamic susceptibility~$\chi_{n n}^{\mathrm{\prime \, (NH)}}(\waitingtime, \omega)$ (\enquote{anti-commutator}).
		(c)~Dynamic susceptibilities, rescaled according to the \ac{fdr}~\protect\eqref{eq:fdrsusceptibility} at early and late waiting times.
		The effective temperatures~$\kb T / \hbar J = \{ \num{4.5}, \num{4.2} \}$ for $J \waitingtime = \{ \num{0.1}, \num{10} \}$, respectively, are determined by \protect\cref{eq:temperatureleastsquares} using the least-squares method.
		The \ac{fdr} is clearly violated at early times, but it is restored at late times when the system has thermalized.%
	}
\end{figure}

Here, we discuss how a linear response to a non-Hermitian perturbation~\cite{Pan2020,Sticlet2022} permits direct experimental observation of the unequal-time anti-commutator.
Combined with a traditional method for measuring the corresponding unequal-time commutator, e.g., standard linear response, this scheme gives access to both sides of the \ac{fdr} independently, allowing one to track a system's evolution towards thermal equilibrium.
We illustrate this possibility by means of numerical simulations at an example motivated by a ground-breaking cold-atom experiment~\cite{Kaufman2016} --- a Bose--Hubbard system that is quenched from a Mott-insulating initial state to the superfluid phase (see \cref{fig:fdr}). 
This analysis provides a blueprint for revealing the \ac{fdr} using experimental abilities that are common in state-of-the-art engineered quantum systems. 

The key to measuring unequal-time anti-commutators is the ability to engineer (effective) non-Hermitian perturbations.
In recent years, a tremendous interest in non-Hermitian physics has emerged~\cite{ElGanainy2018,Ashida2020}, stimulated by the rapid progress in the experimental generation and control of non-Hermitian systems~\cite{Guo2009,Rueter2010,Naghiloo2019,Cao2020,Chen2021,Oeztuerk2021}.
Indeed, non-Hermiticity gives rise to a wealth of new physics with novel (topological) phases and unconventional critical behavior~\cite{Lee2014,ParraMurillo2017,Ashida2017,Yao2018,Nakagawa2018,Hamazaki2019,Nakagawa2020}, bearing a vast potential for applications, e.g., in strongly enhanced quantum sensing~\cite{Wiersig2020,Wiersig2020a} or adiabatic quantum optimization~\cite{Berman2009,Nesterov2012}.
Leveraging on this development, we design a specific protocol to generate effective non-Hermitian dynamics in a system of cold atoms, enabling access to the fluctuation side of the \ac{fdr} (i.e., to the unequal-time anti-commutator).
Our scheme is most conveniently phrased as an application of the quantum Zeno effect~\cite{Misra1977,Facchi2008}, combining outcoupling to an ancillary system with a projection on the Zeno subspace given by the empty ancilla.
In a cold-atom implementation, this can be realized through a coherent or dissipative perturbation in the linear regime, together with the ability of distinguishing zero from non-vanishing ancilla population in post-selection.
While a single step in the Zeno evolution yields the unequal-time anti-commutator in time domain, an extended Zeno evolution, which we propose to implement harnessing engineered dissipation~\cite{Stannigel2014,Schaefer2020}, allows one to probe frequency-resolved responses in the same way as in standard linear response experiments.
To demonstrate the feasibility of our proposal, we benchmark our protocol by numerically solving the full quantum evolution, including the stochastic dynamics underlying the dissipative scheme, and discuss experimental error sources.
We also examine formal relations to dissipative quantum systems, where non-Hermitian dynamics can be generated by post-selecting individual quantum trajectories on the absence of quantum jumps~\cite{Dalibard1992,Molmer1993,Daley2014,Nakagawa2020,Chen2021}, and establish general cross-connections between (non\nobreakdash-)Hermitian linear response and ancilla-based weak measurements of dynamical correlations~\cite{Uhrich2017,Kastner2018} (see Ref.~\cite{Svensson2013} for a comprehensive review on weak measurements).
Our proposed realization of non-Hermitian linear response is feasible even when existing weak measurement protocols are difficult to engineer experimentally, and it excels in regimes where projective protocols fail as a consequence of their restriction to observables with two eigenvalues~\cite{Knap2013,Uhrich2017,Kastner2018,Uhrich2019,Schuckert2020}, as we demonstrate through numerical benchmarks.
Our approach thus opens the door to probing the \ac{fdr} in quantum many-body systems in an unbiased way and for a broad range of observables.

\section{\label{sec:fdr}The fluctuation--dissipation relation}

For a quantum many-body system in thermal equilibrium, the \ac{fdr}~\cite{Kubo1966} links the symmetrized correlation spectrum $S_{BA}(\omega)$ of any two operators $A$ and $B$ across the entire frequency spectrum~$\omega$ to the dissipative part of the dynamic susceptibility $\chi_{BA}^{\prime \prime}(\omega)$ via
\begin{equation}
	\label{eq:FDR}
	S_{BA}(\omega) = \hbar \coth \left( \frac{\hbar \omega}{2 \kb T} \right) \chi_{BA}^{\prime \prime}(\omega) ,
\end{equation}
where $\hbar$ is the reduced Planck constant and $\kb$ is the Boltzmann constant.
This elegant relation requires only a single parameter as input, the global temperature~$T$.
The ease of accessing $\chi_{BA}^{\prime \prime}(\omega)$ can then be exploited to obtain $S_{BA}(\omega)$.
However, when the system is far from equilibrium, the two sides of \cref{eq:FDR} become non-stationary and the \ac{fdr} can be broken~\cite{PineiroOrioli2019,Boguslavski2020}, making it necessary to devise independent handles on both sides of the relation, as has been proposed in Ref.~\cite{Schuckert2020}.
	
One can generalize the definitions of $S$ and $\chi^{\prime \prime}$ to such a non-equilibrium situation by introducing the response function
\begin{equation}
	\label{eq:responsefunctionhermitian}
	\phi_{BA}(t, t^\prime) = \frac{i}{\hbar} \theta(t - t^\prime) \Braket{\Commutator{B(t)}{A(t^\prime)}}_0
\end{equation}
and the symmetrized dynamic correlation function
\begin{equation}
	\label{eq:symmetrizedcorrelationfunction}
	S_{BA}(t, t^\prime) = \frac{1}{2} \Braket{\Anticommutator{B(t)}{A(t^\prime)}}_0 - \Braket{B(t)}_0 \Braket{A(t^\prime)}_0 ,
\end{equation}
defined, respectively, in terms of the unequal-time commutator and anti-commutator of the Heisenberg operators $A(t)$ and $B(t)$.
Here, $\theta(t)$ is the Heaviside step function, ensuring causality of the response, and the subscript in the expectation value~$\braket{\cdots}_0$ signifies that the Heisenberg operators evolve under the (unperturbed) Hamiltonian~$H_0$.
In the context of non-equilibrium quantum field theory, \cref{eq:responsefunctionhermitian,eq:symmetrizedcorrelationfunction} are also known as the spectral function~$\rho$ and the statistical function~$F$, respectively~\cite{Aarts2001}.
\Cref{eq:responsefunctionhermitian} is the non-equilibrium version of Kubo's well-known linear response function~\cite{Kubo1966},
which determines the evolution of the expectation value $\braket{B(t)}$ under the perturbed Hamiltonian $H(t) = H_0 + H_1(t)$, to linear order in the perturbation $H_1(t) = -f(t) A$, according to
\begin{equation}
	\label{eq:linearresponse}
	\Braket{B(t)} = \Braket{B(t)}_0 + \int_{0}^{t} \diff t^\prime \, \phi_{BA}(t, t^\prime) f(t^\prime) .
\end{equation}

In contrast to the usual equilibrium linear response scenario, the initial state is not necessarily stationary with respect to $H_0$.
In this situation, it is common to define the non-equilibrium generalization of the dynamic susceptibility $\chi$ as the Fourier transform of the response function $\phi$ with respect to the relative time~$\Delta t = t - t^\prime$ at fixed central time~$\tau = (t + t^\prime) / 2$~\cite{Aarts2001},
\begin{equation}
	\label{eq:susceptibility}
	\chi_{BA}(\tau, \omega) = \int_{-2 \tau}^{2 \tau} \diff \Delta t \, \phi_{BA} \bigg( \tau + \frac{\Delta t}{2}, \tau - \frac{\Delta t}{2} \bigg) \etothepowerof{i \omega \Delta t} .
\end{equation}
This quantity is commonly decomposed as $\chi_{BA} = \chi_{BA}^\prime + i \chi_{BA}^{\prime \prime}$ into a reactive part $\chi_{BA}^\prime(\omega) = [ \chi_{BA}(\omega) + \chi_{AB}(-\omega)] / 2$ and a dissipative (or absorptive) part $\chi_{BA}^{\prime \prime}(\omega) = [ \chi_{BA}(\omega)- \chi_{AB}(-\omega) ] / 2i$~\cite{Jensen1991}, the latter entering the right-hand side of the \ac{fdr}~\eqref{eq:FDR}.
The correlation spectrum $S(\tau, \omega)$ on its left-hand side can be defined analogously to \cref{eq:susceptibility} as the Fourier transform of \cref{eq:symmetrizedcorrelationfunction}.

For a thermalizing system, we expect $\chi_{BA}(\tau \to \infty, \omega)$ and $S_{BA}(\tau \to \infty, \omega)$ to reach steady values that fulfill the FDR. As such, the restoration of the FDR provides an excellent probe for how and when a quantum many-body system approaches thermal equilibrium~\cite{Schuckert2020}.
On top of that, the \ac{fdr} yields the effective temperature at which the system thermalizes~\cite{Foini2011,Foini2012,Torre2013,Khatami2013,Lenarcic2014,Rossini2014,PineiroOrioli2019,Schuckert2020}.
Remarkably, this independent way of defining temperature does not require any \textit{a priori} assumptions other than the \ac{fdr}.
In our numerical benchmarks, we find good agreement between the effective temperature extracted from the \ac{fdr} and the expected temperature of a thermal ensemble at the equivalent energy density (see \cref{sec:illustration}).

While the commutator in \cref{eq:responsefunctionhermitian} can be accessed rather straightforwardly, for example, by studying how energy is absorbed or how an observable deviates from its equilibrium value following a time-dependent perturbation~\cite{Kubo1966}, the determination of the unequal-time anti-commutator in \cref{eq:symmetrizedcorrelationfunction} is, unfortunately, considerably more challenging.
We now employ a recent extension of linear response theory to non-Hermitian Hamiltonians~\cite{Pan2020,Sticlet2022} as a general way of gaining access to the left-hand side of \cref{eq:FDR}, which enables direct probes of the \ac{fdr}.

\section{\label{sec:nhlrt}Non-Hermitian linear response theory}

Though long established in the context of open quantum systems~\cite{Gardiner2004,Breuer2007}, recent years have seen a surge of interest in quantum systems with non-Hermitian Hamiltonians~\cite{Ashida2020}. Here, we tap into this development by exploiting the linear response to a non-Hermitian perturbation~\cite{Pan2020,Sticlet2022} in order to extract unequal-time anti-commutators.

In contrast to usual linear response theory, we assume that the system is effectively described by a non-Hermitian Hamiltonian~$H(t)= H_0 + H_1(t)$, where $H_0$ is the unperturbed (Hermitian) Hamiltonian and $H_1(t) = - i f(t) A$ is an anti-Hermitian perturbation with a positive semi-definite operator~$A$ and a non-negative time-dependent function~$f(t)$.
For example, such a scenario arises in the quantum trajectories approach to dissipative quantum systems~\cite{Dalibard1992,Molmer1993,Daley2014} if the evolution is conditioned on the absence of quantum jumps~\cite{Naghiloo2019,Nakagawa2020,Chen2021} (see also \cref{sec:comparison}).
In addition, we show in \cref{sec:weakmeasurement} that existing ancilla-based weak measurement protocols for the unequal-time anti-commutator~\cite{Uhrich2017,Kastner2018} can also be rephrased in the framework of non-Hermitian linear response.
In \cref{sec:realization}, we present a scheme for engineering effective non-Hermitian Hamiltonians based on the quantum Zeno effect to probe such responses, even frequency-resolved, for a wide range of observables.

A quantum state described by the density operator~$\rho(t)$ evolves in time under the non-Hermitian Hamiltonian~$H(t)$ according to the von Neumann equation
\begin{equation}
	\label{eq:vonneumannnonhermitian}
	i \hbar \frac{\diff}{\diff t} \rho = H(t) \rho - \rho H^\dagger(t) = \Commutator{H_0}{\rho} + \Anticommutator{H_1(t)}{\rho}
\end{equation}
with initial condition $\rho(0) = \rho_0$.

Using time-dependent perturbation theory, a straightforward calculation (reported in \cref{app:nhlrt}) shows that, to linear order in the perturbation, the unnormalized expecation value of a (Hermitian) observable~$B$ is given by
\begin{equation}
	\label{eq:linearresponsenonhermitianunnormalized}
	\Tr \left[ B \rho(t) \right] = \Braket{B(t)}_0 - \frac{1}{\hbar} \int_{0}^{t} \diff t^\prime \Braket{\Anticommutator{B(t)}{A(t^\prime)}}_0 f(t^\prime) .
\end{equation}
The non-Hermiticity of the perturbed Hamiltonian has the important consequence that the state $\rho(t)$ is no longer normalized: as can be seen by inserting the identity operator for $B$ in \cref{eq:linearresponsenonhermitianunnormalized}, its norm decreases with time, to linear order, as
\begin{equation}
	\label{eq:normdecrease}
	\Tr \left[ \rho(t) \right] = 1 - \frac{2}{\hbar} \int_{0}^{t} \diff t^\prime \Braket{A(t^\prime)}_0 f(t^\prime) .
\end{equation}
Physically, this decrease can be interpreted as the leakage of the wave function into a complementary state space (see also \cref{sec:realization}).
To account for this loss of probability, we consider the normalized expectation value $\braket{B(t)} = \Tr [ B \rho(t) ] / \Tr [\rho(t)]$, describing a post-selected measurement~\cite{Svensson2013}.
Combining \cref{eq:linearresponsenonhermitianunnormalized,eq:normdecrease}, the disconnected correlations drop out to linear order,
and we can write the response as
\begin{equation}
	\label{eq:linearresponsenonhermitian}
	\Braket{B(t)} = \Braket{B(t)}_0 + \int_{0}^{t} \diff t^\prime \, \phi_{BA}^{\mathrm{(NH)}}(t, t^\prime) f(t^\prime)
\end{equation}
with the \enquote{non-Hermitian} response function
\begin{equation}
	\label{eq:responsefunctionnonhermitian}
\begin{split}
	\phi_{BA}^{\mathrm{(NH)}}(t, t^\prime) = -\frac{1}{\hbar} \theta(t - t^\prime) \Big[ &\Braket{\Anticommutator{B(t)}{A(t^\prime)}}_0 \\
	&- 2 \Braket{B(t)}_0 \Braket{A(t^\prime)}_0 \Big] .
\end{split}
\end{equation}
Here, we insert the Heaviside step function~$\theta$ to ensure causality of the response.
Remarkably, the non-Hermitian response function in \cref{eq:responsefunctionnonhermitian} is the desired measurable quantity that gives direct access to the unequal-time anti-commutator~\eqref{eq:symmetrizedcorrelationfunction} by virtue of the relation $\phi_{BA}^{\mathrm{(NH)}}(t, t^\prime) = -2 \theta(t - t^\prime) S_{BA}(t, t^\prime) / \hbar$.

To establish a link between the response function~\eqref{eq:responsefunctionnonhermitian} and the correlation spectrum appearing on the left-hand side of the \ac{fdr}~\eqref{eq:FDR},
we define the \enquote{non-Hermitian} dynamic susceptibility, similarly to \cref{eq:susceptibility}, as the Fourier transform
\begin{equation}
	\label{eq:susceptibilityfixedcentraltime}
	\chi_{BA}^{\mathrm{(NH)}}(\tau, \omega) = \int_{-2 \tau}^{2 \tau} \diff \Delta t \, \phi_{BA}^{\mathrm{(NH)}} \left( \tau + \frac{\Delta t}{2}, \tau - \frac{\Delta t}{2} \right) \etothepowerof{i \omega \Delta t} .
\end{equation}
We can split this quantity as $\chi_{BA}^{\mathrm{(NH)}} = \chi_{BA}^{\prime \, \mathrm{(NH)}} + i \chi_{BA}^{\prime \prime \, \mathrm{(NH)}}$ into the components (for conciseness, we remove the $\tau$~argument from the following formulas)
\begin{subequations}
	\label{eq:susceptibilitycomponents}
\begin{align}
	\label{eq:susceptibilityreactive}
	\chi_{BA}^{\prime \, \mathrm{(NH)}}(\omega) &= \frac{1}{2} \left[ \chi_{BA}^{\mathrm{(NH)}}(\omega) + \chi_{AB}^{\mathrm{(NH)}}(-\omega) \right] , \\
	\label{eq:susceptibilitydissipative}
	\chi_{BA}^{\prime \prime \, \mathrm{(NH)}}(\omega) &= \frac{1}{2i} \left[ \chi_{BA}^{\mathrm{(NH)}}(\omega) - \chi_{AB}^{\mathrm{(NH)}}(-\omega) \right] ,
\end{align}
\end{subequations}
which we refer to, in analogy to their Hermitian counterparts, as the reactive and dissipative parts of the non-Hermitian susceptibility, respectively.
As shown in \cref{app:nhlrt},
the reactive part, \cref{eq:susceptibilityreactive}, gives access to the correlation spectrum via the identity $S_{BA}(\tau, \omega) = - \hbar \chi_{BA}^{\mathrm{\prime \, (NH)}}(\tau, \omega)$.
This allows us to rewrite the \ac{fdr}~\eqref{eq:FDR} in thermal equilibrium as
\begin{equation}
	\label{eq:fdrsusceptibility}
	\chi_{BA}^{\mathrm{\prime \, (NH)}}(\omega) = - \coth \left( \frac{\hbar \omega}{2 \kb T} \right) \chi_{BA}^{\prime \prime}(\omega) ,
\end{equation}
which is expressed entirely in terms of the susceptibilities $\chi_{BA}^{\mathrm{\prime \, (NH)}}$ and $\chi_{BA}^{\prime \prime}$, accessible using non-Hermitian and standard (Hermitian) linear response, respectively.
As such, linear response theory provides an elegant and general framework for independently probing both sides of the \ac{fdr}~\eqref{eq:fdrsusceptibility} out of equilibrium, which works for arbitrary observables in any quantum many-body system.
Compared to projective protocols for measuring unequal-time anti-commutators~\cite{Knap2013,Uhrich2017,Kastner2018,Uhrich2019,Schuckert2020}, which are restricted to observables with two eigenvalues (see also discussion in \cref{sec:discussion:projectiveprotocols}), one of the main assets of linear response theory is its broad applicability.

It is worthwhile emphasizing that the outlined derivation of the response to a non-Hermitian perturbation is by no means restricted to the linear regime only, but, as well known in standard response theory~\cite{Kubo1957}, can be extended to non-linear responses. In fact, by calculating the expansions in \cref{eq:linearresponsenonhermitianunnormalized,eq:normdecrease} to the desired non-linear order, one can in principle access an infinite hierarchy of unequal-time correlations, order by order.
By perturbing the system at multiple sites simultaneously, non-linear responses could therefore also enable access to (global) many-body operators which are expected not to thermalize and consequently violate the \ac{fdr}.

Approaching the problem of measuring dynamical correlations from the (non\nobreakdash-)Hermitian linear response perspective turns out to be fruitful for a number of reasons.
For one, non-Hermitian linear response is completely agnostic to the way the non-Hermitian perturbation is implemented and therefore directly benefits from any advancements in the field of non-Hermitian physics regarding the generation and control of non-Hermitian Hamiltonians.
Furthermore, it provides an ancilla-free interpretation of common ancilla-based weak measurement schemes for the unequal-time anti-commutator~\cite{Uhrich2017,Kastner2018}.
So far, it has not been clear whether ancilla-free formulations of such protocols allow for a meaningful physical interpretation~\cite{Kastner2018}, but, as we show in \cref{sec:weakmeasurement}, this is indeed possible in the light of non-Hermitian linear response.
Conversely, any non-Hermitian perturbation can in principle be realized with the help of an ancilla using only unitary evolution and standard projective measurements, although the required couplings may not always be straightforward to implement.
In \cref{sec:realization}, we present specific ancilla-based schemes with experimentally feasible system--ancilla couplings, providing access to dynamical correlations for a wide range of observables.
Finally, from a linear response point of view, it is natural to study responses to periodic perturbations that directly give access to frequency-resolved susceptibilities.
As explained in \cref{sec:pulsedzeno,sec:continuouszeno}, this becomes practical within our framework also for non-Hermitian perturbations by exploiting the quantum Zeno effect.

\section{\label{sec:illustration}Illustration: quench in a Bose--Hubbard system}

In this section, we demonstrate how (non\nobreakdash-)Hermitian linear response allows one to access both sides of the \ac{fdr}~\eqref{eq:fdrsusceptibility} independently. Such measurements can be used to either probe thermalization or the absence thereof~\cite{Schuckert2020}.
If a system of interest is coupled to a large thermal bath, it will sooner or later always approach thermal equilibrium with the bath temperature~\cite{Breuer2007}, and the \ac{fdr} will eventually hold.
In contrast, the question whether and how a \emph{closed} quantum system thermalizes once it is brought out of equilibrium is much more subtle.
Remarkably, an isolated system can act as its own bath~\cite{Gogolin2016,DAlessio2016}:
a thermalizing subsystem behaves after long times as if it was in thermal equilibrium with the rest of the system at an effective temperature set by the initial state.
According to the eigenstate thermalization hypothesis~\cite{Deutsch1991,Srednicki1994,Gogolin2016,DAlessio2016,Deutsch2018}, this process occurs on the level of individual eigenstates.
Although the precise conditions for its validity are not yet entirely understood, it is believed that (eigenstate) thermalization generally holds for generic states of interacting quantum many-body systems in the bulk of the spectrum.
Important scenarios where thermalization fails (with concomitant violation of the \ac{fdr}) include integrable models~\cite{Khatami2013,DAlessio2016}, many-body localization~\cite{Nandkishore2015,Abanin2019}, as well as Hilbert space fragmentation and the related phenomenon of quantum many-body scars~\cite{Sala2020,Serbyn2021,Regnault2022}.
On top of that, breaking \acp{fdr} is a characteristic signature of far-from-equilibrium systems near a non-thermal fixed point~\cite{PineiroOrioli2019,Boguslavski2020}.
All of these settings represent promising targets for our (non\nobreakdash-)Hermitian linear response scheme to reveal either the validity or the breakdown of the \ac{fdr}.

For illustrative purposes, we focus here on the generic case where the system does thermalize and the \ac{fdr} is expected to hold.
In ground-breaking cold-atom experiments, it has been shown that even in very small interacting quantum systems, expectation values can reach steady states that are consistent with thermal equilibrium~\cite{Kaufman2016}.
We now illustrate how such an experiment could go one step further by demonstrating the validity of the \ac{fdr}.
To this end, we numerically solve the full quantum evolution for the Bose--Hubbard chain describing the experiment in Ref.~\cite{Kaufman2016} (we emphasize that our approach does not depend on such a model choice and can be applied to general quantum systems).
The Bose--Hubbard Hamiltonian is given by
\begin{equation}
	\label{eq:bosehubbard}
	H_0 = -\hbar J \sum_{\ell = 1}^{L} (a_\ell^\dagger a_{\ell + 1} + \mathrm{h.c.}) + \frac{\hbar U}{2} \sum_{\ell = 1}^{L} n_\ell (n_\ell - 1) .
\end{equation}
Here, the optical lattice sites are denoted by $\ell = 1 \dots L$ with associated bosonic annihilation, creation, and number (density) operators $a_{\ell}$, $a_{\ell}^\dagger$, and $n_{\ell}$, respectively. $J$ is the strength of the nearest-neighbor hopping, for which we assume periodic boundary conditions, and $U$ the on-site interaction rate.
In our numerics, we do not truncate the local Hilbert-space dimension and employ an adaptive Krylov subspace method for time evolution~\cite{Lubich2008,Hochbruck2010,Jawecki2020}.
While previous numerical studies of \acp{fdr} in this model have focused on density autocorrelations at large $U / J$ and low fillings~\cite{Schuckert2020}, here we consider quenches into the superfluid regime ($U / J \sim 1$) at unit filling and also explore off-site correlations as a function of distance.

The linear response protocol is illustrated at the top of \cref{fig:fdr}.
We initialize the system of $N = L$ bosons in a Mott-insulating state at $U / J \to \infty$ and then quench it at time $t = 0$ into the superfluid phase at $U / J = \num{1.5625}$~\footnote{The results are insensitive to the precise choice of parameters.}, chosen consistent with the experiment in Ref.~\cite{Kaufman2016}. This quench throws the system heavily out of equilibrium. After a variable waiting time~$\waitingtime$, we either apply a Hermitian or an anti-Hermitian perturbation in order access the desired response functions in {\renewcommand{\crefpairconjunction}{~or~}\cref{eq:responsefunctionhermitian,eq:responsefunctionnonhermitian}}, respectively.
The perturbation is applied as a rectangular pulse of strength~$s$ and duration~$\delta t$, $f(t) = \hbar s \left[ \theta(t - \waitingtime) - \theta(t - \waitingtime - \delta t) \right] / \delta t$. The exact shape of the pulse is unimportant as long as the pulse duration is sufficiently short compared to the characteristic time scales of the system (cf.~\cref{app:perturbationtheory}). In this case, the pulse can be approximated by a $\delta$~function as $f(t) \approx \hbar s \delta(t - \waitingtime)$.
\Cref{fig:fdr:response} shows the time trace of the response to a (non\protect\nobreakdash-)Hermitian perturbation giving access to density autocorrelations ($B = A = n$), computed in a system of $L = 12$ sites for a perturbation of strength~$s = \num{0.05}$ and duration $J \delta t = \num{0.01}$~\footnote{It does not matter at which site the perturbation is applied as the model is translationally invariant for periodic boundary conditions.}.
The thermalization dynamics of the corresponding dynamic susceptibilities~$\chi^{\prime \, \mathrm{(NH)}}$ and $\chi^{\prime \prime}$ is depicted in \cref{fig:fdr:spectrum}.
For the purposes of this section, we evaluate the susceptibilities at fixed waiting time~$\waitingtime$~\cite{PineiroOrioli2019}, i.e.,
\begin{equation}
	\label{eq:susceptibilityfixedwaitingtime}
	\chi_{BA}^{\mathrm{(NH)}}(\waitingtime, \omega) = \int_{-\infty}^{\infty} \diff \Delta t \, \phi_{BA}^{\mathrm{(NH)}}(\waitingtime + \Delta t, \waitingtime) \etothepowerof{i \omega \Delta t} ,
\end{equation}
using an exponential filter of characteristic frequency~$\gamma / J = \num{0.2}$ to ensure convergence of the Fourier integrals (see \cref{app:quench:technical} for technical details).
$\chi^{\prime \, \mathrm{(NH)}}$ is symmetric in $\omega$ and grows as a broad central peak with small wings of opposite sign that gradually disappear as $\waitingtime$ increases, while $\chi^{\prime \prime}$ is anti-symmetric and develops characteristic peaks around non-zero frequencies. To assess whether the two functions satisfy the \ac{fdr}, we use the least-squares method to find the effective temperature~$T$ that best relates the susceptibilities via the \ac{fdr}~\eqref{eq:fdrsusceptibility}, i.e., for a fixed waiting time~$\waitingtime$,
\begin{equation}
	\label{eq:temperatureleastsquares}
	T = \argmin_{\Theta} \int \diff \omega \, \left[ -\chi^{\prime \, \mathrm{(NH)}} \tanh \left( \frac{\hbar \omega}{2 \kb \Theta} \right) - \chi^{\prime \prime} \right]^2 .
\end{equation}
In \cref{fig:fdr:earlylate}, one can see that the \ac{fdr} is clearly violated at early times, i.e., there exists no global value of $T$ such that \cref{eq:fdrsusceptibility} holds, but at later times, the agreement is remarkable and supports the interpretation that the system undergoes thermalization.

\begin{figure}[h!]
	\subfloat{\label{fig:fdrspatial:spectrum}}%
	\subfloat{\label{fig:fdrspatial:temperature}}%
	\subfloat{\label{fig:fdrspatial:errorrel}}%
	\subfloat{\label{fig:fdrspatial:errorabs}}%
	\includegraphics[width=\columnwidth]{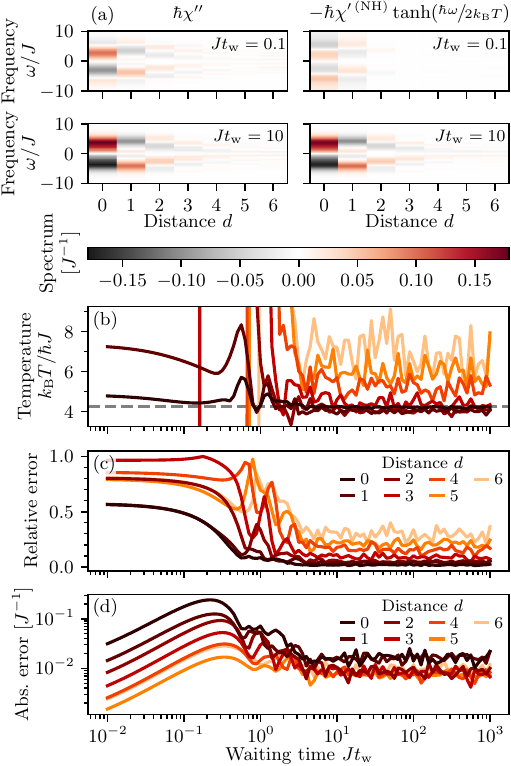}
	\caption{\label{fig:fdrspatial}%
		Thermalization dynamics of density correlations in a Bose--Hubbard system.
		(a)~Comparison of the dynamic susceptibilities~$\chi_{n_\ell n_{\ell + d}}^{\prime \prime}$ and $\chi_{n_\ell n_{\ell + d}}^{\mathrm{\prime \, (NH)}}$, rescaled according to the \ac{fdr}~\protect\eqref{eq:fdrsusceptibility}, for different waiting times~$\waitingtime$ as a function of the spatial distance $d$. The effective temperature~$T$ is determined for each configuration according to \protect\cref{eq:temperatureleastsquares} using the least-squares method. At early times, clear deviations are visible, but for late times, the two quantities agree and the \ac{fdr}~\protect\eqref{eq:fdrsusceptibility} is fulfilled.
		(b)~Least-squares value of the effective temperature~$T$, (c)~relative error, and (d)~absolute error of the \ac{fdr} as a function of the waiting time $\waitingtime$ for several distances $d$. At small distances, after times on the order of $J^{-1}$, the effective temperature approaches a constant value consistent with the prediction $\braket{H_0}_T = E_0$ for a thermal state (gray dashed line), and the relative error becomes small. As the distance increases, the relative error grows, but the absolute deviation becomes small (see also \protect\cref{app:quench}).%
	}
\end{figure}

A distinct feature of the \ac{fdr} in equilibrium is that it holds for any pair of observables $A$ and $B$. To confirm this prediction for our model system, we have computed $\chi^{\prime \, \mathrm{(NH)}}$ and $\chi^{\prime \prime}$ for off-site density correlations corresponding to $A = n_\ell$ and $B = n_{\ell + d}$ as a function of the distance $d$. The results are shown in \cref{fig:fdrspatial:spectrum}, where $\chi^{\prime \, \mathrm{(NH)}}$ is rescaled according to \cref{eq:fdrsusceptibility} with the best-fitting effective temperature~$T$ obtained from \cref{eq:temperatureleastsquares} for each configuration $(\waitingtime, d)$. Qualitatively, it can be seen that the two quantities deviate for early times, but agree well for late times.
To make this statement more quantitative, in \cref{fig:fdrspatial:errorrel,fig:fdrspatial:errorabs}, we show the relative and absolute error of the \ac{fdr}, i.e., the $L^2$ norm of the difference between the left- and right-hand side of \cref{eq:fdrsusceptibility} (see \cref{app:quench:technical}), as a figure of merit measuring how well the \ac{fdr}~\eqref{eq:fdrsusceptibility} is fulfilled at a particular instance of time.
For small distances, the relative error becomes vanishingly small after waiting times on the order of $J^{-1}$, while for larger distances, the error tends to drop later and fluctuates around a non-zero offset.
A similar behavior is exhibited by the effective temperature (see \cref{fig:fdrspatial:temperature}): at small distances and late times, the temperatures are approximately constant and agree with each other, while this is no longer true for larger distances.
Only in the former case, where the relative error is small, the effective temperature can be attributed the physical meaning of the temperature at which the subsystem degrees of freedom thermalize.
This temperature is consistent with the one obtained for a thermal state at the equivalent energy density using the condition $\braket{H_0}_T = E_0$ (gray dashed line at $\kb T / \hbar J = \num{4.27}$ in \cref{fig:fdrspatial:temperature}, calculated for $L = 8$ using exact diagonalization), where $\braket{\cdots}_T$ denotes the expectation value with respect to a canonical ensemble at temperature $T$, and $E_0$ is the energy of the initial state after the quench~\cite{Kaufman2016}.

While global many-body observables are expected to violate the \ac{fdr} due to the purity of the global quantum state, one would expect two-site observables like the off-site density correlations shown in \cref{fig:fdrspatial} to thermalize and thus satisfy the \ac{fdr} for sufficiently large systems and late times.
Although the absolute error gradually decreases with increasing distance due to the lower signal strength (see \cref{fig:fdrspatial:spectrum,fig:fdrspatial:errorabs}), relative discrepancies persist even after very long times.
In \cref{app:quench}, we investigate the behavior of the error as a function of system size and study a similar quench scenario for a \ac{2d} Bose--Hubbard system of $4 \times 4$ lattice sites and $N = 16$ bosons (unit filling).
Our analysis reveals that the relative error at long waiting times decreases with increasing system size. Furthermore, the larger \ac{2d} system exhibits only a minor trend towards larger relative errors as the distance increases, and the \ac{fdr} is overall better fulfilled than in the smaller \ac{1d} chain.
This points to the conclusion that the observed discrepancies of the \ac{fdr} in \cref{fig:fdrspatial} for large distances are likely due to finite-size effects.
Thus, our numerical results indicate that for sufficiently large systems, off-site density correlations fulfill the \ac{fdr} even at long distances, confirming the expectation that subsystems consisting of few degrees of freedom thermalize.

Having illustrated how the \ac{fdr} becomes accessible via (non\nobreakdash-)Hermitian linear response, we now turn to the question of how to realize the corresponding non-Hermitian perturbations experimentally.

\section{\label{sec:realization}Realization of non-Hermitian linear response}

There exists a growing body of work that describes how non-Hermitian physics can be generated in quantum many-body systems~\cite{ElGanainy2018,Ashida2020}.
Non-Hermitian Hamiltonians naturally arise in the context of dissipative quantum systems~\cite{Gardiner2004,Breuer2007}, where they govern the evolution of individual quantum trajectories conditioned on the absence of quantum jumps~\cite{Dalibard1992,Molmer1993,Daley2014}. This way, it is possible to harness natural sources of dissipation in order to explore novel non-Hermitian physics~\cite{Naghiloo2019,Nakagawa2020,Chen2021}.
Over the years, ever better techniques of screening experiments as much as possible from any sources of dissipation have been developed, with the goal of observing clean unitary dynamics in isolated quantum systems. This bears the potential to re-introduce channels of engineered dissipation using specifically designed control schemes.

\begin{figure}[t]
	\subfloat{\label{fig:ancillapostselection:density}}%
	\subfloat{\label{fig:ancillapostselection:correlator}}%
	\includegraphics[width=\columnwidth]{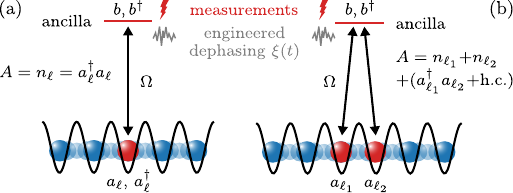}\\
	\vspace{5pt}%
	\subfloat{\label{fig:ancillapostselection:singlezeno}}%
	\subfloat{\label{fig:ancillapostselection:pulsedcontinuouszeno}}%
	\includegraphics[width=\columnwidth]{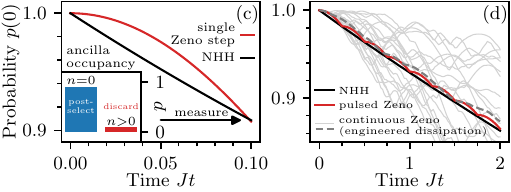}%
	\caption{\label{fig:ancillapostselection}%
		Realization of an effective non-Hermitian Hamiltonian using the quantum Zeno effect, illustrated for an optical lattice.
		(a)~Coupling a single lattice site~$\ell$ to an ancilla gives rise to a perturbation by the density operator~$A = n_\ell$ at that site.
		(b)~A perturbation by the hopping operator~$a_{\ell_1}^\dagger a_{\ell_2} + a_{\ell_2}^\dagger a_{\ell_1}$ can be achieved by coupling two sites~$\ell_1$ and $\ell_2$ simultaneously to an ancilla.
		(c)~Single step in the quantum Zeno evolution.
		The probability~$p(0)$ of detecting no particles in the ancilla gradually decreases over time (red).
		A measurement of the ancilla population, post-selected on the condition that the ancilla is empty (inset), projects the system on the empty-ancilla subspace.
		The coupled evolution plus projection corresponds to an effective non-Hermitian perturbation (NHH, black).
		(d)~When the projective measurement is performed frequently as compared to the strength of the coherent coupling~$\Omega$, the system plus ancilla is kept in the quantum Zeno regime for a prolonged period of time.
		The resulting pulsed Zeno evolution (red) is interpolated by the evolution under an effective non-Hermitian Hamiltonian (NHH, black).
		Alternatively, the repeated measurements can be substituted by strong engineered dissipation on the ancilla.
		The light gray lines show $20$ trajectories corresponding to different realizations of engineered classical dephasing noise~$\xi(t)$ on the ancilla, whose ensemble average (gray dashed line) approximates an effective non-Hermitian evolution.%
	}
\end{figure}

In this section, we propose an ancilla-based protocol that relies entirely on synthetic sources of dissipation in order to realize an effective non-Hermitian Hamiltonian.
The perturbation can selectively be applied as a short pulse or under continuous modulation of its strength, allowing one to probe frequency-dependent responses in the same way as in standard linear response scenarios.
Moreover, our flexible and experimentally feasible choice of system--ancilla coupling gives access to a wide range of observables.

\Cref{fig:ancillapostselection} gives an overview of the scheme, which is most conveniently phrased as an application of the quantum Zeno effect~\cite{Misra1977,Facchi2008}.
Depending on the desired perturbation operator~$A$, the relevant subsystem, e.g., a single site or two neighboring sites in an optical lattice, is coherently coupled to an initially empty ancilla, as depicted in \cref{fig:ancillapostselection:density,fig:ancillapostselection:correlator}.
A measurement of the ancilla population projects the system on the subspace with a definite number of particles in the ancilla.
As will become clear further below, non-Hermitian dynamics is realized by post-selecting those measurement outcomes where the ancilla remains empty (see \cref{fig:ancillapostselection:singlezeno}).
Repeating this measurement frequently gives rise to a quantum Zeno effect: as the measurement frequency tends to infinity, the probability of populating the ancilla vanishes.
If, instead, the measurement frequency is finite, there is a finite probability of populating the ancilla.
As illustrated in \cref{fig:ancillapostselection:pulsedcontinuouszeno}, this leads to a \enquote{pulsed} leakage of probability from the subspace where the ancilla is empty to a complementary subspace with non-vanishing ancilla population.
Instead of the pulsed Zeno effect, we can also use the continuous Zeno effect~\cite{Facchi2008}, which can be realized by substituting the repeated measurements with strong engineered dissipation on the ancilla~\cite{Stannigel2014}. This has the advantage of not requiring any non-destructive measurements during the evolution, but only a single post-selected measurement at the final evolution time.
Both the pulsed Zeno evolution and the ensemble average over many noise realizations in the continuous case can be described by an effective non-Hermitian Hamiltonian~\cite{Facchi2008,Militello2020,Biella2021} (see \cref{fig:ancillapostselection:pulsedcontinuouszeno}), which realizes the desired anti-Hermitian perturbation for measuring the unequal-time anti-commutator.

While our scheme can be implemented on various platforms, for the sake of concreteness, we focus here on bosons in optical lattices, where the ancilla may correspond to an auxiliary lattice site or an additional internal state. A crucial experimental requirement is the ability to distinguish an empty ancilla from one with non-zero population, which enables the projection on the empty-ancilla Zeno subspace. This requirement is met, for instance, by modern quantum gas microscopes, which reach both single-site and single-particle resolution~\cite{Bakr2009,Sherson2010}.

It is instructive to first consider a single step in the Zeno evolution consisting of a short coupling pulse followed by a projection, as depicted in \cref{fig:ancillapostselection:singlezeno}.
It turns out that this scenario corresponds to applying a $\delta$-like perturbation suitable for measuring the time trace of the non-Hermitian response function~\eqref{eq:responsefunctionnonhermitian} like in \cref{sec:illustration}.
Subsequently, we explain how the quantum Zeno effect enables a prolonged evolution under a non-Hermitian Hamiltonian, focusing on the scenario with strong engineered dephasing noise that induces a continuous Zeno effect (cf.~\cref{fig:ancillapostselection:pulsedcontinuouszeno}).
We benchmark variants of our scheme for measurements in both time and frequency domain at the example of the Bose--Hubbard chain introduced in \cref{sec:illustration}.
Bose--Hubbard systems subject to dissipation have been extensively studied with the goal of exploring the rich dynamics of open quantum systems~\cite{Kordas2015,Denis2018,Nakagawa2020}, whereas here, we use engineered dissipation as a tool~\cite{Schaefer2020} to probe dynamical correlations in closed systems via non-Hermitian linear response.
In \cref{sec:discussion}, we compare our approach with other protocols for measuring unequal-time anti-commutators, including ancilla-based weak measurement schemes~\cite{Uhrich2017,Kastner2018} and projective protocols~\cite{Knap2013,Uhrich2017,Kastner2018,Uhrich2019,Schuckert2020}, and discuss potential sources of errors as well as strategies on how to mitigate them.

\subsection{\label{sec:singlezenostep}Non-Hermitian linear response as a single step in the quantum Zeno evolution}

In this subsection, we discuss a single step in the Zeno evolution, which corresponds to applying an effective non-Hermitian $\delta$-like perturbation as in \cref{sec:illustration} and allows one to access the unequal-time anti-commutator in \cref{eq:responsefunctionnonhermitian} in time domain.

\subsubsection{Outline of the scheme}

We consider a system--ancilla coupling Hamiltonian of the form
\begin{equation}
	\label{eq:hamiltoniancoupling}
	H_\cpl = \hbar \Omega (b^\dagger a + a^\dagger b) ,
\end{equation}
where $a$ ($a^\dagger$) and $b$ ($b^\dagger$) represent the bosonic annihilation (creation) operators of the system mode to be probed and the ancilla, respectively, and $\Omega$ is the coupling strength.
In the coupling scheme depicted in \cref{fig:ancillapostselection:density}, the operator~$a$ represents a single lattice site~$\ell$, giving rise to an effective anti-Hermitian perturbation by the density (number) operator~$A = n_\ell = a_\ell^\dagger a_\ell$, as becomes clear below.
The scheme in \cref{fig:ancillapostselection:correlator} couples two lattice sites~$\ell_1$ and $\ell_2$, which may or may not be nearest neighbors, simultaneously to the ancilla. This corresponds to the replacement $a \to a_{\ell_1} + a_{\ell_2}$ in \cref{eq:hamiltoniancoupling} and produces a non-Hermitian perturbation by the operator~$A = n_{\ell_1} + n_{\ell_2} + a_{\ell_1}^\dagger a_{\ell_2} + a_{\ell_2}^\dagger a_{\ell_1}$.
This type of perturbation can therefore be used to access \acp{fdr} for the hopping operator~$a_{\ell_1}^\dagger a_{\ell_2} + a_{\ell_2}^\dagger a_{\ell_1}$, as we demonstrate below for nearest neighbors.
It is possible to consider even more general setups~\cite{Geier2021}, e.g., by adding a relative phase between the two couplings in \cref{fig:ancillapostselection:correlator} using laser-assisted tunneling~\cite{Jaksch2003}, or by coupling a multitude of sites to one or more ancillas, enabling global perturbations by sums of local operators.
The general form of the accessible perturbations is given in \cref{app:generalcoupling}.

A single Zeno step of duration~$\delta t$ corresponds to a unitary evolution described by the time evolution operator $U(\delta t) = \exp ( -i H \delta t / \hbar )$, followed by a projection on the Zeno subspace defined by the performed measurement~\cite{Facchi2008}.
During the coupling, the total Hamiltonian is given by $H = H_0 + H_\cpl$, but for sufficiently short~$\delta t$, it is permissible to neglect the evolution under $H_0$ (for simplicity, we assume the ancilla to have no internal dynamics).
A measurement of the ancilla population projects the system on one of the Zeno subspaces with a fixed number of particles in the ancilla, which can be realized experimentally via post-selection. Prior to the coupling, we require the ancilla to be in the vacuum state. Let $\mathcal{P}$ denote the projection operator on the empty-ancilla subspace.
Then, during one Zeno step, the state~$\rho(\waitingtime)$ at the waiting time~$\waitingtime$ changes, up to a normalization, as
\begin{equation}
	\label{eq:zenostep}
	\rho(\waitingtime) \to \rho^\prime(\waitingtime + \delta t) = \mathcal{P} U(\delta t) \rho(\waitingtime) U^\dagger(\delta t) \mathcal{P} .
\end{equation}
As shown in \cref{app:singlezeno}, to leading order in the effective coupling strength~$s = (\Omega \delta t)^2 / 2$, this process corresponds to the evolution under an effective non-Hermitian Hamiltonian,
\begin{equation}
	\label{eq:singlezenoeffectivenonhermitian}
	\rho^\prime(\waitingtime + \delta t) = \etothepowerof{-i H_\eff \delta t / \hbar} \rho(\waitingtime) \etothepowerof{i H_\eff^\dagger \delta t / \hbar} + \mathcal{O}(s^2) ,
\end{equation}
with $H_\eff = H_0 -i \hbar s A / \delta t$ and $A = a^\dagger a$.

For the purpose of measuring the non-Hermitian linear response in time domain, a single Zeno step is sufficient. The system subsequently evolves unitarily under the unperturbed Hamiltonian~$H_0$ up to the final observation time~${t > \waitingtime + \delta t}$.
The unnormalized expectation value of an observable~$B$ is then given by
\begin{equation}
	\label{eq:ancillameasurementunnormalized}
		\Tr \left[ B \rho^\prime(t) \right] = \Braket{B(t)}_0 - s \Braket{\Anticommutator{B(t)}{A(\waitingtime)}}_0 .
\end{equation}
In the linear regime, the probability of detecting no particles in the ancilla after the coupling reads $p(0) = 1 - 2 s \braket{A(\waitingtime)}_0$, which can be found by inserting the identity operator for $B$ in \cref{eq:ancillameasurementunnormalized}.
Normalizing \cref{eq:ancillameasurementunnormalized} by this probability yields, to leading order in $s$, the conditional expectation value
\begin{equation}
	\label{eq:ancillameasurementnormalized}
	\begin{split}
		\Braket{B(t)}_\mathcal{P} = \Braket{B(t)}_0 - s \Big[ &\Braket{\Anticommutator{B(t)}{A(\waitingtime)}}_0 \\
		&- 2 \Braket{B(t)}_0 \Braket{A(\waitingtime)}_0 \Big] ,
	\end{split}
\end{equation}
representing a post-selected measurement conditioned on the empty ancilla.
As anticipated, a comparison with \cref{eq:linearresponse,eq:responsefunctionnonhermitian} shows that this result effectively corresponds to a linear response after applying the anti-Hermitian perturbation $H_1(t) = -i \hbar s \delta(t - \waitingtime) A$, giving direct access to the symmetrized correlation function~\eqref{eq:symmetrizedcorrelationfunction} via $S_{BA}(t, \waitingtime) = -\left[ \braket{B(t)}_\mathcal{P} - \braket{B(t)}_0 \right] / 2 s$.
As required in \cref{sec:nhlrt}, the perturbation operator~${A = a^\dagger a}$ is indeed positive semi-definite, in line with the physical intuition that the norm of the state can only decrease through outcoupling followed by a projection.

It is instructive to compare the result in \cref{eq:ancillameasurementnormalized} with the one obtained if no projection on the empty-ancilla subspace is performed, e.g., if the measurement apparatus is unable to distinguish an empty ancilla from one with non-vanishing population or the result of the ancilla measurement is ignored.
In this case, a simple average over all ancilla populations is obtained, where, to leading order in the effective coupling strength~$s$, only single occupancies of the ancilla contribute. The unconditional response then reads (see \cref{app:perturbationtheory})
\begin{equation}
\label{eq:ancillameasurementtrace}
\begin{split}
	\Braket{B(t)} = \Braket{B(t)}_0 - s \big\langle &\Anticommutator{B(t)}{a^\dagger(\waitingtime) a(\waitingtime)} \\
	&- 2 a^\dagger(\waitingtime) B(t) a(\waitingtime) \big\rangle_0 .
\end{split}
\end{equation}
The last term stems from a process where a single particle ends up in the ancilla after the coupling. Post-selecting on the empty ancilla eliminates this undesired contribution, yielding a pure non-Hermitian evolution that gives access to the unequal-time anti-commutator.

\subsubsection{\label{sec:timedomainmeasurement}Numerical benchmark: non-Hermitian linear response in time domain}

\begin{figure}[h!]
	\subfloat{\label{fig:ancillameasurementtime:response}}%
	\subfloat{\label{fig:ancillameasurementtime:spectrum}}%
	\subfloat{\label{fig:ancillameasurementtime:response:correlator}}%
	\subfloat{\label{fig:ancillameasurementtime:spectrum:correlator}}%
	\includegraphics[width=\columnwidth]{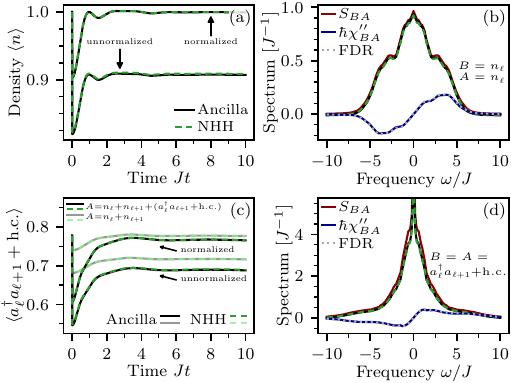}
	\caption{\label{fig:ancillameasurementtime}%
		Simulation of the linear response to a non-Hermitian perturbation generated by a single step in the Zeno evolution of coupling to an ancilla followed by a projection on the empty-ancilla subspace (see \protect\cref{fig:ancillapostselection:singlezeno}).
		(a)~Time trace of the density after applying the perturbation to a single site as in \protect\cref{fig:ancillapostselection:density}.
		The unnormalized and normalized responses correspond to \protect\cref{eq:ancillameasurementunnormalized,eq:ancillameasurementnormalized}, respectively.
		The result agrees well with the response to a non-Hermitian perturbation by the density operator~$A = n_\ell$ (NHH, green dashed line).
		(b)~The correlation spectrum extracted from the response in (a) agrees well with the exact result~$S_{BA}$.
		The \ac{fdr}~\protect\eqref{eq:FDR} between $\chitwoprime_{BA}$ and $S_{BA}$, calculated using the known temperature~$\kb T / \hbar J = \num[round-mode=places, round-precision=2]{4.2684382}$ of the thermal state, is shown for comparison.
		(c)~Time trace of the nearest-neighbor correlator after coupling two neighboring sites simultaneously (black, see \protect\cref{fig:ancillapostselection:correlator}) or individually (gray, \protect\cref{fig:ancillapostselection:density}) to the ancilla.
		Subtracting the latter quantity from the former yields the response to a perturbation by the hopping operator~$A = a_{\ell}^\dagger a_{\ell + 1} + \mathrm{a_{\ell + 1}^\dagger a_{\ell}}$.
		The respective responses agree well with their effective descriptions in terms of non-Hermitian Hamiltonians (NHH, green dashed lines).
		(d)~The extracted correlation spectrum reproduces the exact one to good accuracy.%
	}
\end{figure}

To benchmark our scheme, we numerically solve the full quantum evolution describing a measurement of $S_{BA}(t, \waitingtime)$ for a thermal state~$\rho_T = \exp ( - H_0 / \kb T ) / Z(T)$ in a Bose--Hubbard chain of ${L = 8}$ sites at unit filling and with periodic boundary conditions. Here, $Z(T) = \Tr [ \exp ( - H_0 / \kb T ) ]$ is the canonical partition sum, and the temperature $T$ is chosen such that the mean energy $\braket{H_0}_T = \Tr (H_0 \rho_T)$ corresponds to that of a Mott-insulating state.
A thermal state is an ideal benchmark for our purposes since the temperature~$T$ is known and the \ac{fdr} is satisfied exactly, so any deviations from the \ac{fdr} indicate deficiencies of the method.

In \cref{fig:ancillameasurementtime:response,fig:ancillameasurementtime:response:correlator}, we show the time traces of the responses to perturbations corresponding to the coupling configurations in \cref{fig:ancillapostselection:density,fig:ancillapostselection:correlator}, respectively, i.e., for on-site densities ($A = B = n_\ell$) and nearest-neighbor correlators ($A = n_{\ell} + n_{\ell + 1} + a_{\ell}^\dagger a_{\ell + 1} + a_{\ell + 1}^\dagger a_{\ell}$, $B = a_{\ell}^\dagger a_{\ell + 1} + a_{\ell + 1}^\dagger a_{\ell}$).
From the latter measurement, the response for the combination $A = B = a_{\ell}^\dagger a_{\ell + 1} + a_{\ell + 1}^\dagger a_{\ell}$ can be obtained by subtracting the response of the same observable~$B$ to perturbations~$A$ involving only the densities at the relevant sites.
Experimentally, the nearest-neighbor correlator~$\braket{B} = \braket{a_{\ell}^\dagger a_{\ell + 1} + \mathrm{h.c.}}$ can be measured, e.g., by projecting the system on non-interacting double wells and monitoring the double-well occupancy as a function of time~\cite{Kessler2014,Atala2014}.
The coupling to the ancilla is applied as a rectangular pulse of duration $J \delta t = \num{0.01}$ and its strength is chosen such that the effective coupling becomes $s = \num{0.05}$ for the density and $s = \num{0.02}$ for the correlator, corresponding to a decay of the norm by about $\SI{10}{\percent}$ in both cases.
As can be seen in \cref{fig:ancillameasurementtime:response,fig:ancillameasurementtime:response:correlator}, the simulated ancilla measurement agrees well with the description in terms of the effective non-Hermitian Hamiltonian in \cref{eq:singlezenoeffectivenonhermitian}.
In \cref{fig:ancillameasurementtime:spectrum,fig:ancillameasurementtime:spectrum:correlator}, we compare the correlation spectra extracted from the responses in \cref{fig:ancillameasurementtime:response,fig:ancillameasurementtime:response:correlator}, respectively, with the exact result.
The Fourier integrals have been calculated using exponential filters of characteristic frequencies $\gamma / J = \num{0.1}$ for the density and $\gamma / J = \num{0.05}$ for the correlator.
Due to the sizable static contribution to the response in the latter case, the height of the central peak in \cref{fig:ancillameasurementtime:spectrum:correlator} strongly depends on the choice of $\gamma$, but this is irrelevant for probing \acp{fdr} because the value of the correlation spectrum at $\omega = 0$ is not constrained by the \ac{fdr}~\eqref{eq:FDR}.
Up to small deviations resulting from non-linear effects, which can be reduced at the cost of a lower signal-to-noise ratio (see discussion in \cref{sec:discussion:errors}), our scheme provides an accurate measurement of the correlation spectrum for both densities and correlators.

\subsection{\label{sec:pulsedzeno}Non-Hermitian linear response via the pulsed quantum Zeno effect}

We now explain how to realize a prolonged evolution under a (possibly time-dependent) effective non-Hermitian Hamiltonian, suitable for probing frequency-resolved responses as is common in standard linear response experiments.
To this end, we generalize the coupling Hamiltonian in \cref{eq:hamiltoniancoupling} by allowing for an arbitrary modulation~$g(t)$ of the coupling strength, i.e.,
\begin{equation}
	\label{eq:hamiltoniancouplingtimedependent}
	H_\cpl(t) = g(t) \hbar \Omega \left( b^\dagger a + a^\dagger b \right) .
\end{equation}
We first note that the naive approach of extending the coupling duration in the previous scheme, consisting of a single Zeno step of coupling plus projection, does not yield the desired result.
If the coupling duration in \cref{eq:zenostep} is prolonged up to the final measurement time~$t > \waitingtime$, instead of \cref{eq:ancillameasurementunnormalized}, we obtain to leading order the response
\begin{equation}
	\label{eq:ancillameasurementunnormalizedgeneral}
\begin{split}
	\Tr \left[ B \rho^\prime(t) \right] = &\Braket{B(t)}_0 - \Omega^2 \int_{\waitingtime}^{t} \diff t^\prime g(t^\prime) \int_{\waitingtime}^{t^\prime} \diff t^{\prime \prime} g(t^{\prime \prime}) \\
	&\times \Braket{B(t) a^\dagger(t^\prime) a(t^{\prime \prime}) + a^\dagger(t^{\prime \prime}) a(t^\prime) B(t))}_0 .
\end{split}
\end{equation}
The three-time correlations in the integrand appear because the leading perturbative contribution to the response is of quadratic order in the coupling Hamiltonian (see \cref{app:pulsedzeno} for details).
If $g(t)$ is properly normalized and has compact support on the interval~$[\waitingtime, \waitingtime + \delta t]$ with $\delta t$ sufficiently short as compared to the characteristic time scales of~$H_0$, \cref{eq:ancillameasurementunnormalizedgeneral} reduces to \cref{eq:ancillameasurementunnormalized}, but in general does \emph{not} yield the desired two-time anti-commutator.

The key to obtaining a response as in \cref{eq:linearresponsenonhermitianunnormalized} is to iterate the Zeno step presented in the previous subsection as depicted in \cref{fig:ancillapostselection:pulsedcontinuouszeno}.
Such a repeated series of measurements is the common scenario for the pulsed quantum Zeno effect~\cite{Misra1977,Facchi2008,Biella2021}.
To this end, we split the interval~$[\waitingtime, t]$ into $n$ steps such that $\waitingtime = t_0 < t_1 < \dots < t_n = t$ with $t_{i + 1} - t_{i} = \delta t = (t - \waitingtime) / n$.
The evolution from $t_i$ to $t_{i + 1}$ is described by \cref{eq:zenostep}, corresponding to an individual Zeno step of unitary evolution under the Hamiltonian~$H(t) = H_0 + H_\cpl(t)$, followed by a measurement of the ancilla population that projects the system on the subspace with empty ancilla (realizations where one or more particles are detected in the ancilla are discarded).
Thus, the state evolves, up to a normalization, as
\begin{equation}
	\label{eq:pulsedzeno}
	\rho(\waitingtime) \to \rho^\prime(t) = \pprojector U_{n} \pprojector \cdots \pprojector U_1 \rho(\waitingtime) U_1^\dagger \pprojector \cdots \pprojector U_{n}^\dagger \pprojector ,
\end{equation}
where $U_i = U(t_i, t_{i - 1})$ denotes the time evolution operator from time~$t_{i - 1}$ to $t_i$.
This equation describes the evolution under a continuously applied system--ancilla coupling with intermittent measurements of the ancilla population.
The role of the measurements is to destroy the coherences between the relevant Zeno subspaces, giving rise to a different evolution than in \cref{eq:ancillameasurementunnormalizedgeneral}, where a measurement is performed only once at the final time.
In \cref{app:pulsedzeno}, we show that, to leading order in the coupling and for $\delta t$ sufficiently short as compared to the characteristic time scales of $H_0$ and $g(t)$, the (unnormalized) expectation value of an observable~$B$ after the Zeno evolution is given by
\begin{equation}
	\label{eq:ancillameasurementunnormalizedpulsedzeno}
	\Tr \left[ B \rho^\prime(t) \right] = \Braket{B(t)}_0 - s \sum_{i = 0}^{n - 1} g^2(t_i) \Braket{\Anticommutator{B(t)}{A(t_i)}}_0\,,
\end{equation}
with $A = a^\dagger a$ and $s = (\Omega \delta t)^2 / 2$.
Approximating the sum by an integral, this result coincides with a linear response to the anti-Hermitian perturbation~$H_1(t) = -i f(t) A$ according to \cref{eq:linearresponsenonhermitianunnormalized}, where $f(t) = g^2(t) \Omega^2 \delta t / 2$.
Since the operator $f(t) A$ is positive semi-definite (cf.~\cref{sec:nhlrt}), this effective non-Hermitian Hamiltonian describes a gradual leakage of probability out of the empty-ancilla Zeno subspace (see \cref{fig:ancillapostselection:pulsedcontinuouszeno}).

\subsection{\label{sec:continuouszeno}Non-Hermitian linear response via the continuous quantum Zeno effect}

Unfortunately, implementing the pulsed Zeno effect without destroying the sample during the intermittent measurements poses a prohibitive layer of complexity for many experiments.
For this reason, we instead exploit the \emph{continuous} Zeno effect in what follows~(see \cref{fig:ancillapostselection:pulsedcontinuouszeno}).
This formulation of the Zeno effect arises in the presence of a strong coupling to an external system, which plays the role of a measurement apparatus and leads to wildly fluctuating phases between the relevant Zeno subspaces~\cite{Facchi2008}.
One way of generating such a continuous Zeno effect is by adding engineered classical noise to the system, which has been proposed, e.g., in Ref.~\cite{Stannigel2014} to constrain the dynamics of quantum simulators for lattice gauge theories.
Here, we apply this idea to realize a time-dependent effective non-Hermitian perturbation.

\subsubsection{\label{sec:engineereddissipation}From engineered dissipation to non-Hermitian dynamics}

We consider the ancilla to be subject to classical dephasing noise, as indicated in \cref{fig:ancillapostselection}.
Such a source of noise can be engineered via a rapidly fluctuating effective detuning, e.g., in form of a Zeeman or ac Stark shift, acting on the ancilla only. We assume that the fluctuations are sufficiently fast compared to all relevant physical time scales, such that their effect can be approximated by a Gaussian white-noise process~$\xi(t)$ satisfying $\ensembleaverage{\xi(t)} = 0$ and $\ensembleaverage{\xi(t) \xi(t^\prime)} = \delta(t - t^\prime)$, where $\ensembleaverage{\cdots}$ denotes the ensemble average over all noise realizations.
For example, using lasers to generate an ac Stark shift, this technical requirement can be fulfilled using acousto-optical devices~\cite{Maier2019}.
The evolution of the density operator~$\rho(t)$ can then be described by the stochastic von Neumann equation~\cite{Stannigel2014}
\begin{equation}
	\label{eq:stochasticvonneumann}
	\diff \rho = -\frac{i}{\hbar} \Commutator{H(t)}{\rho} \diff t - i \sqrt{2 \kappa} \Commutator{b^\dagger b}{\rho} \diff W(t) ,
\end{equation}
with dephasing rate~$\kappa > 0$ and Wiener increments $\diff W(t) = \xi(t) \diff t$, subject to the Stratonovich interpretation of stochastic calculus~\cite{Kloeden1992,Gardiner2009} (see also \cref{app:engineereddissipationtonhh}). The deterministic part of \cref{eq:stochasticvonneumann} is governed by the Hamiltonian~$H(t) = H_0 + H_\cpl(t)$, where the coupling Hamiltonian~$H_\cpl(t)$ is given by \cref{eq:hamiltoniancouplingtimedependent}.

By virtue of stochastic calculus, it can be shown (see \cref{app:engineereddissipationtonhh:masterequation}) that the noise-averaged density operator~$\sigma(t) \equiv \ensembleaverage{\rho(t)}$ satisfies the Lindblad master equation
\begin{equation}
	\label{eq:masterequation}
	\frac{\diff}{\diff t} \sigma = -\frac{i}{\hbar} \Commutator{H(t)}{\sigma} - \kappa \left( \Anticommutator{L^\dagger L}{\sigma} - 2 L \sigma L^\dagger \right) ,
\end{equation}
with the Hermitian Lindblad operator $L = b^\dagger b$.
The stochastic differential equation~\eqref{eq:stochasticvonneumann} represents a diffusive unraveling~\cite{Barchielli2009} of the master equation~\eqref{eq:masterequation}.
Such diffusive unravelings typically arise in the theory of continuous measurements, where a quantum system is continuously monitored and the resulting measurement back action gives rise to diffusive quantum trajectories~\cite{Jacobs2006,Wiseman2009}.
By contrast, in our case, there are no actual measurements involved and \cref{eq:stochasticvonneumann} describes a random unitary evolution with pure dephasing~\cite{Hasegawa1980,Semina2014}.
In fact, there exists an infinite number of stochastic unravelings, both diffusive and jump-like, whose ensemble average is described by \cref{eq:masterequation}~\cite{Gisin1992,Dalibard1992,Breuer2007}.
As an alternative to the approach in \cref{eq:stochasticvonneumann} using engineered dephasing, we could also start from \cref{eq:masterequation} with the Lindblad operator $L = b$, describing a spontaneous decay of particles in the ancilla at a decay rate~$\kappa$. As shown in \cref{app:engineereddissipationtonhh}, such a setting gives rise to the same effective non-Hermitian Hamiltonian as considered below.

The quantum Zeno effect is realized in the strong-noise limit $\kappa \to \infty$~\cite{Stannigel2014}.
The strong dissipation leads to an exponential decay of coherences between Zeno subspaces, in analogy to the effect of repeated measurements, and thereby suppresses the build-up of population in the ancilla.
As shown in \cref{app:engineereddissipationtonhh:nhh}, to leading order in perturbation theory, the density operator~$\sigma_\mathcal{P}(t) = \mathcal{P} \sigma(t) \mathcal{P}$, projected on the subspace with no particles in the ancilla, obeys the evolution equation
\begin{equation}
	\label{eq:effectivenonhermitian}
	i \hbar \frac{\diff}{\diff t} \sigma_\mathcal{P} = H_\mathrm{eff}(t) \sigma_\mathcal{P} - \sigma_\mathcal{P} H_\mathrm{eff}^\dagger(t) ,
\end{equation}
generated by the effective non-Hermitian Hamiltonian $H_\mathrm{eff}(t) = H_0 - i f(t) A$ with $A = a^\dagger a$ and $f(t) = g^2(t) \hbar \Omega^2 / \kappa$. As required in \cref{sec:nhlrt}, the perturbation operator~$A$ is positive semi-definite and $f(t)$ is non-negative, describing a leakage of probability out of the empty-ancilla subspace.
In \cref{fig:ancillapostselection:pulsedcontinuouszeno}, we illustrate that this effective non-Hermitian dynamics arises as the ensemble average over stochastic trajectories governed by \cref{eq:stochasticvonneumann}.
The crucial advantage of the continuous Zeno effect over the pulsed formulation, where repeated non-destructive measurements are required, is that a single projection at the final measurement time is sufficient, which can conveniently be realized as a post-selection on measurement outcomes where no particles are detected in the ancilla.

\subsubsection{\label{sec:frequencydomainmeasurement}Numerical benchmark: non-Hermitian linear response in frequency domain}

We now demonstrate how our scheme enables access to the \ac{fdr} directly in frequency domain. From the structure of the general linear response formula~\eqref{eq:linearresponse} it becomes clear that by applying a non-Hermitian perturbation under a suitable periodic modulation~$f(t)$ continuously until the final observation time~$\finaltime$, it is possible to directly measure non-Hermitian dynamic susceptibilities of the form
\begin{equation}
	\label{eq:susceptibilityfixedfinaltime}
	\chi_{BA}^{\mathrm{(NH)}}(\finaltime, \omega) = \int_{-\finaltime}^{\finaltime} \diff \Delta t \, \phi_{BA}^{\mathrm{(NH)}}(\finaltime, \finaltime - \Delta t) \etothepowerof{i \omega \Delta t} .
\end{equation}

It is our goal to extract the reactive part~$\chi^{\prime \mathrm{(NH)}}$ of this quantity from the linear response to the effective non-Hermitian Hamiltonian in \cref{eq:effectivenonhermitian}.
For simplicity, we focus on the common case where~$\chi^{\prime \mathrm{(NH)}}$ corresponds to the real part of \cref{eq:susceptibilityfixedfinaltime}, and we consider Hermitian operators $A$ and $B$ such that the response function~\eqref{eq:responsefunctionnonhermitian} is real. Due to the non-negativity constraint on $f(t)$, it is not possible to modulate the effective coupling around zero. Instead, we choose the modulation in \cref{eq:stochasticvonneumann} as $g(t) = \sqrt{2} \cos[\omega (\finaltime - t) / 2]$, for a fixed final observation time~$\finaltime$, such that $f(t) = [1 + \cos \omega (\finaltime - t)] \hbar \Omega^2 / \kappa$.
According to \cref{eq:linearresponsenonhermitian}, the response is then given by
\begin{equation}
	\label{eq:responsemodulation}
\begin{split}
	\Braket{B(\finaltime)}_\mathcal{P} = &\Braket{B(\finaltime)}_0 + \frac{\hbar \Omega^2}{\kappa} \int_{0}^{\finaltime} \diff t \, \phi_{BA}^{\mathrm{(NH)}}(\finaltime, t) \\
	&+ \frac{\hbar \Omega^2}{\kappa} \int_{0}^{\finaltime} \diff t \, \phi_{BA}^{\mathrm{(NH)}}(\finaltime, t) \cos[\omega (\finaltime - t)] ,
\end{split}
\end{equation}
where $\braket{B(\finaltime)}_\mathcal{P} = \Tr \left[ B \sigma_\mathcal{P}(\finaltime) \right] / \Tr [\sigma_\mathcal{P}(\finaltime)]$ is the conditional expectation value obtained from post-selection.
The first two terms on the right-hand side of \cref{eq:responsemodulation} represent the response to a static non-Hermitian perturbation with $g(t) \equiv 1$ and the last term is proportional to the desired real part of \cref{eq:susceptibilityfixedfinaltime}, which can be seen after changing the integration variable to $\Delta t = \finaltime - t$.
Thus, it is possible to extract the quantity~$\chiprimenh_{BA}(\finaltime, \omega)$ for a given probe frequency~$\omega$ from two linear response measurements, one with a periodic modulation and one with a constant perturbation, the latter being subtracted from the former.

\begin{figure}
	\subfloat{\label{fig:ancillameasurementfrequency:norm}}%
	\subfloat{\label{fig:ancillameasurementfrequency:response}}%
	\subfloat{\label{fig:ancillameasurementfrequency:spectrum}}%
	\includegraphics[width=\columnwidth]{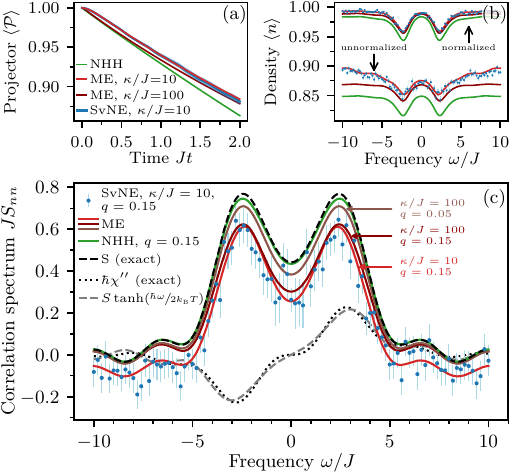}
	\caption{\label{fig:ancillameasurementfrequency}%
		Simulation of the non-Hermitian linear response scheme for measuring the density autocorrelation spectrum in frequency domain. The effective non-Hermitian perturbation is generated through coupling to an ancilla subject to strong engineered dephasing, exploiting the continuous quantum Zeno effect.
		The simulations are based on the stochastic von Neumann equation~\protect\eqref{eq:stochasticvonneumann} (SvNE), the master equation~\protect\eqref{eq:masterequation} (ME), and the effective non-Hermitian Hamiltonian~\protect\eqref{eq:effectivenonhermitian} (NHH).
		The stochastic simulation has been averaged over $200$ realizations, and the error bars show the ensemble standard deviation of the mean.
		(a)~Decrease of the norm resulting from the projection on the empty-ancilla subspace for the probe frequency~$\omega = 0$ and norm decay~$q = \num{0.15}$ up to the final evolution time~$J \finaltime = 2$.
		A stronger dephasing rate~$\kappa$ improves the agreement between the ME (SvNE) and NHH descriptions at early times, while the deviations at later times are due to non-linear effects.
		(b)~Unnormalized and normalized (conditional) responses, corresponding to \cref{eq:linearresponsenonhermitianunnormalized,eq:responsemodulation}, respectively, as a function of frequency for a fixed final time~$J \finaltime = \num{2}$ and norm decay~$q = \num{0.15}$.
		(c)~Correlation spectra~$S_{nn}$ extracted from the responses in (b) according to \protect\cref{eq:responsemodulation}, in comparison with the exact result and the \protect\ac{fdr}~\protect\eqref{eq:FDR} between $S$ and~$\chitwoprime$ (evaluated for the same truncation $J \finaltime = 2$).
		The different combinations of the parameters $\kappa$ and $q$ for the ME simulation show that the agreement with the exact result can be improved by going deeper into the limit of strong dissipation and weak perturbations.%
	}
\end{figure}

To benchmark our protocol, we resort to the previous example of the density autocorrelation spectrum $S_{BA}(\omega) = -\hbar \chiprimenh_{BA}(\omega)$ ($B = A = n$) of a thermal state in a periodic \ac{1d} Bose--Hubbard chain.
For this purpose, we numerically solve the stochastic von Neumann equation~\eqref{eq:stochasticvonneumann} --- the most fundamental equation in our approach --- using stochastic Magnus integration~\cite{Burrage1999,Blanes2009,Kamm2021}. By comparing the results of the stochastic simulation to those obtained based on \cref{eq:masterequation,eq:effectivenonhermitian}, we demonstrate the validity of the approximations underlying the effective description in terms of a non-Hermitian Hamiltonian.

When choosing the final evolution time~$\finaltime$, which determines the cutoff of the integrals in \cref{eq:responsemodulation}, it is important to keep in mind the trade-off between signal-to-noise ratio and accuracy:
while a longer propagation time~$\finaltime$ can yield a more accurate approximation of the Fourier integral in \cref{eq:susceptibilityfixedfinaltime}, the strength of the perturbation must typically be reduced accordingly in order to stay within the linear response regime, which lowers the signal-to-noise ratio.
The optimal balance between these effects depends on several conditions, such as the targeted frequency range, the properties of the response function, and the resolution of the measurement apparatus.
For concreteness, in the following benchmark example we choose $J \finaltime = \num{2}$ for all probed frequencies~$\omega$.
While this truncation affects the form of the extracted spectrum at low frequencies~%
\footnote{For the small system size considered in this benchmark, the integrals in \cref{eq:responsemodulation} do not converge for $\finaltime \to \infty$ due to revivals and are therefore truncated at finite~$\finaltime$. Alternatively, convergence can be enforced, e.g., through an exponential ramp of the modulation amplitude as $g(t) \to \etothepowerof{-\gamma (\finaltime - t) / 2} g(t)$.}%
, it yields an adequate approximation of the Fourier integral at higher frequencies most relevant for probing the \ac{fdr}.

To account for the different sensitivities of the responses at different probe frequencies, we parametrize the perturbation strength in terms of the norm decay~$q$ due to the effective non-Hermitian Hamiltonian. That is, for each frequency~$\omega$, given the fixed final observation time~$\finaltime$ and the dephasing rate~$\kappa$, we adjust the coupling strength~$\Omega$ such that according to \cref{eq:normdecrease} the norm of the state has decreased by the amount~$q$ at the end of the evolution.
For a translationally invariant system at unit filling, we have $\braket{A(t)}_0 = \braket{n(t)}_0 = 1$, and therefore $\Omega = \left[ \kappa q / 2 \finaltime (1 + \sinc (\omega \finaltime / \pi)) \right]^{1 / 2}$, where $\sinc(x) = \sin(\pi x) / \pi x$.

In \cref{fig:ancillameasurementfrequency}, we compare various simulations of the scheme based on \cref{eq:stochasticvonneumann,eq:masterequation,eq:effectivenonhermitian} for a system of $L = 4$ lattice sites.
\Cref{fig:ancillameasurementfrequency:norm} illustrates a typical decay of the norm over time due to the effective non-Hermitian perturbation.
The perturbation strength is adjusted such that by the final evolution time $J \finaltime = 2$ the norm has dropped approximately by an amount~$q = 0.15$, which results in a good signal-to-noise ratio, but lies slightly beyond the onset of the non-linear regime (longer propagation times may require balancing with a reduced perturbation strength).
The simulated unnormalized and normalized (conditional) responses as a function of frequency are shown in \cref{fig:ancillameasurementfrequency:response}, from which we extract the correlation spectra presented in \cref{fig:ancillameasurementfrequency:spectrum}.
The results are compared to the exact correlation spectrum, which is evaluated for the same truncation $J \finaltime = 2$ of the Fourier integral in \cref{eq:susceptibilityfixedfinaltime} to allow for a consistent benchmark.
For the stochastic simulation, we choose the accessible dephasing rate~$\kappa / J = \num{10}$.
As can be seen in \cref{fig:ancillameasurementfrequency}, the stochastic simulation based on \cref{eq:stochasticvonneumann} agrees with the simulation based on the master equation~\eqref{eq:masterequation} within the statistical error bars that show the ensemble standard deviation of the mean for an accessible number of $\num{200}$ realizations.
Moreover, \cref{fig:ancillameasurementfrequency:spectrum} shows that these parameters already yield the correlation spectrum at a reasonable accuracy suitable for certifying the validity of the \ac{fdr}~\eqref{eq:FDR}.
The description in terms of the effective non-Hermitian Hamiltonian in \cref{eq:effectivenonhermitian} is closer to the exact result than the description in terms of the master equation~\eqref{eq:masterequation} for the same parameters, revealing that the linear regime is wider for the former than for the latter, which could be remedied through extrapolation.
In the effective non-Hermitian description~\eqref{eq:effectivenonhermitian}, the coupling strength~$\Omega$ and the dephasing rate~$\kappa$ enter only via the ratio~$\Omega^2 / \kappa$, which is proportional to the norm decay~$q$, while these two parameters enter \cref{eq:stochasticvonneumann,eq:masterequation} individually.
Going deeper into the Zeno limit of large $\kappa$ improves the validity of \cref{eq:effectivenonhermitian} at early times and at higher frequencies, shown in \cref{fig:ancillameasurementfrequency} for $\kappa / J = 100$.
Decreasing at the same time the effective coupling strength, as illustrated in \cref{fig:ancillameasurementfrequency:spectrum} for $q = \num{0.05}$, the agreement between the extracted correlation spectrum and the exact result improves further, which shows that, at the cost of decreasing the signal-to-noise ratio, the exact correlation spectrum can in principle be approximated to arbitrary accuracy.

\section{\label{sec:discussion}Discussion}

In this section, we put our non-Hermitian linear response approach for measuring dynamical correlations and \acp{fdr} in perspective with other schemes.
We first demonstrate that common ancilla-based weak measurement protocols~\cite{Uhrich2017,Kastner2018} fit into this general framework since their ancilla-free formulations can be interpreted as a non-Hermitian linear response.
In addition, we compare the ancilla-based technique for realizing non-Hermitian linear response presented in \cref{sec:realization} with other schemes for accessing non-Hermitian dynamics or measuring dynamical correlations, including non-invasive and projective protocols~\cite{Knap2013,Uhrich2017,Kastner2018,Uhrich2019,Schuckert2020}.
We conclude with a discussion of experimental aspects and potential error sources.

\subsection{\label{sec:weakmeasurement}General relation between non-Hermitian linear response and ancilla-based weak measurements}

To reveal the close connection between non-Hermitian linear response and ancilla-based weak (or non-invasive) measurements of the unequal-time anti-commutator, we first briefly review common protocols of the latter kind.
While these weak measurement protocols have originally been developed for spin systems~\cite{Uhrich2017,Kastner2018}, here we formulate them for general quantum systems and allow for arbitrary durations of the system--ancilla coupling (typically, only short coupling pulses are considered).
For further details on the following points, see \cref{app:weakmeasurement}.

System and ancilla are assumed to be initially in a product state, $\rho_0 = \rho_\sys \otimes \rho_\anc$.
The non-invasive protocol of Refs.~\cite{Uhrich2017,Kastner2018} starts by evolving the system under the unperturbed Hamiltonian~$H_0$ up to a certain waiting time~$\waitingtime$, while the ancilla does not participate in the dynamics.
System and ancilla are then coupled by the Hamiltonian
\begin{equation}
	\label{eq:weakmeasurementhamiltoniancoupling}
	H_\cpl(t) = f(t) A \otimes X ,
\end{equation}
where $A$ and $X$ are Hermitian operators acting on system and ancilla, respectively, and $f(t)$ represents an arbitrary time-dependent modulation.
The form of the coupling Hamiltonian is one of the main differences to our protocol in \cref{sec:singlezenostep} (see discussion in the next subsection).
After a coupled evolution up to time~$t > \waitingtime$, one measures projectively the observables~$B$ on the system and $Y$ on the ancilla, respectively.
Instead of directly correlating the measurement outcomes as proposed in Refs.~\cite{Uhrich2017,Kastner2018}, we consider here conditional expectation values in order to reveal the connection to non-Hermitian linear response.
As derived in \cref{app:weakmeasurement}, the expectation value of~$B$ under the condition that the ancilla measurement of~$Y$ yields the outcome~$y$ is given, to linear order in the coupling, by
\begin{multline}
	\label{eq:weakmeasurementconditionalexpectation}
	\Braket{B(t)}_y = \Braket{B(t)}_0 - \frac{i}{\hbar} \int_{\waitingtime}^t \diff t^\prime f(t^\prime) \\
	\times \left\{ \lambda_y \left[ \Braket{B(t) A(t^\prime)}_0 - \Braket{B(t)}_0 \Braket{A(t^\prime)}_0 \right] - \mathrm{c.c.} \right\}
\end{multline}
with $\lambda_y = \braket{\pprojector_y X}_0 / \braket{\pprojector_y}_0 \in \doublestruck{C}$. Here, $\pprojector_y$ is the projector on the eigenspace of eigenvalue~$y$ and $\mathrm{c.c.}$ denotes the complex conjugate.
The key to access the unequal-time anti-commutator is to choose the ancilla state as well as the operators~$X$ and $Y$ such that the expectation value~$\braket{\pprojector_y X}_0 = \Tr [ \rho_\anc \pprojector_y X ]$ becomes purely imaginary (choosing $\braket{\pprojector_y X}_0$ real instead yields the commutator).
The conditional expectation value of~$B$ is then formally equivalent to the non-Hermitian linear response in \cref{eq:linearresponsenonhermitian}.
In fact, by tracing out the ancilla, it can be shown (see \cref{app:weakmeasurement}) that the coupled evolution of system and ancilla corresponds to the evolution under the effective non-Hermitian Hamiltonian~$H_\eff = H_0 - i s f(t) A$ with $s = i \lambda_y \in \doublestruck{R}$.
Interestingly, this leads to the insight that any such weak measurement protocol for the unequal-time anti-commutator can be interpreted as the linear response to a non-Hermitian perturbation.

The general connection between non-Hermitian linear response and ancilla-based weak measurements is beneficial for both disciplines: particular observables previously accessible only via ancilla-based schemes may be obtainable more efficiently in an ancilla-free way using the tools of non-Hermitian physics, while certain non-Hermitian Hamiltonians difficult to engineer directly may be realized with the help of an ancilla.

\subsection{\label{sec:comparison}Comparison to other schemes}

\subsubsection{Ancilla-based weak measurements}

One of the main challenges of the ancilla-based weak measurement scheme for the unequal-time anti-commutator discussed above is to engineer the ancilla state as well as the observables~$X$ and $Y$ in such a way that $\braket{\pprojector_y X}_0$ becomes purely imaginary.
While Refs.~\cite{Uhrich2017,Kastner2018} discuss suitable configurations for spin systems, it is far less obvious how to choose the setup in an experimentally feasible way on other platforms such as bosons in optical lattices (for example, as discussed in \cref{app:weakmeasurement}, to access density correlations, number non-conserving coupling Hamiltonians may be required, which cannot be realized with massive particles).

By contrast, our ancilla-based scheme in \cref{sec:realization} relies on the system--ancilla coupling in \cref{eq:hamiltoniancoupling}, which is quadratic in the creation and annihilation operators.
Despite its simple form, the coupling can flexibly be adapted to measure the unequal-time anti-commutator of a wide range of previously inaccessible observables such as nearest-neighbor correlators, as discussed in \cref{sec:singlezenostep,app:generalcoupling}.
In addition, our choice of the initial ancilla state in the form of the vacuum is particularly easy to prepare experimentally.
On the formal level, an important difference to the common weak measurement approach is that for our choice of the coupling Hamiltonian~\eqref{eq:hamiltoniancoupling}, the linear order in perturbation theory vanishes and the leading contribution to the response stems from the quadratic order (see \cref{app:perturbationtheory}), where the anti-commutator naturally arises and can be isolated by post-selection on realizations without any particles in the ancilla.

\subsubsection{\label{sec:discussion:projectiveprotocols}Projective protocols}

Projective protocols allow one to probe dynamical correlations of dichotomic observables (observables with two eigenvalues) by performing consecutive projective measurements directly on the system and correlating the outcomes in a suitable way~\cite{Knap2013,Uhrich2017,Kastner2018,Uhrich2019,Schuckert2020}.
As such, compared to schemes based on weak perturbations such as linear response, projective protocols are backaction-free and feature a higher signal-to-noise ratio.
Despite these advantages, the fact that projective protocols work only for dichotomic observables restricts their general applicability.

In Ref.~\cite{Schuckert2020}, it has been analyzed how projective protocols can be applied to approximately dichotomic observables, in particular densities in Bose--Hubbard systems close to the hard-core limit.
In \cref{app:projection_vs_nhlr}, we present numerical benchmarks comparing the performance of projective protocols and non-Hermitian linear response for measuring the unequal-time anti-commutator in Bose--Hubbard systems at various fillings and on-site interactions.
Our analysis shows that projective protocols perform well at low fillings and large on-site interactions, but yield unsatisfactory results when applied beyond this regime, i.e., as soon as multiple occupancies can no longer be neglected.
In particular, the relevant scenario of Bose--Hubbard systems at unit filling and moderate values of~$U / J$, which we study in \cref{sec:illustration} inspired by the experiment of Ref.~\cite{Kaufman2016}, remains beyond the scope of projective protocols.
By contrast, our non-Hermitian linear response approach does not have restrictions on observables regarding the number of eigenvalues and performs well across the entire parameter space explored in \cref{app:projection_vs_nhlr}.
Thus, our scheme allows one to reliably access unequal-time anti-commutators and the associated \acp{fdr} also in regimes outside the range of projective protocols, e.g., in the pair superfluid phase of dipolar bosons~\cite{Dutta2015} or other phases where multiple occupancies play an essential role.

\subsubsection{Dissipative dynamics and post-selected quantum trajectories}

The structure of the second term on the right-hand side of \cref{eq:ancillameasurementtrace}, characterizing the unconditional response after coupling to the ancilla without measuring the ancilla population, resembles to the \enquote{recycling term} in Lindblad master equations~\cite{Breuer2007,Daley2014}.
In fact, the short coupling pulse to the ancilla can be viewed as an effective dissipative perturbation, $\rho(\waitingtime) \to \rho(\waitingtime + \delta t) = \rho(\waitingtime) + \delta t \mathcal{D}[\rho(\waitingtime)]$, where $\mathcal{D}[\rho] = \gamma \left( 2 a \rho a^\dagger - \anticommutator{a^\dagger a}{\rho} \right)$ is the Lindblad dissipator with dissipation rate~$\gamma = s / \delta t$. This yields \cref{eq:ancillameasurementtrace} for the expectation value of an observable $B$ after a unitary evolution up to time~$t$.
In the quantum trajectories picture~\cite{Dalibard1992,Molmer1993,Daley2014}, the Lindblad dissipator~$\mathcal{D}[\rho]$ generates an evolution under the non-Hermitian Hamiltonian~$H_1 = -i \hbar \gamma a^\dagger a$, subject to quantum jumps described by the \enquote{recycling term} $2 \gamma a \rho a^\dagger$. By post-selecting on the absence of quantum jumps, it is possible to isolate the pure non-Hermitian evolution~\cite{Naghiloo2019,Nakagawa2020,Chen2021}. From this point of view, the projection on the empty-ancilla subspace in \cref{eq:zenostep} can be interpreted as a post-selection on the absence of quantum jumps, i.e., particles hopping to the ancilla. This allows us to eliminate the undesired contribution in \cref{eq:ancillameasurementtrace} due to the \enquote{recycling term} and obtain instead the result in \cref{eq:ancillameasurementnormalized}, reflecting a purely non-Hermitian perturbation that gives access to the unequal-time anti-commutator.

\subsection{\label{sec:discussion:errors}Experimental considerations and error sources}

Many experimental setups such as quantum gas microscopes permit the simultaneous readout of all site populations in a single shot~\cite{Bakr2009,Sherson2010}. This is convenient for simultaneously measuring the responses of different observables~$B$, e.g., $B = n_\ell$ for $\ell = 1 \dots L$, to a fixed perturbation~$A$ determined by the coupling scheme.
In addition, for the single Zeno step and the continuous Zeno evolution in \cref{sec:singlezenostep,sec:continuouszeno}, respectively, the measurement of the ancilla population can be deferred up to the final observation time~$t$ and measured along with the other site populations (cf.~Ref.~\cite{Uhrich2017}).
The projection on the empty-ancilla subspace is then achieved by post-selecting those realizations where no particles are detected in the ancilla.
Since the effective coupling~$s$ needs to be chosen sufficiently weak to stay within the regime of linear response, the fidelity of the post-selection is typically high (see \cref{fig:ancillapostselection:singlezeno}).
However, there is the usual linear response trade-off between maximizing the measurement signal (large $s$) and staying within the perturbative regime where the linear approximation is valid (small $s$).

One can distinguish two types of detection errors: false positives, i.e., at least one particle is detected in the ancilla, but there is actually none, and false negatives, i.e., no particles are detected, but there is at least one.
Let $\alpha$ be the false positive rate and let $\beta$ be the false negative rate.
If the measurement is post-selected on the condition that no particles are detected in the ancilla, which may in some cases be erroneous, the conditional state in \cref{eq:zenostep} is replaced by $\rho^\prime = (1 - \alpha) \mathcal{P} \rho \mathcal{P} + \beta \mathcal{Q} \rho \mathcal{Q}$, where $\rho$ is the state right after the coupling and before the projection, and $Q = \mathds{1} - \mathcal{P}$ is the projector on the subspace with a non-vanishing ancilla population (for simplicity, we do not distinguish different false-negative probabilities within the $Q$~subspace since the error due to single occupancies dominates in the linear regime).
This shows that false positives lower the measurement fidelity, while false negatives contribute a systematic error to the result, arising from the inadvertent projection on a complementary subspace (see the discussion in the context of \cref{eq:ancillameasurementtrace} and \cref{app:singlezeno}).

\section{\label{sec:conclusion}Conclusion}

In this work, we have demonstrated that non-Hermitian linear response enables access to the unequal-time anti-commutator as the missing piece for the direct observation of the \ac{fdr} in quantum systems. As an illustration, we have discussed how a Bose--Hubbard system after a global quench reaches thermal equilibrium, and we have derived techniques to generate the required non-Hermitian dynamics in cold-atom systems coupled to an ancillary mode by exploiting the quantum Zeno effect. This proposal provides a concrete scenario for the direct observation of the \ac{fdr} and an unbiased way of probing thermalization dynamics in state-of-the-art experiments on synthetic quantum matter.

Our non-Hermitian linear response approach is completely agnostic to specific platforms and implementations, and as such can be applied in any non-Hermitian system. It is independent of microscopic details such as interactions, geometry, or particle statistics, and can thus be used with bosons, fermions, or spins alike.
Higher orders in the response may be used to access nested unequal-time anti-commutators of increasing order.
Moreover, we have shown that common ancilla-based weak measurement protocols for dynamical correlations fit in the same framework, as these can be interpreted in the light of (non\nobreakdash-)Hermitian linear response.
Our proposed ancilla-based realization of non-Hermitian linear response permits the extraction of dynamical correlations for a wide range of previously inaccessible observables beyond density correlations, even frequency-resolved.
While we have focused on lattice systems, our protocol can immediately be applied to continuous systems, e.g., via spatially focused laser beams, giving access to dynamical correlations of the field operator coarse-grained over a small region in space~\cite{Mora2003}.
The discussed framework thus provides an array of possibilities to experimentally --- and also numerically~\cite{PineiroOrioli2019,Boguslavski2020} --- characterize quantum systems in and out of thermal equilibrium.

\appendix

\begin{acknowledgments}
We thank J.~Berges, M.~Gärttner, S.~Lannig, M.~K.~Oberthaler, A.~{Pi{\~n}eiro Orioli}, J.~Reichstetter, and A.~Salzinger for discussions.
This project has received funding from the European Research Council (ERC) under the European Union’s Horizon 2020 research and innovation programme (ERC StG StrEnQTh, Grant Agreement No.\ $804305$).
We further acknowledge support by Provincia Autonoma di Trento.
This work is part of the Collaborative Research Centre \mbox{ISOQUANT} (project ID $273811115$) and has been supported by Q@TN, the joint lab between the University of Trento, FBK --- Fondazione Bruno Kessler, INFN --- National Institute for Nuclear Physics, and CNR --- National Research Council.
The authors acknowledge support by the state of Baden-Württemberg through bwHPC.
\end{acknowledgments}

\section{\label{app:nhlrt}Non-Hermitian linear response theory}

\subsection{\label{app:nhlrt:derivation}Derivation of the non-Hermitian linear response formula}

To derive \cref{eq:linearresponsenonhermitianunnormalized}, we first transform to the interaction picture with respect to the unperturbed (Hermitian) Hamiltonian~$H_0$. The von Neumann equation~\eqref{eq:vonneumannnonhermitian} then reads
\begin{equation}
	\label{eq:vonneumannnonhermitianinteractionpicture}
	\frac{\diff}{\diff t} \tilde{\rho} = -\frac{i}{\hbar} \Anticommutator{\tilde{H}_1(t)}{\tilde{\rho}} ,
\end{equation}
where $\tilde{\rho}(t) = \etothepowerof{i H_0 t / \hbar} \rho(t) \etothepowerof{-i H_0 t / \hbar}$ is the density operator and $\tilde{H}_1(t) = -i f(t) \tilde{A}(t)$ with $\tilde{A}(t) = \etothepowerof{i H_0 t / \hbar} A \etothepowerof{-i H_0 t / \hbar}$ is the anti-Hermitian perturbation in the interaction picture.
This equation can equivalently be expressed in integral form as
\begin{equation}
	\label{eq:vonneumannnonhermitianintegral}
	\tilde{\rho}(t) = \tilde{\rho}(0) -\frac{i}{\hbar} \int_{0}^{t} \diff t^\prime \, \Anticommutator{\tilde{H}_1(t^\prime)}{\tilde{\rho}(t^\prime)}
\end{equation}
with $\tilde{\rho}(0) = \rho_0$.
To linear order in the perturbation, we can replace $\tilde{\rho}(t^\prime)$ in the integrand by $\rho_0$, yielding
\begin{equation}
	\label{eq:vonneumannnonhermitianlinear}
	\tilde{\rho}(t) = \rho_0 -\frac{1}{\hbar} \int_{0}^{t} \diff t^\prime \, \Anticommutator{\tilde{A}_1(t^\prime)}{\rho_0} f(t^\prime) .
\end{equation}
The expectation value of an observable~$B$ can be computed in the interaction picture as $\braket{B(t)} = \Tr [ \tilde{B}(t) \tilde{\rho}(t) ]$, where $\tilde{B}(t) = \etothepowerof{i H_0 t / \hbar} B \etothepowerof{-i H_0 t / \hbar}$. Inserting \cref{eq:vonneumannnonhermitianlinear} into this expression and using the cyclic property of the trace leads to the result in \cref{eq:linearresponsenonhermitianunnormalized}.

\subsection{\label{app:nhlrt:susceptibility}Connection between correlation spectrum and non-Hermitian dynamic susceptibility}

To show that $S_{BA}(\tau, \omega) = - \hbar \chi_{BA}^{\mathrm{\prime \, (NH)}}(\tau, \omega)$, we first note that the symmetric correlation function~\eqref{eq:symmetrizedcorrelationfunction} obeys the symmetry relation $S_{BA}(t, t^\prime) = S_{AB}(t^\prime, t)$.
In what follows, we use the short-hand notation $S_{BA}(\tau, \Delta t) = S_{BA}(t = \tau + \Delta t / 2, t^\prime = \tau - \Delta t / 2)$. Then, the aforementioned identity reads $S_{BA}(\tau, \Delta t) = S_{AB}(\tau, -\Delta t)$.
This allows us to express the correlation spectrum as
\begin{equation}
	\begin{split}
		S_{BA}(\tau, \omega) = &\int_{-2 \tau}^{2 \tau} \diff \Delta t \, S_{BA}(\tau, \Delta t) \etothepowerof{i \omega \Delta t} \\
		= &\int_{0}^{2 \tau} \diff \Delta t \, S_{BA}(\tau, \Delta t) \etothepowerof{i \omega \Delta t} \\
		&+ \int_{0}^{2 \tau} \diff \Delta t \, S_{AB}(\tau, \Delta t) \etothepowerof{-i \omega \Delta t} .
	\end{split}
\end{equation}
Using further \cref{eq:responsefunctionnonhermitian}, noting that the Heaviside step function allows us to extend the integration domain to negative $\Delta t$, as well as the definition of the generalized susceptibility, \cref{eq:susceptibility}, we arrive at
\begin{equation}
	\begin{split}
		S_{BA}(\tau, \omega) = &-\frac{\hbar}{2} \int_{-2 \tau}^{2 \tau} \diff \Delta t \, \phi_{BA}^\mathrm{(NH)}(\tau, \Delta t) \etothepowerof{i \omega \Delta t} \\
		&-\frac{\hbar}{2} \int_{-2 \tau}^{2 \tau} \diff \Delta t \, \phi_{AB}^\mathrm{(NH)}(\tau, \Delta t) \etothepowerof{-i \omega \Delta t} \\
		= &-\frac{\hbar}{2} \left[ \chi_{BA}^\mathrm{(NH)}(\tau, \omega) + \chi_{AB}^\mathrm{(NH)}(\tau, -\omega) \right] \\
		= &- \hbar \chi_{BA}^{\mathrm{\prime \, (NH)}}(\tau, \omega) .
	\end{split}
\end{equation}

By contrast, if we consider the Fourier transform at fixed waiting time~$\waitingtime$ as in \cref{eq:susceptibilityfixedwaitingtime}, the integrand $S_{BA}(\waitingtime + \Delta t, \waitingtime)$ does not possess a symmetry with respect to the relative time~$\Delta t$ in general. Thus, out of equilibrium, we generally have $S_{BA}(\waitingtime, \omega) \neq - \hbar \chi_{BA}^{\mathrm{\prime \, (NH)}}(\waitingtime, \omega)$, but this relation is restored once the system reaches a stationary state.

\section{\label{app:quench}Fluctuation--dissipation relations after a quench in a Bose--Hubbard system}

In this Appendix, we provide details on our analysis in \cref{sec:illustration} of \acp{fdr} following a quench in a Bose--Hubbard system.
In particular, we investigate in more detail the behavior of the deviation from the \ac{fdr} for off-site density correlations as a function of distance.
To this end, we study the error as a function of system size and compare our results in \cref{sec:illustration} to a similar analysis for a Bose--Hubbard system in \ac{2d}.

\subsection{\label{app:quench:technical}Technical details on the computation of dynamical susceptibilities}

The susceptibilities appearing in the \ac{fdr}~\eqref{eq:fdrsusceptibility} can be found by varying both the waiting time~$\waitingtime$ and the observation time~$t = \waitingtime + \Delta t$, and computing the Fourier transform with respect to the relative time~$\Delta t$ at fixed central time~$\tau$ according to \cref{eq:susceptibility,eq:susceptibilityfixedcentraltime}.
However, since at early times~$\tau$ the integration domains of the Fourier integrals in \cref{eq:susceptibility,eq:susceptibilityfixedcentraltime} are limited to only a short range of~$\Delta t$, we consider instead the non-equilibrium generalization of the susceptibility as the Fourier transform of the response function at fixed waiting time~$\waitingtime$~\cite{PineiroOrioli2019} according to \cref{eq:susceptibilityfixedwaitingtime} (and analogously for its Hermitian counterpart).
This form has the advantage that the integration domain is unbounded, and it is often more efficient to compute as the integrand is directly obtained from the linear response at fixed waiting times.
In practice, the integral in \cref{eq:susceptibilityfixedwaitingtime} needs to be regulated appropriately, for example, by truncating it once the correlations have decayed sufficiently or by means of a frequency filter accounting for a finite spectral resolution in experiments.
Unless stated otherwise, we follow the latter approach, using an exponential filter of characteristic frequency $\gamma / J = \num{0.2}$, which amounts to the replacement $\etothepowerof{i \omega \Delta t} \to \etothepowerof{(i \omega - \gamma) \Delta t}$ in the Fourier integral~\eqref{eq:susceptibilityfixedwaitingtime}.

Note that out of equilibrium, the susceptibilities in \cref{eq:susceptibilityfixedcentraltime,eq:susceptibilityfixedwaitingtime} generally disagree, and, in particular, $S_{BA}(\waitingtime, \omega) \neq - \hbar \chi_{BA}^{\mathrm{\prime \, (NH)}}(\waitingtime, \omega)$ in general (cf.~\cref{app:nhlrt:susceptibility}). Once the system has reached a stationary state, the response functions depend only on the relative time~$\Delta t$ and the different conventions become equivalent, provided the integration domains are chosen appropriately.
For our purposes, we probe the \ac{fdr} out of equilibrium in the form of \cref{eq:fdrsusceptibility}, expressed in terms of the susceptibilities obtainable from the (non\nobreakdash-)Hermitian linear response at fixed waiting time as in \cref{eq:susceptibilityfixedwaitingtime}.

Moreover, the susceptibility components in \cref{eq:susceptibilitycomponents} are in general complex~\cite{Jensen1991} and require measurements of both the response of $B$ to a perturbation by $A$ and vice versa.
Sufficient conditions for them to be real include the case $B = A^\dagger$, or, if $A$ and $B$ are Hermitian, the property $\braket{B(t) A(t^\prime)} = \braket{A(t) B(t^\prime)}$. The latter is fulfilled, for instance, if $A$ and $B$ are on-site observables in a system that is invariant under both translations and reflections, as it is the case for density--density correlations in a Bose--Hubbard chain with periodic boundary conditions.
Thus, for our model system and our choice of observables, the reactive and dissipative parts of the dynamic susceptibility are real and correspond, respectively, to the real and imaginary parts of the susceptibility~\eqref{eq:susceptibilityfixedcentraltime} (and similarly for the Hermitian susceptibility in \cref{eq:susceptibility}).

We quantify deviations from the \ac{fdr}~\eqref{eq:fdrsusceptibility} by the absolute error
\begin{equation}
	\epsilon_{\mathrm{abs}} = \left\lVert -\chi^{\prime \, \mathrm{(NH)}} \tanh \left( \frac{\hbar \omega}{2 \kb T} \right) - \chi^{\prime \prime} \right\rVert_2 ,
\end{equation}
where $\lVert \cdot \rVert_2$ denotes the $L^2$ norm, which we define by
\begin{align}
	\left\lVert f \right\rVert_2^2 = \frac{1}{|\Omega|} \int_{\Omega} \diff \omega \, |f(\omega)|^2 .
\end{align}
In our numerical benchmarks, we choose the fixed integration domain~$\Omega = [-10 \, J, 10 \, J]$.
The relative error is obtained by normalizing the absolute error with respect to the sum of the individual norms,
\begin{equation}
	\epsilon_{\mathrm{rel}} = \frac{\epsilon_{\mathrm{abs}}}{\lVert \chi^{\prime \, \mathrm{(NH)}} \tanh (\hbar \omega / 2 \kb T )\rVert_2 + \lVert \chi^{\prime \prime} \rVert_2} .
\end{equation}

\subsection{\label{app:quench:bh1d}One-dimensional Bose--Hubbard system}

\begin{figure}
	\includegraphics[width=\columnwidth]{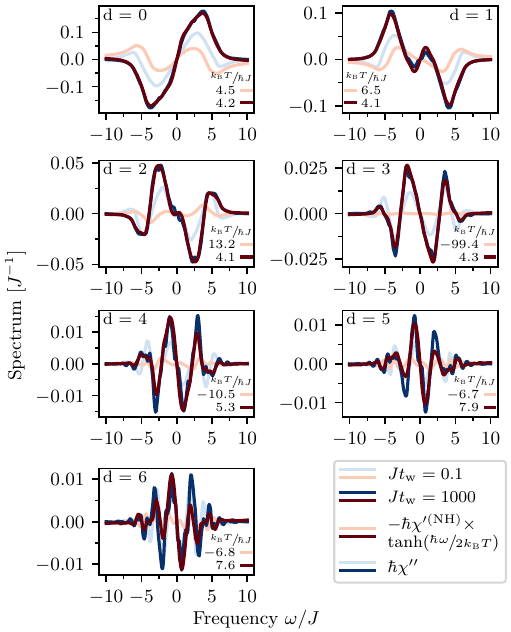}
	\caption{\label{fig:fdrspatialdetail}%
		\Acp{fdr} for off-site density correlations at early and late waiting times~$\waitingtime$ for several distances~$d$.
		The dynamic susceptibility~$\chitwoprime$ (blue) is compared to $\chiprimenh$ (red), rescaled according to the \ac{fdr}~\protect\eqref{eq:fdrsusceptibility} using the least-squares result for the effective temperature indicated in the plots.
		At small distances and late times, the curves overlap well, while at larger distances discrepancies persist even after long times.%
	}
\end{figure}

To better understand the significance of the increase of the relative error as a function of distance in \cref{fig:fdrspatial:errorrel}, we show in \cref{fig:fdrspatialdetail} the extracted \acp{fdr} at early and late waiting times for the individual distances.
The data are the same as in \cref{fig:fdrspatial} and the susceptibility~$\chiprimenh$ has been rescaled according to the \ac{fdr}~\eqref{eq:fdrsusceptibility} using the indicated effective temperature~$T$ obtained from the least-squares fit in \cref{eq:temperatureleastsquares}.
At early waiting times, there is no global value of $T$ to make $\chiprimenh$ and $\chitwoprime$ overlap, and the \ac{fdr} is clearly violated (in some cases, an attempted fit can even yield unphysical negative temperatures).
By contrast, at late waiting times, $\chiprimenh$ and $\chitwoprime$ fulfill the \ac{fdr} and at small distances, the extracted effective temperatures are consistent with the temperature~$\kb T / \hbar J = \num{4.27}$ of a thermal state at the same energy density as the initial state (calculated for $L = 8$ using exact diagonalization).
At larger distances, some peaks in \cref{fig:fdrspatialdetail} exhibit clear deviations which persist even after very long times and contribute to the increased relative error in \cref{fig:fdrspatial:errorrel}.

For an ergodic system in the thermodynamic limit, it is generally expected that a two-site subsystem, regardless of the distance between the two sites in real space, eventually thermalizes and thus satisfies the \ac{fdr}.
The observed deviations in \cref{fig:fdrspatial,fig:fdrspatialdetail} may therefore be an artifact of the finite system size.
Apart from that, numerical errors induced in the course of the data analysis, such as integration and truncation errors in the evaluation of Fourier integrals or distortions caused by the frequency filter, may contribute to the deviation.
We have checked that improving on the latter points does not alter the picture qualitatively.

\begin{figure}
	\includegraphics[width=\columnwidth]{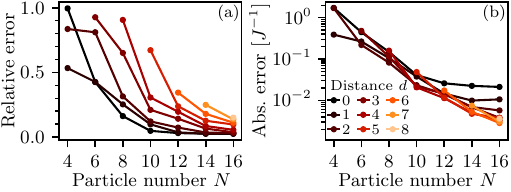}
	\caption{\label{fig:errorfinitesize}%
		Finite-size behavior of the deviations from the \ac{fdr}.
		(a)~Relative error and (b)~absolute error as a function of the particle number~$N$ (equal to the number of lattice sites~$L$ at unit filling) at waiting time~$J \waitingtime = \num{10}$ for all possible distances in the periodic chain.
		Both errors clearly decrease with increasing system size.%
	}
\end{figure}

To study the influence of finite-size effects, we have calculated the error as a function of the particle number~$N$ (corresponding to the number of lattice sites~$L$ at unit filling) up to $N = L = 16$.
\Cref{fig:errorfinitesize} shows the relative and absolute errors at the moderate waiting time~$J\waitingtime = \num{10}$ for all possible distances~$d$ in the respective systems.
Note that for a periodic chain of length~$L$, the maximum distance is~$d = \lfloor L / 2 \rfloor$.
The Fourier integrals have been truncated at $J \Delta t = 30$ using an exponential filter of characteristic frequency~$\gamma / J = 0.1$.
Both the relative and the absolute errors for all distances decrease as the system size increases until the relative error saturates at a value close to zero.
Although the exponential growth of the Hilbert-space dimension makes an exact numerical treatment of even larger systems inaccessible, the clear trend in \cref{fig:errorfinitesize} suggests that the deviations from the \ac{fdr} in \cref{fig:fdrspatial,fig:fdrspatialdetail} at large distances for $L = 12$ are likely due to finite-size effects.
Our analysis thus confirms the expectation that the two-site subsystems relevant for off-site density correlations thermalize and thus fulfill the \ac{fdr}, provided the system is not too small.

\subsection{\label{app:quench:bh2d}Two-dimensional Bose--Hubbard system}

To show that our results for the \ac{1d} Bose--Hubbard chain are generic, we study the analogous quench scenario in a \ac{2d} Bose--Hubbard system.
We consider a system of $4 \times 4$ lattice sites with $N = 16$ particles (unit filling) and periodic boundary conditions in each direction.
The larger system size compared to the \ac{1d} setting above allows us to support the conjecture that the \ac{fdr} is better fulfilled as the system size increases.
As before, we initialize the system in a Mott-insulating state and quench at time $t = 0$ into the superfluid phase at $U / J = \num{1.5625}$.

\begin{figure}
	\subfloat{\label{fig:fdrspatial2d:spectrum}}%
	\subfloat{\label{fig:fdrspatial2d:temperature}}%
	\subfloat{\label{fig:fdrspatial2d:errorrel}}%
	\subfloat{\label{fig:fdrspatial2d:errorabs}}%
	\includegraphics[width=\columnwidth]{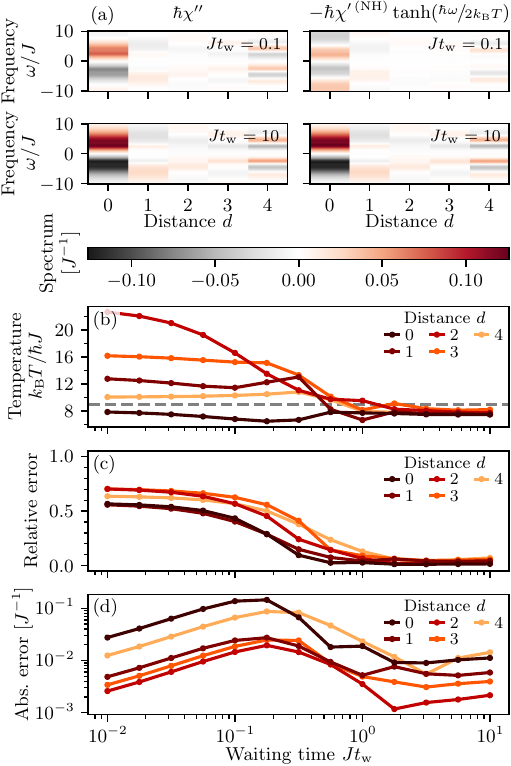}
	\caption{\label{fig:fdrspatial2d}%
		Same as \protect\cref{fig:fdrspatial}, but for the \ac{2d} Bose--Hubbard system.
		After waiting times~$\waitingtime$ on the order of $J^{-1}$, the relative error (c) of the \ac{fdr} remains small for all distinct lattice distances~$d$.%
	}
\end{figure}

In \cref{fig:fdrspatial2d}, we present the same analysis for the \ac{2d} system as carried out in \cref{fig:fdrspatial} for the \ac{1d} chain (cf.~\cref{sec:illustration}).
Due to the periodic boundary conditions and the isotropy of the hopping, the curves fall into five classes corresponding to distances~$d = 0 \dots 4$ between the perturbed and probed lattice site.
Note that these distances do not correspond to the physical distances in the $4 \times 4$ lattice, but to the minimum number of hopping events connecting the two sites.
Similarly to the \ac{1d} setting, the \ac{fdr} is violated at short waiting times, indicated by the large relative error in \cref{fig:fdrspatial2d:errorrel}.
After times on the order of $J^{-1}$, the errors decrease dramatically and the \ac{fdr} is fulfilled for all accessible distances with only a minor trend towards larger relative errors for larger distances.
The effective temperatures in \cref{fig:fdrspatial2d:temperature} for the individual distances reach approximately constant values that mutually agree up to deviations of about ten percent or less.
Furthermore, the effective temperatures are close to the temperature~$\kb T / \hbar J = \num[round-mode=places, round-precision=2]{8.90659}$ of a thermal state at the same energy density as the initial state (calculated for a $3 \times 3$ system using exact diagonalization).

\begin{figure}
	\includegraphics[width=\columnwidth]{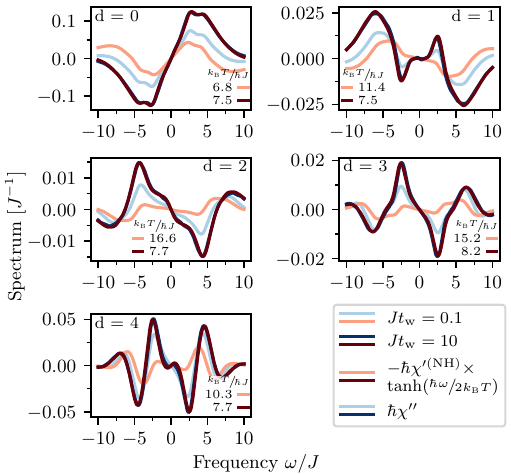}
	\caption{\label{fig:fdrspatial2ddetail}%
		Same as \protect\cref{fig:fdrspatialdetail}, but for the \ac{2d} Bose--Hubbard system.
		The data are the same as in \protect\cref{fig:fdrspatial2d}.
		At long waiting times, the dynamic susceptibility~$\chitwoprime$ (blue) agrees well with $\chiprimenh$ (red) after the latter is rescaled according to the \ac{fdr}~\protect\eqref{eq:fdrsusceptibility}.%
	}
\end{figure}

The fulfillment of the \ac{fdr} at late times is further illustrated in \cref{fig:fdrspatial2ddetail} (analogously to \cref{fig:fdrspatialdetail}), where the agreement between $\chitwoprime$ and the rescaled $\chiprimenh$ is remarkable.
This supports our conclusion in \cref{app:quench:bh1d} that the deviations observed for smaller systems are likely due to finite-size effects, while sufficiently large systems thermalize as expected.

\section{\label{app:perturbationtheory}Derivation of non-Hermitian linear response via the pulsed quantum Zeno effect}

In this Appendix, we derive how an effective non-Hermitian perturbation can be generated by coupling the system to an ancilla and exploiting the quantum Zeno effect (see \cref{fig:ancillapostselection}).
We first discuss in detail a single step in the Zeno evolution, which corresponds to a $\delta$-like non-Hermitian perturbation that allows one to access the unequal-time anti-commutator in time domain.
Furthermore, we derive the general form of perturbation operators that can be realized this way.
Finally, we explain how the pulsed Zeno effect, generated by repeatedly projecting the coupled system on the subspace with no particles in the ancilla, allows one to realize an effective evolution under a non-Hermitian Hamiltonian for an extended period of time.

\subsection{\label{app:singlezeno}Single Zeno step}

In what follows, we use time-dependent perturbation theory to derive \cref{eq:ancillameasurementnormalized,eq:ancillameasurementtrace}, describing, respectively, the conditional and unconditional response after a single step in the Zeno evolution of coupling to the ancilla followed by a projection on the empty-ancilla subspace, as depicted in \cref{fig:ancillapostselection:singlezeno}.

The protocol starts by evolving the initial state $\rho_0$ under the Hamiltonian~$H_0$ up to the waiting time~$\waitingtime$, at which the perturbation is applied. Before the coupling, the state is given by $\rho(\waitingtime) = \etothepowerof{-i H_0 \waitingtime / \hbar} \rho_0 \etothepowerof{i H_0 \waitingtime / \hbar}$.
In \cref{sec:singlezenostep}, we have approximated a $\delta$-like perturbation as a rectangular pulse of duration~$\delta t$.
Here, we consider a slightly more general scenario where we allow for an arbitrarily shaped pulse~$g(t)$ as in \cref{eq:hamiltoniancouplingtimedependent}.
The corresponding total Hamiltonian reads $H(t) = H_0 + g(t) H_\mathrm{cpl}$ with $H_\cpl$ given by \cref{eq:hamiltoniancoupling}.
It is convenient to work in the interaction picture with respect to the unperturbed Hamiltonian~$H_0$.
Time evolution is then governed by the von Neumann equation
\begin{equation}
	\label{eq:vonneumanninteractionpicture}
	i \hbar \frac{\diff}{\diff t} \tilde{\rho}(t) = \Commutator{g(t) \tilde{H}_\mathrm{cpl}(t)}{\tilde{\rho}(t)} ,
\end{equation}
where $\tilde{\rho}(t) = \etothepowerof{i H_0 t / \hbar} \rho(t) \etothepowerof{-i H_0 t / \hbar}$ and $\tilde{H}_\mathrm{cpl}(t) = \hbar \Omega \left[ \tilde{a}^\dagger(t) b + b^\dagger \tilde{a}(t) \right]$ with $\tilde{a}(t) = \etothepowerof{i H_0 t / \hbar} a \etothepowerof{-i H_0 t / \hbar}$ denote, respectively, the density operator and the coupling Hamiltonian in the interaction picture.
Rewriting \cref{eq:vonneumanninteractionpicture} as an integral equation and substituting the left-hand side into the right-hand side, we arrive at
\begin{widetext}
\begin{equation}
	\label{eq:vonneumannintegral}
	\tilde{\rho}(t) = \tilde{\rho}(\waitingtime) - \frac{i}{\hbar} \int_{\waitingtime}^{t} \diff t^{\prime} \, g(t^{\prime}) \Commutator{\tilde{H}_\mathrm{cpl}(t^{\prime})}{\tilde{\rho}(\waitingtime)}
- \frac{1}{\hbar^2} \int_{\waitingtime}^{t} \diff t^{\prime} \, g(t^{\prime}) \int_{\waitingtime}^{t^{\prime}} \diff t^{\prime \prime} \, g(t^{\prime \prime}) \Commutator{\tilde{H}_\mathrm{cpl}(t^{\prime})}{\Commutator{\tilde{H}_\mathrm{cpl}(t^{\prime \prime})}{\tilde{\rho}(t^{\prime \prime})}} .
\end{equation}
\end{widetext}
For the discussion of a single step in the quantum Zeno evolution, we consider a pulse~$g(t)$ with compact support on the interval ${[\waitingtime, \waitingtime + \delta t]}$, normalized such that $\int_{\waitingtime}^{\waitingtime + \delta t} \diff t \, g(t) = \delta t$.
If the pulse duration~$\delta t$ is much shorter than the characteristic time scales of~$H_0$, we can approximate $\tilde{\rho}(t^{\prime \prime}) \approx \tilde{\rho}(\waitingtime)$ and $\tilde{H}_\mathrm{cpl}(t^\prime) \approx \tilde{H}_\mathrm{cpl}(t^{\prime \prime}) \approx \tilde{H}_\mathrm{cpl}(\waitingtime)$ in the integrands, yielding, up to second order in $\delta t$, the result
\begin{equation}
	\label{eq:vonneumannintegralsecondorder}
\begin{split}
	\tilde{\rho}(\waitingtime + \delta t) \approx {} &\tilde{\rho}(\waitingtime) - \frac{i}{\hbar} \delta t \Commutator{\tilde{H}_\mathrm{cpl}(\waitingtime)}{\tilde{\rho}(\waitingtime)} \\
	&\begin{split}
	- \frac{\delta t^2}{2 \hbar^2} \big( &\Anticommutator{\tilde{H}_\mathrm{cpl}^2(\waitingtime)}{\tilde{\rho}(\waitingtime)} \\
	&- 2 \tilde{H}_\mathrm{cpl}(\waitingtime) \tilde{\rho}(\waitingtime) \tilde{H}_\mathrm{cpl}(\waitingtime) \big) .
	\end{split}
\end{split}
\end{equation}

We require the ancilla to be empty before the coupling. More specifically, we assume that the combined state of system and ancilla at the waiting time~$\waitingtime$ is given by the product state $\tilde{\rho}(\waitingtime) = \tilde{\rho}_\mathrm{S}(\waitingtime) \otimes \tilde{\rho}_\mathrm{A}$, where the ancilla is in the pure vacuum state $\tilde{\rho}_\mathrm{A} = \ketbra{0}{0}$.
(For concreteness, we focus here on bosonic systems, but the derivation for fermions proceeds analogously and yields the same result for the unequal-time anti-commutator.)
Inserting this state into \cref{eq:vonneumannintegralsecondorder}, we obtain
\begin{equation}
	\label{eq:postancillacoupling}
\begin{split}\raisetag{12ex}
	\tilde{\rho}(&\waitingtime + \delta t) = {} \tilde{\rho}_\mathrm{S}(\waitingtime) \otimes \ketbra{0}{0} \\
	&- i \Omega \delta t \left[ \tilde{a}(\waitingtime) \tilde{\rho}_\mathrm{S}(\waitingtime) \otimes \ketbra{1}{0} - \mathrm{h.c.} \right] \\
	&\begin{split}
	- \frac{\left( \Omega \delta t \right)^2}{2}
	\Big[
		&\Anticommutator{\tilde{n}(\waitingtime)}{\tilde{\rho}_\mathrm{S}(\waitingtime)} \otimes \ketbra{0}{0} \\
		&- 2 \tilde{a}(\waitingtime) \tilde{\rho}_\mathrm{S}(\waitingtime) \tilde{a}^\dagger(\waitingtime) \otimes \ketbra{1}{1} \\
		&+ \sqrt{2} \left( \tilde{a}^2(\waitingtime) \tilde{\rho}_\mathrm{S}(\waitingtime) \otimes \ketbra{2}{0} + \mathrm{h.c.} \right)
	\Big] ,
	\end{split}
\end{split}
\end{equation}
where $\tilde{n}(\waitingtime) = \tilde{a}^\dagger(\waitingtime) \tilde{a}(\waitingtime)$ is the number operator and $\mathrm{h.c.}$ denotes the Hermitian conjugate.

After coupling the system to the ancilla, a single step in the Zeno evolution is completed by measuring the population of the ancilla, projecting the state on a subspace with a definite number of particles in the ancilla.
Let $\pprojector_n = \mathds{1} \otimes \ketbra{n}{n}$ be the projection operator on the subspace with $n$ particles in the ancilla.
Since $\commutator{\pprojector_n}{H_0} = 0$, the measurement can optionally be deferred up to the final observation time (cf.\ Ref.~\cite{Uhrich2017}).
The projected states read
\begin{subequations}
	\label{eq:projectedstates}
\begin{align}
	\label{eq:projectedstatezero}
	\begin{split}
	\pprojector_0 \tilde{\rho}(\waitingtime + \delta t) \pprojector_0 &=
	\left( \tilde{\rho}_\mathrm{S}(\waitingtime) - s \Anticommutator{\tilde{n}(\waitingtime)}{\tilde{\rho}_\mathrm{S}(\waitingtime)} \right) \\
	&\quad\otimes \ketbra{0}{0} ,
	\end{split} \\
	\label{eq:projectedstateone}
	\pprojector_1 \tilde{\rho}(\waitingtime + \delta t) \pprojector_1 &= 2 s \tilde{a}(\waitingtime) \tilde{\rho}_\mathrm{S}(\waitingtime) \tilde{a}^\dagger(\waitingtime) \otimes \ketbra{1}{1} ,
\end{align}
\end{subequations}
where $s = (\Omega \delta t)^2 / 2$ is the effective coupling strength, and $\pprojector_n \tilde{\rho}(\waitingtime + \delta t) \pprojector_n = 0$ for $n \ge 2$, up to second order in $\delta t$.
The probability of detecting $n$ particles in the ancilla is then given by $p(n) = \Tr [ \pprojector_n \tilde{\rho}(\waitingtime + \delta t) \pprojector_n ]$, which yields
\begin{subequations}
	\label{eq:ancillaprobabilities}
\begin{align}
	p(0) &= 1 - 2 s \Braket{n(\waitingtime)}_0 , \\
	p(1) &= 2 s \Braket{n(\waitingtime)}_0 ,
\end{align}
\end{subequations}
and $p(n \ge 2) = 0$, up to second order in $\delta t$.
Here, we have used $\Tr [ \tilde{\rho}(\waitingtime) \tilde{n}(\waitingtime) ] = \Tr [ \rho_0 n(\waitingtime) ] = \braket{n(\waitingtime)}_0$.

Remarkably, the result in \cref{eq:projectedstatezero} can, to leading order in the coupling, be expressed as the evolution under an effective non-Hermitian Hamiltonian,
\begin{equation}
	\pprojector_0 \tilde{\rho}(\waitingtime + \delta t) \pprojector_0 = \etothepowerof{-i \tilde{H}_\eff(\waitingtime) \delta t} \tilde{\rho}(\waitingtime) \etothepowerof{i \tilde{H}_\eff^\dagger(\waitingtime) \delta t} ,
\end{equation}
with $\tilde{H}_\eff(t) = -i \hbar s \tilde{A}(t) / \delta t$ and perturbation operator~$A = n = a^\dagger a$.

According to Lüders' rule~\cite{Lueders1950}, the conditional state, given that $n$ particles have been detected in the ancilla, is obtained by normalizing the projected states~\eqref{eq:projectedstates} by the respective probabilities~\eqref{eq:ancillaprobabilities}, i.e., $\tilde{\rho}(\waitingtime + \delta t | n) = {\pprojector_n \tilde{\rho}(\waitingtime + \delta t) \pprojector_n} / p(n)$. Up to leading order in $s$, we find
\begin{subequations}
	\label{eq:conditionalstates}
	\begin{align}
		\label{eq:conditionalstatezero}
		\begin{split}
		\tilde{\rho}(\waitingtime + \delta t | 0) &=
		\tilde{\rho}(\waitingtime) - s \big[ \Anticommutator{\tilde{n}(\waitingtime)}{\tilde{\rho}(\waitingtime)}\\
		&\quad\phantom{\tilde{\rho}(\waitingtime) - s \big[} - 2 \Braket{n(\waitingtime)}_0 \tilde{\rho}(\waitingtime) \big] ,
		\end{split} \\
		\label{eq:conditionalstateone}
		\tilde{\rho}(\waitingtime + \delta t | 1) &= \frac{\tilde{a}(\waitingtime) \tilde{\rho}(\waitingtime) \tilde{a}^\dagger(\waitingtime)}{\Braket{n(\waitingtime)}_0} ,
	\end{align}
\end{subequations}
where we have discarded (or traced out) the ancilla and omitted the subscripts indicating system density operators.
By contrast, if the ancilla population is not measured or if the measurement outcomes are ignored, the state after the coupling is instead described by the unconditional density operator
\begin{equation}
	\label{eq:unconditionalstate}
\begin{split}
\tilde{\rho}(\waitingtime + \delta t) &= \sum_{n} p(n) \tilde{\rho}(\waitingtime + \delta t | n) \\
&= \tilde{\rho}(\waitingtime) - s \big( \Anticommutator{\tilde{n}(\waitingtime)}{\tilde{\rho}(\waitingtime)} \\
&\quad\phantom{\tilde{\rho}(\waitingtime) - s \big(}- 2 \tilde{a}(\waitingtime) \tilde{\rho}(\waitingtime) \tilde{a}^\dagger(\waitingtime) \big) ,
\end{split}
\end{equation}
which can be obtained directly from \cref{eq:postancillacoupling} after tracing out the ancilla.

For times~$t > \waitingtime + \delta t$, the coupling is switched off and the system evolves solely under the Hamiltonian~$H_0$. According to \cref{eq:vonneumanninteractionpicture}, this evolution is trivial in the interaction picture, such that $\tilde{\rho}(t) = \tilde{\rho}(\waitingtime + \delta t)$.
The (unnormalized) expectation value of an observable~$B$ with respect to the state~\eqref{eq:projectedstatezero} projected on the empty-ancilla subspace reads
\begin{equation}
\begin{split}
\Tr \left[ \tilde{B}(t) \pprojector_0 \tilde{\rho}(t) \pprojector_0 \right] &= \Tr \left[ \tilde{B}(t) \tilde{\rho}(\waitingtime) \right] \\
&\hphantom{{}={}} - s \Tr \left[ \tilde{B}(t) \Anticommutator{\tilde{n}(\waitingtime)}{\tilde{\rho}(\waitingtime)} \right] \\
&= \Tr \left[ B(t) \rho_0 \right] \\
&\hphantom{{}={}} - s \Tr \left[ B(t) \Anticommutator{n(\waitingtime)}{\rho_0} \right] \\
&= \Braket{B(t)}_0 - s \Braket{\Anticommutator{B(t)}{n(\waitingtime)}}_0 ,
\end{split}
\end{equation}
where, in the second step, we have transformed from the interaction picture to the Heisenberg picture, and in the last step, we have used the cyclic property of the trace.
This is the result reported in \cref{eq:ancillameasurementunnormalized} in the main text.
For the conditional state~\eqref{eq:conditionalstatezero}, we recover \cref{eq:ancillameasurementnormalized}, which describes a post-selected measurement conditioned on the empty ancilla.

If we were to post-select on the condition that a single particle is detected in the ancilla, corresponding to the conditional state~\cref{eq:conditionalstateone}, we would instead obtain
\begin{equation}
	\Tr \left[ \tilde{B}(t) \tilde{\rho}(t | 1) \right] = \frac{\Braket{a^\dagger(\waitingtime) B(t) a(\waitingtime)}_0}{\Braket{n(\waitingtime)}_0} .
\end{equation}
This quantity contributes a systematic error in the case of faulty detection with false negatives (see the discussion in \cref{sec:comparison}). From the derivation in this Appendix it becomes clear that, up to second order in $\delta t$, i.e., up to linear order in the effective coupling strength~$s$, the error is dominated by single occupancies of the ancilla site.
Finally, the unconditional expectation value~\eqref{eq:ancillameasurementtrace} follows from \cref{eq:unconditionalstate}. This shows that post-selection is essential in order to remove the undesired contribution in form of the \enquote{recycling term} due to \cref{eq:conditionalstateone}, enabling access to the unequal-time anti-commutator.

\subsection{\label{app:generalcoupling}General system--ancilla coupling}

The coupling schemes in \cref{fig:ancillapostselection:density,fig:ancillapostselection:correlator} are designed to realize non-Hermitian perturbations by the density operator and the hopping operator, respectively.
We now consider the general situation where an arbitrary number of system modes is coupled to up to $M$ ancillary modes.
This scenario is described by the general coupling Hamiltonian
\begin{equation}
	\label{eq:hamiltoniancouplinggeneral}
	H_\cpl = \sum_{m = 1}^{M} \hbar \Omega_m \left( b_m^\dagger \alpha_m + \alpha_m^\dagger b_m \right) ,
\end{equation}
where the operator
\begin{equation}
	\alpha_m = \sum_{\ell} \lambda_{m \ell} a_\ell
\end{equation}
is a linear combination of system modes~$a_\ell$ with coefficients $\lambda_{m \ell} \in \doublestruck{C}$, coupled to the $m$-th ancilla with coupling strength~$\Omega_m \ge 0$.
The configuration in \cref{fig:ancillapostselection:density}, a single lattice site~$\ell^*$ coupled to a single ancilla, is recovered for $M = 1$ and $\lambda_{1 \ell} = \delta_{\ell \ell^*}$, while \cref{fig:ancillapostselection:correlator}, two sites~$\ell_1$ and $\ell_2$ simultaneously coupled to a single ancilla, corresponds to $M = 1$ and $\lambda_{1 \ell} = \delta_{\ell \ell_1} + \delta_{\ell \ell_2}$.

As before, we consider a short coupling pulse of duration~$\delta t$ such that the state after the coupling is given by \cref{eq:vonneumannintegralsecondorder}.
Subsequently, a measurement of the individual ancilla occupancies is performed and the state is conditioned on the outcome of that measurement (as mentioned above, the measurement may also be deferred up to the final observation time).
Given the outcome $(n_1, \dots, n_M)$, the post-measurement state, up to a normalization, reads $\pprojector_{n_1 \dots n_M} \tilde{\rho}(\waitingtime + \delta t) \pprojector_{n_1 \dots n_M}$, where $\pprojector_{n_1 \dots n_M} = \doublestruck{1} \otimes \ketbra{n_1 \cdots n_M}{n_1 \cdots n_M}$ is the projection operator on the subspace with a definite ancilla population corresponding to the measurement outcome.

Up to leading order in the coupling, only processes where at most a single particle ends up in one of the ancillas contribute.
Let $\pprojector_0 = \pprojector_{0 \dots 0}$ denote the projector on the subspace with all ancillas empty.
The projector on the subspace with a single particle in the $m$-th ancilla and all others empty can then be expressed as $\pprojector_1^{(m)} = b_m^\dagger \pprojector_0 b_m$.
Using the bosonic commutation relations
\begin{subequations}
	\label{eq:commutationrelationsbosons}
\begin{align}
	\commutator[\big]{\alpha_m}{b_{m^\prime}} &= \commutator[\big]{\alpha_m}{b_{m^\prime}^\dagger} = 0 , \\
	\commutator[\big]{b_m}{b_{m^\prime}^\dagger} &= \delta_{m m^\prime} ,
\end{align}
\end{subequations}
we find the (unnormalized) post-measurement states
\begin{subequations}
	\label{eq:projectedstatesgeneral}
\begin{gather}
	\begin{multlined}
	\pprojector_0 \tilde{\rho}(\waitingtime + \delta t) \pprojector_0 = \\ \tilde{\rho}(\waitingtime) - \sum_{m = 1}^{M} s_m \Anticommutator{\tilde{\alpha}_m^\dagger(\waitingtime) \tilde{\alpha}_m(\waitingtime)}{\tilde{\rho}(\waitingtime)} ,
	\end{multlined} \\
	\pprojector_1^{(m)} \tilde{\rho}(\waitingtime + \delta t) \pprojector_1^{(m)} = 2 s_m \tilde{a}_m(\waitingtime) \tilde{\rho}(\waitingtime) \tilde{a}_m^\dagger(\waitingtime) ,
\end{gather}
\end{subequations}
where we have traced out the ancillas and introduced the effective coupling strengths~$s_m = (\Omega_m \delta t)^2 / 2$.
We note that this result holds for fermions as well, where instead of \cref{eq:commutationrelationsbosons} the corresponding fermionic anti-commutation relations apply.
The respective probabilities of finding no particles in any ancilla or a single particle in the $m$-th ancilla read
\begin{subequations}
	\label{eq:ancillaprobabilitiesgeneral}
\begin{align}
	P_0 &= 1 - 2 \sum_{m = 1}^{M} s_m \Braket{\alpha_m^\dagger(\waitingtime) \alpha_m(\waitingtime)}_0 , \\
	P_1^{(m)} &= 2 s_m \Braket{\alpha_m^\dagger(\waitingtime) \alpha_m(\waitingtime)}_0 .
\end{align}
\end{subequations}

A comparison of \cref{eq:projectedstatesgeneral,eq:ancillaprobabilitiesgeneral} with \cref{eq:projectedstates,eq:ancillaprobabilities} shows that the coupling to the $m$-th ancilla in the general coupling Hamiltonian~\eqref{eq:hamiltoniancouplinggeneral} generates an effective non-Hermitian perturbation by the operator
\begin{equation}
	A_m = \alpha_m^\dagger \alpha_m = \sum_{\ell \ell^\prime} \lambda_{m \ell}^* \lambda_{m \ell^\prime} a_l^\dagger a_{\ell^\prime} .
\end{equation}
Coupling to multiple ancillas simultaneously can be used to realize perturbation by (arbitrarily weighted) sums of the operators~$A_m$.
This demonstrates that our scheme enables flexible access to unequal-time correlations and \acp{fdr} for a wide range of observables, two specific examples of which, namely densities and nearest-neighbor correlators, we have illustrated in \cref{sec:realization}.

\subsection{\label{app:pulsedzeno}Prolonged non-Hermitian evolution via the pulsed quantum Zeno effect}

To gain a deeper understanding of how the pulsed quantum Zeno effect enables a prolonged evolution under an effective non-Hermitian Hamiltonian, we consider in detail two consecutive steps in the Zeno evolution, using a similar formalism as in Ref.~\cite{Facchi2008}, and investigate the role of the projective measurement after the first step.

To this end, let $\pprojector$ denote the projection operator on the empty-ancilla subspace~$\mathscr{H}_{\pprojector}$ and $\qprojector = \doublestruck{1} - \pprojector$ the projector on the complementary subspace~$\mathscr{H}_{\qprojector} = \mathscr{H}_{\pprojector}^\perp$ with at least one particle in the ancilla.
It is convenient to write the density operator~$\rho$ on the total Hilbert space~$\mathscr{H} = \mathscr{H}_{\pprojector} \oplus \mathscr{H}_{\qprojector}$ in the form
\begin{equation}
	\rho = \begin{pmatrix}
		\rho_{\pprojector \pprojector} & \rho_{\pprojector \qprojector} \\
		\rho_{\qprojector \pprojector} & \rho_{\qprojector \qprojector}
	\end{pmatrix} ,
\end{equation}
where $\rho_{\pprojector \pprojector} = \pprojector \rho \pprojector$ and $\rho_{\qprojector \qprojector} = \qprojector \rho \qprojector$ are the populations of $\mathscr{H}_{\pprojector}$ and $\mathscr{H}_{\qprojector}$, respectively, and $\rho_{\pprojector \qprojector} = \pprojector \rho \qprojector = \rho_{\qprojector \pprojector}^\dagger$ are the coherences between these two subspaces.
Similarly, the time evolution operator from time $t_0$ to $t$ corresponding to \cref{eq:vonneumanninteractionpicture} can be expressed as
\begin{equation}
	U(t, t_0) = \begin{pmatrix}
		U_{\pprojector \pprojector}(t, t_0) & U_{\pprojector \qprojector}(t, t_0) \\
		U_{\qprojector \pprojector}(t, t_0) & U_{\qprojector \qprojector}(t, t_0)
	\end{pmatrix}
\end{equation}
with $U_{\pprojector \qprojector}(t, t_0) = \pprojector U(t, t_0) \qprojector$.

Let us denote the initial state $\tilde{\rho}(\waitingtime)$ of the Zeno evolution by $\rho^{0}$ and set $t_0 = \waitingtime$.
To keep the notation simple, for the purposes of this subsection, we omit the tilde indicating interaction picture operators and use the abbreviation $\mathcal{H}(t) \equiv g(t) \tilde{H}_\cpl(t)$.
Since the ancilla is initially empty, we have $\rho^{0}_{\pprojector \qprojector} = \rho^{0}_{\qprojector \qprojector} = 0$.
The unitary evolution from time~$t_0$ to $t_1$ in the presence of the system--ancilla coupling changes the state as
\begin{equation}
	\label{eq:firstzenounitary}
\begin{split}
	\rho^{0} &=
	\begin{pmatrix}
		\rho_{\pprojector \pprojector}^{0} & 0 \\
		0 & 0
	\end{pmatrix} \\
	\overset{U(t_1, t_0)}{\longrightarrow}
	\rho^{1} &=
	\begin{pmatrix}
		U_{\pprojector \pprojector}^1 \rho_{\pprojector \pprojector}^{0} (U_{\pprojector \pprojector}^1)^\dagger & U_{\pprojector \pprojector}^1 \rho_{\pprojector \pprojector}^{0} (U_{\qprojector \pprojector}^1)^\dagger \\
		U_{\qprojector \pprojector}^1 \rho_{\pprojector \pprojector}^{0} (U_{\pprojector \pprojector}^1)^\dagger & U_{\qprojector \pprojector}^1 \rho_{\pprojector \pprojector}^{0} (U_{\qprojector \pprojector}^1)^\dagger
	\end{pmatrix} ,
\end{split}
\end{equation}
where $U_{\pprojector \qprojector}^i = \pprojector U(t_i, t_{i - 1}) \qprojector$.
From \cref{eq:vonneumannintegral}, using $\pprojector^2 = \pprojector$, $\qprojector^2 = \qprojector$, $\pprojector \qprojector = \qprojector \pprojector = 0$, and $\pprojector \mathcal{H}(t) \pprojector = 0$, we obtain the populations and coherences of $\rho^1$, up to quadratic order in the coupling, as
\begin{align}
	\begin{split}
	\rho_{\pprojector \pprojector}^1 &= \rho_{\pprojector \pprojector}^0 - \frac{1}{\hbar^2} \int_{t_0}^{t_1} \diff t^\prime \int_{t_0}^{t^\prime} \diff t^{\prime \prime} \\
	&\hphantom{{}={} \rho_{\pprojector \pprojector}^0 {}-{}} \times \left[ \mathcal{H}_{\pprojector \qprojector}(t^\prime) \mathcal{H}_{\qprojector \pprojector}(t^{\prime \prime}) \rho_{\pprojector \pprojector}^0 + \mathrm{h.c.} \right] ,
	\end{split} \\
	\begin{split}
	\rho_{\pprojector \qprojector}^1 &= \frac{i}{\hbar} \int_{t_0}^{t_1} \rho_{\pprojector \pprojector}^0 H_{\pprojector \qprojector}(t^\prime) \\
	&\hphantom{{}={}}- \frac{1}{\hbar^2} \int_{t_0}^{t_1} \diff t^\prime \int_{t_0}^{t^\prime} \diff t^{\prime \prime} \, \rho_{\pprojector \pprojector}^0 \mathcal{H}_{\pprojector \qprojector}(t^{\prime \prime}) \mathcal{H}_{\qprojector \qprojector}(t^{\prime}) ,
	\end{split} \\
	\rho_{\qprojector \qprojector}^1 &= \frac{1}{\hbar^2} \int_{t_0}^{t_1} \diff t^\prime \int_{t_0}^{t_1} \diff t^{\prime \prime} \, \mathcal{H}_{\qprojector \pprojector}(t^\prime) \rho_{\pprojector \pprojector}^0 \mathcal{H}_{\pprojector \qprojector}(t^{\prime \prime}) ,
\end{align}
where $\mathcal{H}_{\pprojector \qprojector}(t) = \mathcal{H}_{\qprojector \pprojector}^\dagger(t) = \pprojector \mathcal{H}(t) \qprojector = g(t) \hbar \Omega \pprojector \tilde{a}^\dagger(t) b \qprojector$.
Measuring the ancilla population projects the state on the subspace with a definite number of particles in the ancilla. Without registering the measurement outcome, this yields the unconditional state~$\pprojector \rho^1 \pprojector + \qprojector \rho^1 \qprojector$.
Crucially, the measurement process destroys any coherences~$\rho_{\pprojector \qprojector}$ and $\rho_{\qprojector \pprojector}$ between the Zeno subspaces~$\mathscr{H}_\pprojector$ and $\mathscr{H}_\qprojector$.
We are interested in measurement outcomes where no particles are detected in the ancilla.
Conditioning the state on this outcome corresponds to a projection on the empty-ancilla subspace~$\mathscr{H}_\pprojector$,
\begin{equation}
	\label{eq:firstzenoprojection}
	\rho^{1} =
	\begin{pmatrix}
		\rho_{\pprojector \pprojector}^1 & \rho_{\pprojector \qprojector}^1 \\
		\rho_{\qprojector \pprojector}^1 & \rho_{\qprojector \qprojector}^1
	\end{pmatrix}
	\overset{\pprojector}{\longrightarrow}
	\pprojector \rho^{1} \pprojector =
	\begin{pmatrix}
		\rho_{\pprojector \pprojector}^{1} & 0 \\
		0 & 0
	\end{pmatrix} .
\end{equation}

The second Zeno step proceeds analogously to \cref{eq:firstzenounitary,eq:firstzenoprojection}: the state first evolves unitarily from time $t_1$ to $t_2$ in the presence of the system--ancilla coupling and is then projected on the empty-ancilla subspace~$\mathscr{H}_\pprojector$,
\begin{equation}
\begin{split}
	\pprojector \rho^{1} \pprojector
	\overset{U(t_2, t_1)}{\longrightarrow}
	\rho^{2} =
	\begin{pmatrix}
	\rho_{\pprojector \pprojector}^2 & \rho_{\pprojector \qprojector}^2 \\
	\rho_{\qprojector \pprojector}^2 & \rho_{\qprojector \qprojector}^2
	\end{pmatrix} \\
	\overset{\pprojector}{\longrightarrow}
	\pprojector \rho^{2} \pprojector =
	\begin{pmatrix}
		\rho_{\pprojector \pprojector}^{2} & 0 \\
		0 & 0
	\end{pmatrix} ,
\end{split}
\end{equation}
with
\begin{equation}
	\label{eq:secondzenoprojection}
\begin{split}
	\rho_{\pprojector \pprojector}^{2} &= U_{\pprojector \pprojector}(t_2, t_1) \rho_{\pprojector \pprojector}^{1} U_{\pprojector \pprojector}^\dagger(t_2, t_1) \\
	&= U_{\pprojector \pprojector}^2 U_{\pprojector \pprojector}^1 \rho_{\pprojector \pprojector}^{0} (U_{\pprojector \pprojector}^1)^\dagger (U_{\pprojector \pprojector}^2)^\dagger \\
	&= \rho_{\pprojector \pprojector}^0 - \frac{1}{\hbar^2} \bigg( \int_{t_0}^{t_1} \diff t^\prime \int_{t_0}^{t^\prime} \diff t^{\prime \prime} + \int_{t_1}^{t_2} \diff t^\prime \int_{t_1}^{t^\prime} \diff t^{\prime \prime} \bigg) \\
	&\hphantom{{}={} \rho_{\pprojector \pprojector}^0 {}-{}} \times \left[ \mathcal{H}_{\pprojector \qprojector}(t^\prime) \mathcal{H}_{\qprojector \pprojector}(t^{\prime \prime}) \rho_{\pprojector \pprojector}^0 + \mathrm{h.c.} \right] ,
\end{split}
\end{equation}
up to leading order in the coupling.
It is instructive to compare this result to the one obtained if no measurement is performed after the first step.
The state then receives additional contributions from the coherences, yielding, to leading order in the coupling,
\begin{equation}
	\label{eq:secondzenoprojectionnomeasurement}
\begin{split}
	\rho_{\pprojector \pprojector}^{2 \, \prime} &= \pprojector U(t_2, t_1) \rho^1 U^\dagger(t_2, t_1) \pprojector \\
	&= U_{\pprojector \pprojector}^2 \rho_{\pprojector \pprojector}^1 (U_{\pprojector \pprojector}^2)^\dagger + U_{\pprojector \pprojector}^2 \rho_{\pprojector \qprojector}^1 (U_{\pprojector \qprojector}^2)^\dagger \\
	&\hphantom{{}={}}+ U_{\pprojector \qprojector}^2 \rho_{\qprojector \pprojector}^1 (U_{\pprojector \pprojector}^2)^\dagger + U_{\pprojector \qprojector}^2 \rho_{\qprojector \qprojector}^1 (U_{\pprojector \qprojector}^2)^\dagger \\
	&= \rho_{\pprojector \pprojector}^2 - \int_{t_0}^{t_1} \diff t^\prime \int_{t_1}^{t_2} \diff t^{\prime \prime} \\
	&\hphantom{{}= \rho_{\pprojector \pprojector}^2 -{}} \times \left[ \rho_{\pprojector \pprojector}^0 \mathcal{H}_{\pprojector \qprojector}(t^\prime) H_{\qprojector \pprojector}(t^{\prime \prime}) + \mathrm{h.c.} \right] \\
	&= \rho_{\pprojector \pprojector}^0 - \int_{t_0}^{t_2} \diff t^\prime \int_{t_0}^{t^\prime} \diff t^{\prime \prime} \\
	&\hphantom{{}= \rho_{\pprojector \pprojector}^0 -{}} \times \left[ \mathcal{H}_{\pprojector \qprojector}(t^\prime) H_{\qprojector \pprojector}(t^{\prime \prime}) \rho_{\pprojector \pprojector}^0 + \mathrm{h.c.} \right] .
\end{split}
\end{equation}
The result in the last line could have been directly obtained from \cref{eq:vonneumannintegral} for $t = t_2$ by applying the projector~$\pprojector$ on both sides.
This is evident because without the projection after the first step, the system plus ancilla evolves unitarily from time~$t_0$ to $t_2$.
However, \cref{eq:secondzenoprojectionnomeasurement} explicitly exposes the crucial effect of the measurement after the first step: the last term in the second-to-last line is precisely the contribution from the coherences~$\rho_{\pprojector \qprojector}^1$ and $\rho_{\qprojector \pprojector}^1$, which is missing in \cref{eq:secondzenoprojection} since the coherences have been destroyed by the measurement.

Iterating the Zeno evolution for $n$ steps up to time~$t_n$ (including projections after each step), the resulting state is given, to leading order in the coupling, by
\begin{equation}
\begin{split}
	\rho_{\pprojector \pprojector}^{n} &= \rho_{\pprojector \pprojector}^0 - \frac{1}{\hbar^2} \sum_{i = 0}^{n - 1} \int_{t_i}^{t_{i + 1}} \diff t^\prime \int_{t_i}^{t^\prime} \diff t^{\prime \prime} \\
	&\hphantom{{}= \rho_{\pprojector \pprojector}^0 -{}} \times \left[ \mathcal{H}_{\pprojector \qprojector}(t^\prime) H_{\qprojector \pprojector}(t^{\prime \prime}) \rho_{\pprojector \pprojector}^0 + \mathrm{h.c.} \right] .
\end{split}
\end{equation}
If the duration~$t_{i + 1} - t_i$ of each Zeno step is sufficiently short as compared to the time scales of the unperturbed Hamiltonian as well as the modulation~$g(t)$, the integrand in each integral is approximately constant, yielding \cref{eq:ancillameasurementunnormalizedpulsedzeno}. As discussed in \cref{sec:pulsedzeno}, this result can in turn be interpolated by a continuous evolution under an effective non-Hermitian Hamiltonian (see \cref{fig:ancillapostselection:singlezeno}).

By contrast, if the state is only projected at the final observation time, but no projections are performed during the evolution as in \cref{eq:secondzenoprojectionnomeasurement}, we obtain, to leading order in the coupling,
\begin{equation}
	\begin{split}
		\rho_{\pprojector \pprojector}^{n \, \prime} &= \rho_{\pprojector \pprojector}^0 - \frac{1}{\hbar^2} \int_{t_0}^{t_n} \diff t^\prime \int_{t_0}^{t^\prime} \diff t^{\prime \prime} \\
		&\hphantom{{}= \rho_{\pprojector \pprojector}^0 -{}} \times \left[ \mathcal{H}_{\pprojector \qprojector}(t^\prime) H_{\qprojector \pprojector}(t^{\prime \prime}) \rho_{\pprojector \pprojector}^0 + \mathrm{h.c.} \right] ,
	\end{split}
\end{equation}
which corresponds to \cref{eq:ancillameasurementunnormalizedgeneral} in the main text.
Since the evolution time $t_n - t_0$ may be on the same order or longer than the characteristic time scales of the unperturbed Hamiltonian, it is not possible to approximate the integrand as constant here.
Consequently, this procedure does not yield the desired two-time anti-commutator in general.

As these discussions show, exploiting the Zeno effect allows us to apply effective non-Hermitian perturbations for an extended period of time.
The essential mechanism is the destruction of the coherences between the Zeno subspaces due to the intermittent measurements.
As explained in \cref{sec:continuouszeno,app:engineereddissipationtonhh}, this effect can be mimicked if the ancilla is exposed to strong (engineered) dissipation, which represents an alternative way of realizing non-Hermitian linear response via the quantum Zeno effect.

\section{\label{app:engineereddissipationtonhh}Derivation of the effective non-Hermitian Hamiltonian from engineered dissipation}

In this Appendix, we use stochastic calculus to derive the Lindblad master equation~\eqref{eq:masterequation} by noise averaging the stochastic von~Neumann equation~\eqref{eq:stochasticvonneumann}, which describes the engineered dephasing scenario in \cref{sec:engineereddissipation}.
We then consider the strong noise limit and show how the continuous quantum Zeno effect gives rise to the evolution under an effective non-Hermitian Hamiltonian.

\subsection{\label{app:engineereddissipationtonhh:masterequation}Derivation of the master equation}

The Gaussian white-noise process~$\xi(t)$ considered in \cref{sec:engineereddissipation} can be viewed as the idealization of a smooth physical noise process with finite correlation time, arising, for example, from a rapidly fluctuating electric or magnetic field.
As such, it is appropriate to interpret the stochastic von Neumann equation~\cref{eq:stochasticvonneumann} as \iac{sde} in Stratonovich form, which obeys the rules of ordinary calculus~\cite{Kloeden1992,Gardiner2009}. In addition, in the form of \cref{eq:stochasticvonneumann}, unitary evolution of each realization ($\diff \Tr [ \rho(t) ] / \diff t = 0$) is only guaranteed if the Stratonovich interpretation is used~\cite{Hasegawa1980}.

The master equation~\cref{eq:masterequation} can be derived from the stochastic von Neumann equation~\cref{eq:stochasticvonneumann} by averaging over all noise realizations. However, in the Stratonovich interpretation, the Wiener increments~$\diff W(t)$ and the stochastic variable~$\rho(t)$ are not statistically independent at equal times, i.e., $\ensembleaverage{\rho(t) \diff W(t)} \neq 0$ in general.
To arrive at \cref{eq:masterequation}, it is therefore advantageous to convert \cref{eq:stochasticvonneumann} to an Itô \ac{sde}~\cite{Kloeden1992,Gardiner2009}.
According to the conversion rules, the linear Stratonovich \ac{sde} $\diff \rho = L_0(t) \rho \diff t + L_1(t) \rho \diff W$ is equivalent to the linear Itô \ac{sde} $\diff \rho = \left[ L_0(t) + L_1^2(t) / 2 \right] \rho \diff t + L_1(t) \rho \diff W$. In the case of \cref{eq:stochasticvonneumann}, $L_0$ and $L_1$ are given by the Liouvillian superoperators $L_0(t) \rho = -i \commutator{H(t)}{\rho} / \hbar$ and $L_1 \rho = -i \sqrt{2 \kappa} \commutator{b^\dagger b}{\rho}$, respectively.
Thus \cref{eq:stochasticvonneumann} is equivalent to the Itô \ac{sde}
\begin{equation}
\begin{split}
	\label{eq:stochasticvonneumannito}
	\diff \rho = &-\frac{i}{\hbar} \Commutator{H(t)}{\rho} \diff t - \kappa \left( \Anticommutator{L^\dagger L}{\rho} - 2 L \rho L^\dagger \right) \diff t \\
	&-i \sqrt{2 \kappa} \Commutator{b^\dagger b}{\rho} \diff W
\end{split}
\end{equation}
with $L = b^\dagger b$.
Since the solution of an Itô \ac{sde} is non-anticipating~\cite{Kloeden1992,Gardiner2009}, we have $\ensembleaverage{\rho(t) \diff W(t)} = 0$.
Therefore, taking the ensemble average of \cref{eq:stochasticvonneumannito}, the stochastic term vanishes, and the noise-averaged density operator $\sigma(t) = \ensembleaverage{\rho(t)}$ obeys the master equation~\eqref{eq:masterequation}.

\subsection{\label{app:engineereddissipationtonhh:nhh}Derivation of the effective non-Hermitian Hamiltonian}

To derive the effective non-Hermitian Hamiltonian governing the evolution in \cref{eq:effectivenonhermitian}, following Ref.~\cite{Stannigel2014}, we consider the strong noise limit of \cref{eq:masterequation} projected on the empty-ancilla subspace.
It is convenient to work in the interaction picture, i.e, in a rotating frame with respect to the unperturbed Hamiltonian~$H_0$.
\Cref{eq:masterequation} then reads
\begin{equation}
	\label{eq:masterequationinteractionpicture}
	\frac{\diff}{\diff t} \tilde{\sigma} = -\frac{i}{\hbar} \Commutator{\tilde{H}_\mathrm{cpl}(t)}{\tilde{\sigma}} - \kappa \left( \Anticommutator{L^\dagger L}{\tilde{\sigma}} - 2 L \tilde{\sigma} L^\dagger \right) ,
\end{equation}
where $\tilde{\sigma}(t) = \etothepowerof{i H_0 t / \hbar} \sigma(t) \etothepowerof{-i H_0 t / \hbar}$ and $\tilde{H}_\mathrm{cpl}(t) = g(t) \hbar \Omega \left[ \tilde{a}^\dagger(t) b + b^\dagger \tilde{a}(t) \right]$ with $\tilde{a}(t) = \etothepowerof{i H_0 t / \hbar} a \etothepowerof{-i H_0 t / \hbar}$. The operators $b$ and $b^\dagger$ as well as the Lindblad operators remain unchanged as they act on the ancilla only and therefore commute with $H_0$.

We now use the projection operator on the empty-ancilla subspace~$\pprojector=\mathcal{P}_0$ as well as its complement $\qprojector = \mathds{1} - \pprojector$ to derive coupled equations for the populations~$\tilde{\sigma}_{\pprojector \pprojector} = \pprojector \tilde{\sigma} \pprojector$ and $\tilde{\sigma}_{\qprojector \qprojector} = \qprojector \tilde{\sigma} \qprojector$ of the two subspaces, as well as for their coherences~$\tilde{\sigma}_{\pprojector \qprojector} = \pprojector \tilde{\sigma} \qprojector$ and $\tilde{\sigma}_{\qprojector \pprojector} = \qprojector \tilde{\sigma} \pprojector$.
The projection operators are Hermitian and satisfy the properties $\pprojector^2 = \pprojector$, $\qprojector^2 = \qprojector$, $\pprojector \qprojector = \qprojector \pprojector = 0$, as well as $\commutator{\pprojector}{H_0} = \commutator{\qprojector}{H_0} = 0$, the latter following from the fact that $H_0$ does not change the number of particles in the ancilla. Furthermore, since $\pprojector$ projects on the empty-ancilla subspace, we have $b \pprojector = \pprojector b^\dagger = 0$.
Applying the projectors $\pprojector$ and $\qprojector$ to \cref{eq:masterequationinteractionpicture} from the left and from the right yields the coupled system of equations
\begin{subequations}
	\label{eq:masterequationprojected}
\begin{align}
	\label{eq:masterequationprojectedpp}
	\frac{\diff}{\diff t} \tilde{\sigma}_{\pprojector \pprojector} = &-\frac{i}{\hbar} \left( \tilde{H}_{\pprojector \qprojector} \tilde{\sigma}_{\qprojector \pprojector} - \tilde{\sigma}_{\pprojector \qprojector} \tilde{H}_{\qprojector \pprojector} \right) , \\
	\label{eq:masterequationprojectedpq}
	\begin{split}
	\frac{\diff}{\diff t} \tilde{\sigma}_{\pprojector \qprojector} = &-\frac{i}{\hbar} \left( \tilde{H}_{\pprojector \qprojector} \tilde{\sigma}_{\qprojector \qprojector} - \tilde{\sigma}_{\pprojector \pprojector} \tilde{H}_{\pprojector \qprojector} \right) \\ &+\frac{i}{\hbar} \tilde{\sigma}_{\pprojector \qprojector} \qprojector \tilde{H}_\mathrm{cpl} \qprojector
	- \kappa \tilde{\sigma}_{\pprojector \qprojector} \qprojector L^\dagger L \qprojector ,
	\end{split} \\
	\label{eq:masterequationprojectedqp}
	\begin{split}
	\frac{\diff}{\diff t} \tilde{\sigma}_{\qprojector \pprojector} = &-\frac{i}{\hbar} \left( \tilde{H}_{\qprojector \pprojector} \tilde{\sigma}_{\pprojector \pprojector} - \tilde{\sigma}_{\qprojector \qprojector} \tilde{H}_{\qprojector \pprojector} \right) \\ &-\frac{i}{\hbar} \qprojector \tilde{H}_\mathrm{cpl} \qprojector \tilde{\sigma}_{\qprojector \pprojector}
	- \kappa \qprojector L^\dagger L \qprojector \tilde{\sigma}_{\qprojector \pprojector} ,
	\end{split} \\
	\label{eq:masterequationprojectedqq}
	\begin{split}
	\frac{\diff}{\diff t} \tilde{\sigma}_{\qprojector \qprojector} = &-\frac{i}{\hbar} \left( \tilde{H}_{\qprojector \pprojector} \tilde{\sigma}_{\pprojector \qprojector} - \tilde{\sigma}_{\qprojector \pprojector} \tilde{H}_{\pprojector \qprojector} \right) \\
	&-\frac{i}{\hbar} \Commutator{\qprojector \tilde{H}_\mathrm{cpl} \qprojector}{\tilde{\sigma}_{\qprojector \qprojector}} \\
	&- \kappa \qprojector \left( \Anticommutator{L^\dagger L}{\tilde{\sigma}_{\qprojector \qprojector}} - 2 L \tilde{\sigma}_{\qprojector \qprojector} L^\dagger \right) \qprojector ,
	\end{split}
\end{align}
\end{subequations}
where the operators $\tilde{H}_{\pprojector \qprojector}(t) = g(t) \hbar \Omega \pprojector \tilde{a}^\dagger(t) b \qprojector$ and $\tilde{H}_{\qprojector \pprojector}(t) = g(t) \hbar \Omega \qprojector b^\dagger \tilde{a}(t) \pprojector$ mix the two subspaces.
In deriving \cref{eq:masterequationinteractionpicture}, we have considered the engineered dephasing scenario described by the stochastic von Neumann equation~\eqref{eq:stochasticvonneumann}, in which case the Lindblad operator~$L = b^\dagger b$ is Hermitian and the projectors commute with~$L$. In the alternative setting, where the ancilla is subject to spontaneous decay, the Lindblad operator is given by $L = b$ and does not commute with the projectors. In this case, \cref{eq:masterequationprojectedpp,eq:masterequationprojectedpq,eq:masterequationprojectedqp} receive an additional contribution from the \enquote{recycling terms} $2 \kappa b \tilde{\sigma}_{\qprojector \qprojector} b^\dagger$, whose effect is to incoherently remove particles from the ancilla.
Since these terms are proportional to~$\tilde{\sigma}_{\qprojector \qprojector}$, which is initially zero and whose growth is suppressed by the Zeno effect, their presence does not change the following line of arguments.
Nonetheless, it is possible to get rid of these terms completely by keeping track of all the modes the ancilla decays to and post-selecting on the condition that the ancilla plus these additional modes are empty. To see this, we can assume that the ancilla decays only to a single mode with associated annihilation and creation operators~$c$ and $c^\dagger$.
The corresponding Lindblad operator~$L = c^\dagger b$ now conserves the number of particles in the ancilla plus the extra mode. Consequently, the contribution from the \enquote{recycling terms} to \cref{eq:masterequationprojectedpp,eq:masterequationprojectedpq,eq:masterequationprojectedqp} vanishes due to the action of the projector~$\pprojector$.

We now consider the strong noise limit of \cref{eq:masterequationprojected}.
The terms on the right-hand side of the equations for the coherences \eqref{eq:masterequationprojectedpq} and \eqref{eq:masterequationprojectedqp} rotate at characteristic frequencies of the unperturbed Hamiltonian~$H_0$ via $\tilde{a}(t) = \etothepowerof{i H_0 t / \hbar} a \etothepowerof{-i H_0 t / \hbar}$ as well as via the modulation function~$g(t)$, whose role is to probe dynamic correlations in the system at a given frequency. In contrast, the terms proportional to the dissipation rate~$\kappa$ cause a damping of the coherences. If $\kappa$ is sufficiently large, in particular, if it is much larger than the characteristic frequencies of $H_0$, we can make the approximation that the coherences are instantaneously damped to a momentary equilibrium state given by~$\diff \tilde{\sigma}_{\pprojector \qprojector} / \diff t \approx 0$ (and analogously for $\tilde{\sigma}_{\qprojector \pprojector}$).
This allows us to adiabatically eliminate the fast incoherent dynamics and to solve \cref{eq:masterequationprojectedpq,eq:masterequationprojectedqp} for the coherences. To leading order in $\Omega / \kappa$, we find
\begin{subequations}
	\label{eq:coherences}
\begin{align}
	\tilde{\sigma}_{\pprojector \qprojector} &= - \frac{i}{\hbar \kappa} \left( \tilde{H}_{\pprojector \qprojector} \tilde{\sigma}_{\qprojector \qprojector} - \tilde{\sigma}_{\pprojector \pprojector} \tilde{H}_{\pprojector \qprojector} \right) \left( \qprojector L^\dagger L \qprojector \right)^{-1} , \\
	\tilde{\sigma}_{\qprojector \pprojector} &= - \frac{i}{\hbar \kappa} \left( \qprojector L^\dagger L \qprojector \right)^{-1} \left( \tilde{H}_{\qprojector \pprojector} \tilde{\sigma}_{\pprojector \pprojector} - \tilde{\sigma}_{\qprojector \qprojector} \tilde{H}_{\qprojector \pprojector} \right) ,
\end{align}
\end{subequations}
where $\left( \cdots \right)^{-1}$ denotes the Moore--Penrose pseudoinverse.
To leading order in $\Omega / \kappa$, we can furthermore neglect the terms proportional to~$\tilde{\sigma}_{\qprojector \qprojector}$, which is initially zero and grows, according to \cref{eq:masterequationprojectedqq,eq:coherences}, only slowly at a rate $\Omega^2 / \kappa$. This suppression of the growth of population in the ancilla is precisely a manifestation of the Zeno effect.
Thus, plugging \cref{eq:coherences} into \cref{eq:masterequationprojectedpp}, we obtain
\begin{equation}
	\frac{\diff}{\diff t} \tilde{\sigma}_{\pprojector \pprojector} = - \frac{i}{\hbar} \Anticommutator{\tilde{H}_\mathrm{eff}(t)}{\tilde{\sigma}_{\pprojector \pprojector}}
\end{equation}
with the effective non-Hermitian Hamiltonian
\begin{equation}
	\tilde{H}_\mathrm{eff}(t) = -i g^2(t) \frac{\hbar \Omega^2}{\kappa} \pprojector \tilde{a}^\dagger(t) b \left( \qprojector L^\dagger L \qprojector \right)^{-1} b^\dagger \tilde{a}(t) \pprojector .
\end{equation}
Due to the action of the projector~$\pprojector$ in this expression, the pseudoinverse acts only on states with exactly one particle in the ancilla, where it reduces to a multiplication by unity. Thus, the effective non-Hermitian Hamiltonian takes the simple form $\tilde{H}_\mathrm{eff}(t) = -i g^2(t) \hbar \Omega^2 \pprojector \tilde{a}^\dagger(t) \tilde{a}(t) \pprojector / \kappa$.
Finally, \cref{eq:effectivenonhermitian} follows after transforming back to the non-rotating frame.

\section{\label{app:weakmeasurement}Connection between ancilla-based weak measurements of dynamical correlations and (non-)Hermitian linear response}

Ancilla-based weak measurement schemes for dynamical correlations can be adapted to probe either the unequal-time commutator or anti-commutator through a suitable choice of the ancilla state, the system--ancilla coupling, and the projective measurement performed on the ancilla~\cite{Uhrich2017,Kastner2018}.
It has been shown that those variants that probe the unequal-time commutator can be cast into an ancilla-free formulation~\cite{Kastner2018}, giving rise, e.g., to rotation-based protocols~\cite{Knap2013,Uhrich2017,Kastner2018,Uhrich2019,Schuckert2020}.
For weak perturbations, e.g., small rotation angles, these ancilla-free schemes correspond in fact to (standard) linear response.
By contrast, the interpretation of ancilla-based weak measurement protocols that target the unequal-time anti-commutator is far less obvious.
For instance, Ref.~\cite{Kastner2018} poses the question of whether an ancilla-free measurement of this quantity is possible in general.
Here, we show that, indeed, any ancilla-based weak measurement protocol for the unequal-time anti-commutator can be described in an ancilla-free way as a non-Hermitian linear response, exposing the close connection between these frameworks.

To this end, we consider a general ancilla-based weak measurement that uses only projective measurements of standard (Hermitian) operators on the ancilla.
The following derivation proceeds in analogy to the one for spin systems presented in Refs.~\cite{Uhrich2017,Kastner2018}, but here we consider a more general scenario: we do not specify the type of system, work with general mixed states, and consider arbitrary durations of the system--ancilla coupling.
Let us denote the initial state of system and ancilla by $\rho_\sys$ and $\rho_\anc$, respectively, and assume the combined system to be in a product state initially, $\rho_0 = \rho_\sys \otimes \rho_\anc$.
The target system evolves under the Hamiltonian~$H_0$, while we assume the ancilla to have no internal dynamics.
System and ancilla are coupled via the general coupling Hamiltonian
\begin{equation}
	\label{eq:weakmeasurementhamiltoniancoupling:app}
	H_\cpl(t) = f(t) A \otimes X
\end{equation}
with a time-dependent function~$f(t)$ and Hermitian operators~$A$ and $X$ acting on system and ancilla, respectively.
The total Hamiltonian of the combined system then reads $H(t) = H_0 \otimes \doublestruck{1} + H_\cpl(t)$.
It is convenient to work in the interaction picture, $\tilde{\rho}(t) = \etothepowerof{i H_0 t / \hbar} \rho(t) \etothepowerof{-i H_0 t / \hbar}$.
The von~Neumann equation
\begin{equation}
	i \hbar \frac{\diff}{\diff t} \tilde{\rho}(t) = \Commutator{\tilde{H}_\cpl(t)}{\tilde{\rho}(t)} .
\end{equation}
can equivalently be expressed in integral form as
\begin{equation}
\begin{split}
	\tilde{\rho}(t) &= \tilde{\rho}(0) - \frac{i}{\hbar} \int_0^t \diff t^\prime \Commutator{\tilde{H}_\cpl(t^\prime)}{\tilde{\rho}(t^\prime)} \\
	&\simeq \tilde{\rho}(0) - \frac{i}{\hbar} \int_0^t \diff t^\prime \Commutator{\tilde{H}_\cpl(t^\prime)}{\tilde{\rho}(0)} ,
\end{split}
\end{equation}
where $\tilde{H}_\cpl(t) = f(t) \tilde{A}(t) \otimes X$ is the interaction-picture coupling Hamiltonian with $\tilde{A}(t) = \etothepowerof{i H_0 t / \hbar} A \etothepowerof{-i H_0 t / \hbar}$.
In the last line, we have assumed the coupling to be sufficiently weak such that we can replace $\tilde{\rho}(t^\prime)$ in the integral, to linear order in~$H_\cpl$, by $\tilde{\rho}(0)$.
Note that the validity of this linear approximation is not necessarily restricted to short times~$t$, but can also be ensured for longer times by a sufficiently weak coupling strength~$f(t)$.

After a coupled evolution up to time~$t$, during which system and ancilla become entangled, we measure projectively the observable~$B \otimes Y$, where $B$ and $Y$ are Hermitian operators acting on system and ancilla, respectively, and post-select on the outcome of the ancilla measurement.
Although in practice system and ancilla are often measured simultaneously, it is instructive to treat this process as a consecutive measurement of the ancilla first and the system second.
Without loss of generality, we assume the observable~$Y$ to have a discrete spectrum of (real) eigenvalues~$\set{y}$. Let $\pprojector_y$ denote the projector on the eigenspace of the eigenvalue~$y$.
After obtaining this outcome, according to Lüders' rule~\cite{Lueders1950}, the state collapses to
\begin{equation}
	\tilde{\rho}_y(t) = \frac{1}{p(y)} \pprojector_y \tilde{\rho}(t) \pprojector_y ,
\end{equation}
where $p(y) = \Tr \left[ \pprojector_y \tilde{\rho}(t) \pprojector_y \right]$ is the probability of measuring the outcome~$y$.
For the coupling Hamiltonian~\eqref{eq:weakmeasurementhamiltoniancoupling:app}, the unnormalized post-measurement state reads
\begin{equation}
	\label{eq:weakmeasurementconditionalstateunnormalized}
\begin{split}
	\pprojector_y \tilde{\rho}(t) \pprojector_y {}={} &\rho_\sys \otimes \pprojector_y \rho_\anc \pprojector_y
	- \frac{i}{\hbar} \int_0^t \diff t^\prime \, f(t^\prime) \\
	&\times \left[ \tilde{A}(t^\prime) \rho_\sys \otimes \pprojector_y X \rho_\anc \pprojector_y - \mathrm{h.c.} \right] ,
\end{split}
\end{equation}
where $\mathrm{h.c.}$ denotes the Hermitian conjugate, while the probability of measuring~$y$ becomes
\begin{equation}
	\label{eq:weakmeasurementancillaprobability}
	p(y) = \Braket{\pprojector_y}_0 -i \int_0^t \diff t^\prime f(t^\prime) \Braket{A(t^\prime)}_0 \left[ \Braket{\pprojector_y X}_0 - \Braket{X \pprojector_y}_0 \right] .
\end{equation}
Here, $\braket{O(t)}_0$ denotes the expectation value of the Heisenberg operator~$O(t)$, evolving under the unperturbed Hamiltonian~$H_0$, with respect to the initial state~$\rho_0 = \rho_\sys \otimes \rho_\anc$.
Note that expectation values involving only ancilla operators are time independent since we assumed the ancilla to have no internal dynamics.
Using $(1 + x)^{-1} = 1 - x + \mathcal{O}(x^2)$, we obtain the normalized, conditional post-measurement state, to linear order in the coupling, as
\begin{multline}
	\tilde{\rho}_y(t) {}={} \rho_\sys \otimes \frac{\pprojector_y \rho_\anc \pprojector_y}{\Braket{\pprojector_y}_0}
	- \frac{i}{\hbar} \int_0^t \diff t^\prime \, f(t^\prime) \\
	\times \left\{ \left[ \tilde{A}(t^\prime) \rho_\sys \otimes \frac{\pprojector_y X \rho_\anc \pprojector_y}{\Braket{\pprojector_y}_0} - \Braket{A(t^\prime)}_0 \frac{\Braket{\pprojector_y X}_0}{\Braket{\pprojector_y}_0} \right] - \mathrm{h.c.} \right\} .
\end{multline}

Next, we are interested in the conditional expectation value of the system observable~$B$, given that the measurement of $Y$ on the ancilla yields the outcome~$y$.
In a first step, we trace out the ancilla,
\begin{equation}
\begin{split}
	\Tr_\anc \left[ \tilde{\rho}_y(t) \right] {}={} &\rho_\sys - \frac{i}{\hbar} \int_0^t \diff t^\prime \, f(t^\prime) \\
	&\times \left\{ \frac{\Braket{\pprojector_y X}_0}{\Braket{\pprojector_y}_0} \left[ \tilde{A}(t^\prime) - \Braket{A(t^\prime)}_0 \right] \rho_\sys - \mathrm{h.c.} \right\} .
\end{split}
\end{equation}
This yields the conditional expectation value
\begin{widetext}
\begin{equation}
	\label{eq:weakmeasurementconditionalexpectation:app}
\Braket{B(t)}_y = \Tr \left[ \tilde{B}(t) \tilde{\rho}_y(t) \right]
= \Braket{B(t)}_0 - \frac{i}{\hbar} \int_0^t \diff t^\prime f(t^\prime)
\left\{ \frac{\Braket{\pprojector_y X}_0}{\Braket{\pprojector_y}_0} \left[ \Braket{B(t) A(t^\prime)}_0 - \Braket{B(t)}_0 \Braket{A(t^\prime)}_0 \right] - \mathrm{c.c.} \right\} ,
\end{equation}
\end{widetext}
where $\mathrm{c.c.}$ denotes the complex conjugate.
With this result at hand, we can choose the ancilla state~$\rho_\anc$ as well as the ancilla operators~$X$ and $Y$ such that the integrand contains either the unequal-time commutator or the anti-commutator of the system observables~$A$ and $B$.
If $\braket{\pprojector_y X}_0 = \Tr [\pprojector_y X \rho_\anc]$ is real, i.e., $\braket{\pprojector_y X}_0 / \braket{\pprojector_y}_0 = -s$ with $s \in \doublestruck{R}$, \cref{eq:weakmeasurementconditionalexpectation:app} gives access to the unequal-time commutator,
\begin{equation}
	\label{eq:weakmeasurementcommutator}
	\Braket{B(t)}_y = \Braket{B(t)}_0 + \frac{i}{\hbar} s \int_0^t \diff t^\prime f(t^\prime) \Braket{\Commutator{B(t)}{A(t^\prime)}}_0 .
\end{equation}
This expression coincides with Kubo's linear response formula (cf.~\cref{eq:linearresponse,eq:responsefunctionhermitian}) up to a constant factor in the response function.
There are two special cases worth discussing.
First, if $X = \doublestruck{1}$, $\braket{\pprojector_y X}_0 = \braket{\pprojector_y}$ is always real and the scheme always yields the unequal-time commutator. This is not surprising: for $X = \doublestruck{1}$, system and ancilla always remain in a product state and the coupling Hamiltonian~\eqref{eq:weakmeasurementhamiltoniancoupling:app} corresponds to a Hermitian perturbation on the target system only, which is exactly the linear response scenario.
Second, it is instructive to consider the unconditional expectation value~$\braket{B(t)} = \sum_y \Braket{B(t)}_y p(y)$, which corresponds to not measuring the ancilla at all or disregarding the outcome of the ancilla measurement.
By combining \cref{eq:weakmeasurementancillaprobability,eq:weakmeasurementconditionalexpectation:app}, and using the completeness relation~$\sum_y \pprojector_y = \doublestruck{1}$, we find, to linear order,
\begin{equation}
	\Braket{B(t)} = \Braket{B(t)}_0 - \frac{i}{\hbar} \Braket{X}_0 \int_0^t \diff t^\prime f(t^\prime) \Braket{\Commutator{B(t)}{A(t^\prime)}}_0 ,
\end{equation}
which again always yields the unequal-time commutator.
These two examples illustrate two essential ingredients for extracting the unequal-time anti-commutator from ancilla-based weak measurements: firstly, the coupling must entangle system and ancilla, and secondly, it is necessary to correlate the measurement on the target system in some way with the outcome of the ancilla measurement, e.g., through post-selection.

In order to extract the unequal-time anti-commutator from \cref{eq:weakmeasurementconditionalexpectation:app}, $\braket{\pprojector_y X}_0$ must be purely imaginary, i.e., $\braket{\pprojector_y X}_0 / \braket{\pprojector_y}_0 = -i s$ with $s \in \doublestruck{R}$, yielding
\begin{equation}
	\label{eq:weakmeasurementanticommutator}
\begin{split}
	\Braket{B(t)}_y {}={} &\Braket{B(t)}_0 - \frac{1}{\hbar} s \int_0^t \diff t^\prime f(t^\prime) \\
	&\times \left[ \Braket{\Anticommutator{B(t)}{A(t^\prime)}}_0 - 2 \Braket{B(t)}_0 \Braket{A(t^\prime)}_0 \right] .
\end{split}
\end{equation}
This expression corresponds, up to a constant factor in the response function, directly to the non-Hermitian linear response scenario in \cref{eq:linearresponsenonhermitian,eq:responsefunctionnonhermitian}.

\Cref{eq:weakmeasurementcommutator,eq:weakmeasurementanticommutator} demonstrate the fact that any ancilla-based weak measurement designed to probe the (anti\nobreakdash-)commutator can effectively be described as a (non\nobreakdash-)Hermitian linear response.
To make this connection even more explicit, we trace out the ancilla in the unnormalized post-measurement state~\eqref{eq:weakmeasurementconditionalstateunnormalized},
\begin{multline}
	\Tr_\anc \left[ \pprojector_y \tilde{\rho}(t) \pprojector_y \right] \\
	= \Braket{\pprojector_y}_0 \rho_\sys - \frac{i}{\hbar} \int_0^t \diff t^\prime f(t^\prime) \left[ \Braket{\pprojector_y X}_0 \tilde{A}(t^\prime) \rho_\sys - \mathrm{h.c.} \right] .
\end{multline}
In analogy to Ref.~\cite{Kastner2018}, to linear order, this result can be re-written in terms of generalized measurement (or Kraus) operators~\cite{Wiseman2009,Nielsen2010} as
\begin{equation}
	\Tr_\anc \left[ \pprojector_y \tilde{\rho}(t) \pprojector_y \right] = M_y \rho_\sys M_y^\dagger
\end{equation}
with
\begin{equation}
	\label{eq:weakmeasurementkrausoperator}
	M_y = \sqrt{\Braket{\pprojector_y}_0} \exp \left\{ -\frac{i}{\hbar} \frac{\Braket{\pprojector_y X}_0}{\Braket{\pprojector_y}_0} \int_0^t \diff t^\prime \, f(t^\prime) \tilde{A}(t^\prime) \right\} .
\end{equation}
It is easy to verify that, to linear order, these operators fulfill the completeness relation $\sum_y M_y^\dagger M_y = \doublestruck{1}$.
The measurement operator~$M_y$ describes the effect of the system--ancilla coupling, conditioned on the outcome~$y$ of the ancilla measurement, without explicitly referencing the ancilla.
This ancilla-free description corresponds to the evolution under the effective Hamiltonian~$H_\eff(t) = H_0 + H_1(t)$ with
\begin{equation}
	H_1(t) = \frac{\Braket{\pprojector_y X}_0}{\Braket{\pprojector_y}_0} f(t) A .
\end{equation}
For real $\Braket{\pprojector_y X}$, this evolution is unitary~\cite{Kastner2018} and corresponds to standard (Hermitian) linear response, giving access to the unequal-time commutator.
Remarkably, the case of purely imaginary $\Braket{\pprojector_y X}$, which according to \cref{eq:weakmeasurementanticommutator} probes the unequal-time anti-commutator, corresponds to an anti-Hermitian perturbation and maps directly to the non-Hermitian linear response scenario described in \cref{sec:nhlrt}.
This shows that for any ancilla-based weak measurement of dynamical correlations there is a corresponding ancilla-free linear response description.

Conversely, any linear response protocol can, at least in principle, be realized via an ancilla-based weak measurement.
While obvious for standard (Hermitian) linear response, in the non-Hermitian case, the challenge is to choose the ancilla state as well as the ancilla operators~$X$ and $Y$ appropriately such that $\Braket{\pprojector_y X} \in i \doublestruck{R}$.
To see that this is always possible in general, let $Y$ be any Hermitian operator on a (complex) Hilbert space of dimension two or higher with at least two distinct eigenvalues~$y_1$ and $y_2$. Then, take the ancilla state to be the equal superposition of the corresponding eigenstates, $\ket{\phi} = (\ket{y_1} + \ket{y_2}) / \sqrt{2}$, with $\rho_\anc = \ketbra{\phi}{\phi}$. Now, let $\pprojector_y = \pprojector_{y_1}$ be the projector on the eigenspace with eigenvalue~$y_1$ and set $X = -i (\ketbra{y_1}{y_2} - \ketbra{y_2}{y_1})$. Then, we have $\braket{\pprojector_y X}_0 / \braket{\pprojector_y}_0 = -i$, as desired.
Clearly, for a given anti-Hermitian perturbation, the choice of $\rho_\anc$, $X$, and $Y$ is not unique, and the challenge consists in finding the configuration that is most convenient for the desired application.
Appropriate choices for spin systems have been discussed, for instance, in Refs.~\cite{Uhrich2017,Kastner2018}, but their experimental realization on other platforms, e.g., bosons in optical lattices, is unfortunately not straightforward.
To illustrate this, assume we are interested in perturbations by the density operator~$A = n$ and consider the above scenario of achieving purely imaginary~$\braket{\pprojector_y X}_0$ for a bosonic ancilla, which translates to $Y = b^\dagger b$ (measurement of the occupancy), $X = -i (b - b^\dagger)$, and $\ket{\phi} = (\ket{0} + \ket{1}) / \sqrt{2}$.
However, neither the superposition of Fock states~$\ket{\phi}$ nor the coupling Hamiltonian~$H_\cpl \propto n \otimes X$, which would be cubic in boson operators, can be realized with massive, non-relativistic particles.
More generally, in order to probe unequal-time anti-commutators involving the density~$A = n$ through a particle number measurement~$Y = b^\dagger b$ on a bosonic ancilla, the operator $X$ cannot be diagonal in the Fock basis, as this would imply $\braket{P_y X} \in \doublestruck{R}$, regardless of the ancilla state.
In other words, a particle number non-conserving coupling Hamiltonian would be required in such a setting.
In our proposal of \cref{sec:realization}, this difficulty does not arise because the leading contribution to the response is quadratic in the coupling Hamiltonian, which enables non-Hermitian perturbations for a wide range of observables including densities and correlators with experimentally feasible system--ancilla couplings.

All in all, we have established a general connection between ancilla-based weak measurement protocols for dynamical correlations and (non\nobreakdash-)Hermitian linear response theory.
In particular, our results pave the road to measuring the left-hand side of the \ac{fdr}~\eqref{eq:FDR}, i.e., the unequal-time anti-commutator, via non-Hermitian linear response in an ancilla-free fashion, harnessing the rapidly developing toolbox of non-Hermitian physics~\cite{ElGanainy2018,Ashida2020}.

\section{\label{app:projection_vs_nhlr}Comparison between projective protocols and non-Hermitian linear response}

In this Appendix, we assess to what extent projective protocols~\cite{Knap2013,Uhrich2017,Kastner2018,Uhrich2019,Schuckert2020} represent a good alternative for measuring unequal-time anti-commutators of observables that are not strictly dichotomic.
To this end, we first formulate the protocol for general observables~$A$ and $B$, where $A$ has precisely two eigenvalues.
We then investigate with the help of numerical benchmarks at the example of the Bose--Hubbard model how well the scheme reproduces the exact density autocorrelation spectrum in comparison to non-Hermitian linear response as a function of both the filling and the on-site interaction.

\subsection{General projective protocol}

We begin by briefly reviewing the projective protocol for measuring unequal-time anti-commutators~\cite{Knap2013,Uhrich2017,Kastner2018,Uhrich2019,Schuckert2020}.
Here, we formulate the scheme for a general Hermitian operator~$A$ with two distinct eigenvalues~$a_1, a_2 \in \doublestruck{R}$.
Let $\pprojector_1$ and $\pprojector_2$ be the projection operators on the corresponding eigenspaces such that $A = a_1 \pprojector_1 + a_2 \pprojector_2$ with $\pprojector_1 + \pprojector_2 = \doublestruck{1}$.
This allows us to express both projectors entirely through the operator~$A$ and the known eigenvalues,
\begin{subequations}
	\label{eq:projectordichotomic}
\begin{align}
	\label{eq:projectordichotomic:1}
	\pprojector_1 &= \frac{1}{a_1 - a_2} \left( A - a_2 \doublestruck{1} \right) , \\
	\label{eq:projectordichotomic:2}
	\pprojector_2 &= \frac{1}{a_2 - a_1} \left( A - a_1 \doublestruck{1} \right) ,
\end{align}
\end{subequations}
which would not be possible if $A$ had more than two eigenvalues.
The protocol starts by evolving the initial state~$\rho_0$ (under the target Hamiltonian~$H_0$) to the waiting time~$\waitingtime$.
Then, the observable~$A$ is measured projectively and the state is conditioned on the outcome~$a_1$ or $a_2$ of this measurement, yielding the conditional post-measurement states
\begin{equation}
	\rho(\waitingtime | a_i) = \frac{1}{p(a_i)} \pprojector_i \rho(\waitingtime) \pprojector_i
\end{equation}
with $i \in \set{1, 2}$, where $p(a_i) = \Tr [ \pprojector_i \rho(\waitingtime) \pprojector_i ]$ is the probability of obtaining the measurement outcome~$a_i$ at time~$\waitingtime$.
Subsequently, the conditional state is evolved to the final observation time~$t \ge \waitingtime$.
The conditional expectation values of an observable~$B$ then reads
\begin{multline}
	\Braket{B(t)}_{a_i} = \Tr \left[ B \rho(t | a_i, \waitingtime) \right] \\
	= \frac{\Braket{ A(\waitingtime) B(t) A(\waitingtime) - a_j \Anticommutator{B(t)}{A(\waitingtime)} + a_j^2 B(t) }}{p(a_i) (a_i - a_j)^2} ,
\end{multline}
with $(i, j) \in \set{(1, 2), (2, 1)}$, where we have switched to the Heisenberg picture.
Solving for the unequal-time anti-commutator, we obtain
\begin{multline}
	\label{eq:projectiveprotocolgeneral}
	\Braket{\Anticommutator{B(t)}{A(\waitingtime)}} = \Braket{B(t)} \left( a_1 + a_2 \right) \\
	+ \left[ \Braket{B(t)}_{a_1} p(a_1) - \Braket{B(t)}_{a_2} p(a_2) \right] \left( a_1 - a_2 \right) .
\end{multline}
The probabilities~$p(a_i)$ can be expressed through the expectation value~$\braket{A(\waitingtime)}$ with the help of \cref{eq:projectordichotomic}, yielding
\begin{multline}
	\label{eq:projectiveprotocolgeneral:expectation}
	\Braket{\Anticommutator{B(t)}{A(\waitingtime)}} = \Braket{B(t)}_{a_1} \left[ \Braket{A(\waitingtime)} - a_2 \right] \\
	+ \Braket{B(t)}_{a_2} \left[ \Braket{A(\waitingtime)} - a_1 \right] + \Braket{B(t)} \left( a_1 + a_2 \right) .
\end{multline}
This result states that the desired unequal-time anti-commutator of $A$ and $B$ can be extracted from a measurement of the unconditional expectation value~$\braket{B(t)}$ (without previous projective measurement) as well as the conditional expectation values~$\braket{B(t)}_{a_1}$ and $\braket{B(t)}_{a_2}$, given that the outcomes~$a_1$ and $a_2$ have been obtained from the projective measurement of $A$ at the waiting time~$\waitingtime$, respectively.

A few remarks are in order.
The projective measurement of $A$ at time~$\waitingtime$ can be deferred up to the final observation time~$t$ with the help of an ancilla using shelving techniques~\cite{Schuckert2020}.
This way, the need for non-destructive measurements can be avoided.
Furthermore, it is worth emphasizing that there is no restriction on the number of eigenvalues of the operator~$B$, i.e., the dichotomic constraint applies only to $A$.

\subsection{Numerical benchmark: projective protocols versus non-Hermitian linear response}

We now specialize the projective protocol in \cref{eq:projectiveprotocolgeneral:expectation} to density correlations in a Bose--Hubbard system.
In the hard-core limit~$U / J \to \infty$, multiple occupancies of the same lattice site are prohibited.
The density~$n_\ell$ at site~$\ell$ then becomes a dichotomic observable with only two eigenvalues~$0$ and $1$.
We thus recover the protocol reported in Ref.~\cite{Schuckert2020},
\begin{equation}
	\label{eq:projectiveprotocoldensity}
\begin{split}
	\Braket{\Anticommutator{n_{\ell_2}(t)}{n_{\ell_1}(\waitingtime)}} &= \Braket{n_{\ell_2}(t)} + \Braket{n_{\ell_2}(t)}_1 \Braket{n_{\ell_1}(\waitingtime)} \\
	&\hphantom{{}={}} - \Braket{n_{\ell_2}(t)}_0 \left( 1 - \Braket{n_{\ell_1}(\waitingtime)} \right) .
\end{split}
\end{equation}

For soft-core bosons, \cref{eq:projectiveprotocoldensity} does not hold in general since the density operator~$A = n_{\ell_1}$ can take more than two eigenvalues.
However, the projective protocols in \cref{eq:projectiveprotocolgeneral,eq:projectiveprotocolgeneral:expectation} can still be used to measure the \emph{exact} unequal-time anti-commutator for an arbitrary observable~$B$ and any dichotomic observable~$A$.
For instance, a possible choice of $A$ is the parity~$\Pi_\ell$ of the particle number at site~$\ell$, which in conventional quantum gas microscopes is even more easily accessible than the density itself due to pairwise atom loss caused by the near-resonant imaging light~\cite{Schaefer2020}.
If we associate the eigenvalues $a_{\mathrm{even}} = 0$ and $a_{\mathrm{odd}} = 1$ with even and odd parity, respectively, the operator~$\Pi_\ell = a_{\mathrm{even}} \pprojector_{\ell, \mathrm{even}} + a_{\mathrm{odd}} \pprojector_{\ell, \mathrm{odd}}$ coincides with the density in the hard-core limit.
Thus, in the regime where multiple occupancies can be neglected, we can approximate the density--density anti-commutator in \cref{eq:projectiveprotocoldensity} by the (exactly obtainable) quantity~$\braket{\anticommutator{n_{\ell_2}(t)}{\Pi_{\ell_1}(\waitingtime)}}$.

An alternative strategy to approximate $\braket{\anticommutator{n_{\ell_2}(t)}{n_{\ell_1}(\waitingtime)}}$ for soft-core bosons is to take \cref{eq:projectiveprotocoldensity} literally and compute the conditional expectation values~$\braket{n_{\ell_2}(t)}_0$ and $\braket{n_{\ell_2}(t)}_1$ from only those realizations where the projective measurement of $n_{\ell_1}(\waitingtime)$ yields the outcomes $0$ and $1$, respectively, discarding realizations with higher occupancies.
By contrast, $\braket{n_{\ell_2}(t)}$ still represents the (full) unperturbed expectation value.
This way, the asymptotic behavior of the unequal-time anti-commutator for $t \gg \waitingtime$ is correctly reproduced: two local observables $A$ and $B$ typically become uncorrelated in an ergodic system after sufficiently long times and the anti-commutator reduces to the disconnected product~$2 \braket{B(t)} \braket{A(\waitingtime)}$.
For the Bose--Hubbard model, the conditional expectation values of the local densities in \cref{eq:projectiveprotocoldensity} are expected to eventually re-equilibrate to their unperturbed value~$\braket{n_{\ell_2}(t)}$, such that the right-hand side indeed becomes $2 \braket{n_{\ell_2}(t)} \braket{n_{\ell_1}(\waitingtime)}$.
As long as the system is sufficiently close to the hardcore limit, we can expect \cref{eq:projectiveprotocoldensity} to reproduce the unequal-time anti-commutator for any $t \ge \waitingtime$ to good accuracy.
In what follows, we analyze how well this approximation works for on-site densities ($B = A = n_\ell$) as a function of the filling and the on-site interaction.

\begin{figure}
	\subfloat{\label{fig:projectionVSnhlr_lowfilling:probability_time}}%
	\subfloat{\label{fig:projectionVSnhlr_lowfilling:probability_particles}}%
	\subfloat{\label{fig:projectionVSnhlr_lowfilling:anticommutator}}%
	\subfloat{\label{fig:projectionVSnhlr_lowfilling:spectrum}}%
	\subfloat{\label{fig:projectionVSnhlr_lowfilling:errorrel}}%
	\subfloat{\label{fig:projectionVSnhlr_lowfilling:errorabs}}%
	\includegraphics[width=\columnwidth]{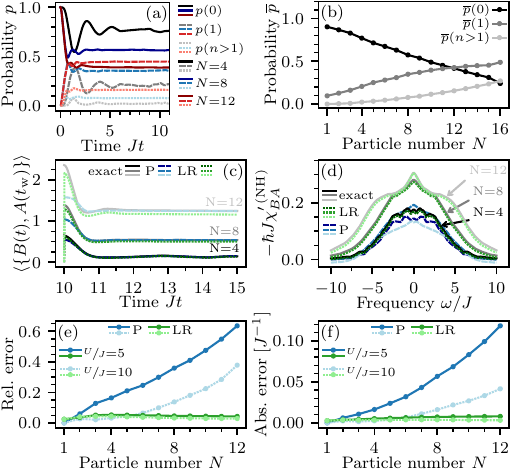}
	\caption{\label{fig:projectionVSnhlr_lowfilling}%
		Comparison of the projective protocol (P) and non-Hermitian linear response (LR) for extracting the unequal-time anti-commutator of the on-site density ($B = A = n_{2, 2}$) in a $4 \times 4$ Bose--Hubbard system with on-site interaction~$U / J = 5$ as a function of the filling.
		(a)~Time trace of the probability~$p(n)$ of finding $n$ particles at site~$(2, 2)$.
		(b)~Time-averaged probability~$\mean{p}(n)$ as a function of the particle number~$N$. For small $N$, higher occupancies are negligible and the on-site density~$n_{2, 2}$ is approximately dichotomic.
		(c)~Time trace of the unequal-time anti-commutator and (d)~correlation spectrum extracted from simulations of the different measurement schemes at $J \waitingtime = 10$.
		(e)~Relative error and (f)~absolute error of the correlation spectra in (d) with respect to the exact results.
		The projective protocol (P) yields good accuracy at low fillings where multiple occupancies are suppressed, but fails as the filling approaches unity.
		Increasing the on-site interaction~$U$ extends the regime of validity.
		The non-Hermitian linear response scheme (LR) performs well irrespective of the filling and the value of $U / J$.%
	}
\end{figure}

In \cref{fig:projectionVSnhlr_lowfilling}, we compare the performance of the projective protocol in \cref{eq:projectiveprotocoldensity} to that of the non-Hermitian linear response scheme discussed in \cref{sec:illustration} for a \ac{2d} Bose--Hubbard system as a function of the filling.
To this end, we vary the number of particles~$N$ on a square lattice with open boundary conditions consisting of $4 \times 4$ sites, labeled by a pair of indices~$(\ell_x, \ell_y)$ with $\ell_x, \ell_y \in \set{1, \dots, 4}$.
We initialize the system in a single Fock state where the particles are distributed to maximize their mutual distances without initially occupying the interior site~$(2, 2)$, at which we probe the density correlations.
\Cref{fig:projectionVSnhlr_lowfilling:probability_time} shows the probability~$p(n)$ of finding zero, one, or more than one particle at the probe site for $U / J = 5$ as a function of time.
The initial oscillations quickly damp and the probabilities become approximately stationary.
In \cref{fig:projectionVSnhlr_lowfilling:probability_particles}, we show the probability~$\mean{p}(n) = t^{-1} \int_0^t \diff t^\prime \, p(n, t^\prime)$, time-averaged up to time~$J t = 10$, as a function of the particle number~$N$.
For small~$N$, higher occupancies~$n > 1$ can be neglected and the density operator at the probe site is approximately dichotomic.
\Cref{fig:projectionVSnhlr_lowfilling:anticommutator,fig:projectionVSnhlr_lowfilling:spectrum} show, respectively, the time trace of the unequal-time anti-commutator ($B = A = n_{2, 2}$) and the reactive part of the non-Hermitian dynamic susceptibility~$\chiprimenh$ (correlation spectrum) at the waiting time $J \waitingtime = 10$ for several values of~$N$.
The exact results are compared to those extracted using the projective protocol and the non-Hermitian linear response scheme.
For the latter, we have used a rectangular pulse of duration~$J \delta t = \num{0.01}$ and a perturbation strength~$s = 0.05$ as in \cref{sec:illustration}.
The Fourier integral in \cref{eq:susceptibilityfixedcentraltime} has been computed using an exponential filter of characteristic frequency~$\gamma / J = 0.2$ (see \cref{app:quench:technical}).
While for small~$N$ the projective protocol correctly reproduces both the exact time trace and the exact spectrum, there are sizable deviations as the number of particles~$N$ (and thus the contribution of higher occupancies) grows.
By contrast, the non-Hermitian linear response scheme reproduces the exact results to good accuracy regardless of the filling.
In \cref{fig:projectionVSnhlr_lowfilling:errorrel,fig:projectionVSnhlr_lowfilling:errorabs}, we show, respectively, the $L^2$ norm of the relative error~$\lVert \chiprimenh_{\mathrm{sim}} - \chiprimenh_{\mathrm{exact}} \rVert_2 / \lVert \chiprimenh_{\mathrm{exact}} \rVert_2$ and the absolute error~$\lVert \hbar \chiprimenh_{\mathrm{sim}} - \hbar \chiprimenh_{\mathrm{exact}} \rVert_2$ of the spectra extracted from the simulated measurement protocols in \cref{fig:projectionVSnhlr_lowfilling:spectrum}.
For the projective protocol, both errors increase with increasing particle number, while the errors remain small for the non-Hermitian linear response scheme.
For larger on-site interactions~$U$, higher occupancies are suppressed more strongly, which delays the rise of the error curve for the projective protocol as the filling increases: given a certain acceptable tolerance for the relative error of, say, less than $\SI{20}{\percent}$, the projective protocol for $U / J = 5$ ($U / J = 10$) yields acceptable results for up to $N = 4$ ($N = 9$) particles.

\begin{figure}
	\subfloat{\label{fig:projectionVSnhlr_unitfilling:probability_time}}%
	\subfloat{\label{fig:projectionVSnhlr_unitfilling:probability_interaction}}%
	\subfloat{\label{fig:projectionVSnhlr_unitfilling:anticommutator}}%
	\subfloat{\label{fig:projectionVSnhlr_unitfilling:spectrum}}%
	\subfloat{\label{fig:projectionVSnhlr_unitfilling:errorrel}}%
	\subfloat{\label{fig:projectionVSnhlr_unitfilling:errorabs}}%
	\includegraphics[width=\columnwidth]{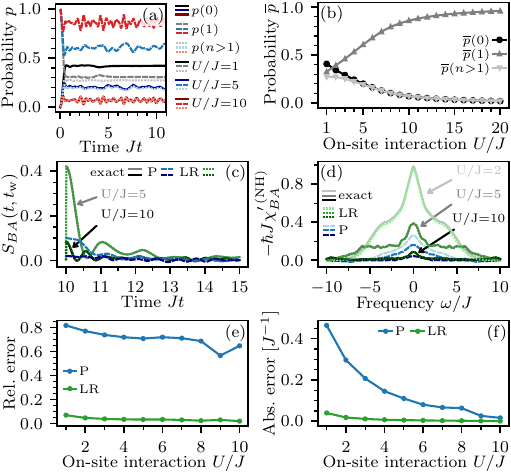}
	\caption{\label{fig:projectionVSnhlr_unitfilling}%
		Same as \protect\cref{fig:projectionVSnhlr_lowfilling}, but for a \ac{1d} Bose--Hubbard system at unit filling as a function of the on-site interaction~$U$.
		Despite the suppression of higher occupancies at large~$U$ (b), the relative error~(e) of the projective protocol~(P) remains sizable.
		By contrast, the non-Hermitian linear response scheme (LR) yields good results for any value of $U$.%
	}
\end{figure}

We now investigate the performance of projective protocols for the scenario in \cref{sec:illustration}, i.e., a quench in a \ac{1d} Bose--Hubbard chain of length~$L = 12$ with periodic boundary conditions at unit filling, initially prepared in a Mott-insulating state.
Since $\braket{n_\ell(t)} \equiv 1$ in this case, the projective protocol in \cref{eq:projectiveprotocoldensity} reduces to $\braket{\anticommutator{n_{\ell_2}(t)}{n_{\ell_1}(\waitingtime)}} \approx 1 + \braket{n_{\ell_2}(t)}_1$.
If we evaluate this expression at $t = \waitingtime$, the right-hand side takes the value~$2$ and therefore the connected anti-commutator extracted from the projective protocol vanishes.
This behavior is qualitatively different from that of the true anti-commutator, which is maximal at $t = \waitingtime$.
Consequently, \cref{eq:projectiveprotocoldensity} represents a rather poor approximation of the unequal-time anti-commutator in this scenario, especially at small $U / J$.
To obtain a slightly better approximation, we resort to the projective protocol in \cref{eq:projectiveprotocolgeneral}, which is no longer equivalent to \cref{eq:projectiveprotocolgeneral:expectation} if $A$ has more than two eigenvalues.
However, unlike \cref{eq:projectiveprotocolgeneral:expectation}, the protocol in \cref{eq:projectiveprotocolgeneral} does not reproduce the correct asymptotic behavior of the anti-commutator for $t \gg \waitingtime$ if $A$ is not dichotomic.
This can be fixed by replacing $\braket{B(t)} (a_1 + a_2) $ on the right-hand side by $\alpha \braket{B(t)}$ with $\alpha = 2 \braket{A(\waitingtime)} - (a_1 - a_2) [p(a_1) - p(a_2)]$.
Since $\braket{B(t)}$ is usually stationary in the regime of interest, this replacement merely contributes a constant offset to the time trace of the anti-commutator, which ensures $\braket{\anticommutator{B(t)}{A(\waitingtime)}} \to 2 \braket{B(t)} \braket{A(\waitingtime)}$ for $t \gg \waitingtime$ and avoids spurious static peaks in the correlation spectrum.

In \cref{fig:projectionVSnhlr_unitfilling}, we present a similar analysis as in \cref{fig:projectionVSnhlr_lowfilling} for the \ac{1d} system at unit filling as a function of the on-site interaction~$U$.
In \cref{fig:projectionVSnhlr_unitfilling:probability_interaction}, it can be seen be seen that there is a significant contribution from states with higher occupancies at small on-site interactions~$U / J$.
As expected, the projective protocol does not perform well in this regime, while the non-Hermitian linear response scheme yields good results.
As we move to larger $U / J$, we enter the Mott-insulating regime where single occupancies dominate and the dynamics is governed by particle--hole excitations~\cite{Huber2007}.
Although the probability of higher occupancies~$p(n > 1)$ diminishes with increasing $U / J$, its contribution remains on the same order as that of the probability for vacancies~$p(0)$.
Thus, the density is nowhere well approximated by a dichotomic observable since the neglected higher occupancies (particle excitations on top of the Mott insulator) are of equal importance as vacancies (hole excitations).
This explains why the absolute error of the projective protocol in \cref{fig:projectionVSnhlr_unitfilling:errorabs} decreases substantially with increasing $U / J$, while the relative error in \cref{fig:projectionVSnhlr_unitfilling:errorrel} decreases only slowly and remains comparatively large even at large $U / J$.
We have checked that the error behaves similarly if we approximate the density--density unequal-time anti-commutator by replacing $A = n_{\ell_1}$ with the parity $\Pi_{\ell_1}$, as discussed above.
Thus, as opposed to non-Hermitian linear response, projective protocols are not well suited for probing unequal-time anti-commutators and the associated \acp{fdr} for densities at unit filling.

Our numerical benchmarks suggest that projective protocols generally work well at low fillings and large on-site interactions, where multiple occupancies can be neglected.
However, they do not represent a good alternative to measure unequal-time anti-commutators and \acp{fdr} in regimes where the relevant observables are not approximately dichotomic.
In our example of the Bose--Hubbard model, this limitation unfortunately applies to a major part of the physical parameter space, including the relevant setting of a system at unit filling and moderate on-site interaction strengths.
In order to explore these regimes of interest, we must therefore resort to alternative methods like non-Hermitian linear response, which performs well across the entire parameter space.

\bibliography{references}

\end{document}